\documentclass[a4paper, 11pt]{article}
\usepackage{CJKutf8}
\usepackage[T1]{fontenc}
\usepackage[utf8]{inputenc}
\usepackage{jheppub}
\usepackage{graphicx}
\usepackage{subcaption}
\usepackage{pifont}
\usepackage{cancel}
\usepackage{slashed}
\usepackage{soul}
\usepackage{tikz}
\usepackage{mathalfa,amssymb}
\usepackage{bbm}
\usepackage{bbold}
\usepackage{comment}
\usepackage{amsmath}
\usepackage{mathtools}
\usepackage{amsthm}
\usepackage{xcolor}
\usepackage[english]{babel}
\usepackage{multirow}
\usepackage{enumitem}
\usepackage[toc,page]{appendix}
\usepackage{hyperref}

\hypersetup{
	colorlinks,
	citecolor=blue,
    filecolor=black,
    linkcolor=blue }

\def\l{\left(}
\def\r{\right)}
\def\p{\partial}
\def\d{\delta}

\def\1{\text{\scalebox{\ss}{(1)}}}
\def\scri{\mathcal{I}}

\def\o{\omega}

\def\dt{\text{d}}

\def\dt{\text{d}}

\def\dr{\text{dr}}
\def\mc{\mathcal }

\newcommand{\pah}[1]{{\color{black}{{#1}}}}

\title{Longitudinal electric corner modes in continuum quantum gauge theory with matter}

\author[a,b]{Gon\c{c}alo Ara\'ujo-Regado,} 
\author[a]{Philipp A.\ H\"ohn,}
\author[c]{Alok Laddha,}
\author[a]{and Bilyana Tomova}

\affiliation[a]{Qubits and Spacetime Unit, Okinawa Institute of Science and Technology Graduate University (\begin{CJK}{UTF8}{min}沖縄科学技術大学院大学\end{CJK}), Onna, Okinawa 904 0495, Japan}
\affiliation[b]{University of Vienna, Faculty of Physics, Boltzmanngasse 5, 1090 Vienna, Austria}
\affiliation[c]{Chennai Mathematical Institute H1, SIPCOT IT Park, Siruseri, Kelambakkam 603103, India}

\emailAdd{goncalo.araujo@oist.jp}
\emailAdd{philipp.hoehn@oist.jp}
\emailAdd{oranjok@gmail.com}
\emailAdd{b.tomova@oist.jp}

\abstract{Subregions in gauge theories give rise to edge modes and corner charges supported on their boundaries. While this is well-established classically and in quantum lattice gauge theories, the fate of corner modes remains poorly understood in continuum quantum field theories owing to their distributional nature. In this paper, we revisit this challenge and propose a new corner operator for a regional Cauchy slice algebra generated by local gauge-invariant observables of scalar QED in Minkowski spacetime. The corner mode is given by the longitudinal electric field sourced by the matter outside the subregion and thus constitutes a centre for the regional algebra. Our main goal is to show that there exists a perturbative GNS state on this algebra that quantises this mode. Pulling back the Fock vacuum to the slice, we construct this state by renormalising the divergences arising from the lack of temporal smearing. We clarify that the inclusion of the new centre mode necessitates a further renormalisation \emph{on the corner} of the region, thereby physically distinguishing it and breaking the translation invariance of the state. Our work shows how the requirements of locality, renormalisability, and positivity together put strong constraints on the existence of a continuum quantum field theory which supports corner modes as genuine quantum operators. We finish with comments about the von Neumann completion of the regional algebra, as well as possible physical interpretations of the renormalised vacuum state.}

\begin{document}
\maketitle

\section{Introduction}

There is by now a vast literature on the nature of so-called ``edge modes'' in gauge theories. The construction of edge (or corner) modes started with the observation in \cite{Buividovich:2008gq, Donnelly:2011hn,Casini:2013rba, Donnelly:2014gva, Ghosh:2015iwa, Donnelly:2016auv, Soni:2015yga} that gauge symmetry places new constraints on the partitioning of a system into local subsystems across spatial cuts, leading to features like local superselection sectors labeled by the eigenvalues of corner supported charges, which commute with all the gauge-invariant observables inside the subregion. It was argued in \cite{Donnelly:2014fua, Donnelly:2015hxa} that taking into account the contribution from this superselection structure to the entanglement entropy of a subregion leads to a match with the contact term obtained from Euclidean methods \cite{Kabat:1995eq, Callan:1994py}\footnote{This is necessary for consistency with black hole entropy and thermodynamics \cite{PhysRevD.50.2700, Jacobson:1994iw, Iyer:1995kg, Larsen:1995ax, Solodukhin:2011gn, Fursaev:1994ea}.} and with the conformal anomaly of the 4D Maxwell field appearing in the logarithmically divergent term of the entropy \cite{Dowker:2010bu}. 

Generically, in Lorentz-invariant four-dimensional continuum gauge theories, these corner supported charges have been rigorously defined  only classically \cite{Donnelly:2016auv,Geiller:2019bti,Carrozza:2021gju,Araujo-Regado:2024dpr,Ball:2024hqe,Gomes:2016mwl,Riello:2021lfl,Gomes:2018dxs,Gomes:2019xto,Hoehn:2025pmx,Shimizu:2026zug,Shimizu:2026wjz,Ball:2024gti,Borsboom:2026xrd}. In the quantum theory, they have been  studied as operators mostly in the context of lattice gauge theories \cite{Casini:2013rba,Soni:2015yga,Delcamp:2016eya,Araujo-Regado:2025ejs,VanAcoleyen:2015ccp}. The main reason is that, in the continuum theory, ``operators'' obtained from smearing the elementary fields on a lower-dimensional spatial surface have divergent correlation functions even in free theory and hence, cannot be realised as well-defined quantum operators in a local QFT \cite{Haag:1992hx,streater2016pct,halvorson2006algebraic}.  Due to this obstruction, local corner charge operators are also not well-defined as they live on a spatial codimension-two surface. Thus, they are generically not part of the local algebra of gauge-invariant operators. The local algebra in the continuum therefore contains no central elements that commute with all others and thereby could define superselection sectors (more precisely, the local algebra is thus a type III$_1$ \emph{factor} \cite{Yngvason:2004uh,Haag:1992hx}).

Although an algebraic definition of local centre operators for theories with localisable charges is well-understood \cite{Doplicher:1969tk, Doplicher:1969kp, Buchholz:1985ii, Doplicher:1990pn, Longo:1994xe,Casini:2019kex}, the interplay between non-localisable charge superselection in theories with massless particles, like QED, and local algebras is on much less firm footing, generically involving the use of non-local operators supported on spacelike cones \cite{Buchholz:1981fj}.\footnote{This is intimately tied to the concept of ``infraparticles'' developed in \cite{osti_4667842, Buchholz:1982ea, Buchholz:1986uj, Mund:2021zhx}.} Recent developments in Maxwell theory in algebraic QFT \cite{Fewster:2025ijg} have extended the local algebra to include operators smeared all the way to the corner for an \emph{isolated} region, thereby recovering the electric flux charges. However, it is not clear at this point if these quasi-local operators are well-defined in the global theory, i.e.\ when embedding the region into the global theory and especially how they play against the usual algebraic QFT axioms, like the isotony or microcausality property of local nets. By contrast, it was shown in \cite{Hoehn:2025pmx} that one can define local algebras including edge modes that obey isotony and microcausality provided one invokes certain non-local boundary conditions and a relational notion of locality. 

There are also a few special lower-dimensional continuum QFTs, in which the algebra of local operators supported in a region can be enhanced to include edge or corner modes that feature a finite covariance.
One example is  Chern-Simons theory on $D^{2}\, \times\, {\bf R}$, where for suitably chosen boundary conditions, the algebra of observables includes the infinite-dimensional Ka\v{c}-moody algebra generated by boundary currents, e.g.\ see \cite{Dunne:1998qy} and references therein.
The second example of a continuum QFT whose algebra of subregion observables include edge mode operators is the so called non-compact quantum Lifshitz model, which is a $(2+1)$-dimensional theory whose ground state probability measure is a partition function of a two-dimensional CFT \cite{Ardonne:2003wa}. In this model, edge modes living on the boundary of a spatial subsystem have finite correlators, after smearing only along the boundary \cite{Angel-Ramelli:2019nji}.
Nevertheless, our understanding of quantum fields that are supported on the boundary of a subregion thus remains far from complete, especially in the continuum.

Furthermore, even on the lattice there is a large ambiguity with regards to which algebra to associate with a given subregion, leading to so-called ``electric centre'', ``magnetic centre'', as well as centreless constructions, differing by choices on the boundary of the region \cite{Casini:2013rba,Soni:2015yga,Delcamp:2016eya,Araujo-Regado:2025ejs,VanAcoleyen:2015ccp}.\footnote{See \cite{Araujo-Regado:2025ejs} for how to relate all these choices from the perspective of so-called quantum reference frames.} This leads to a lack of universality in the associated entanglement entropies, which survives the continuum limit. On the other hand, universal entanglement and correlation measures like relative entropy and mutual information in the continuum are independent of such scheme-dependent choices on the lattice \cite{Casini:2013rba, Casini:2015dsg}, in particular leading to a universal log-divergent coefficient. Unless one succeeds in giving a continuum QFT definition of these corner supported centres, this casts doubt on their physical relevance.

In fact, it was argued in \cite{Casini:2019nmu} that to get the anomaly coefficient for the 4D Maxwell field, one actually needs to couple it to massive charged particles and that the correction to the log-coefficient of the mutual information arises due to particle-antiparticle fluctuations on either side of the thin shell dividing the two regions, when the separation becomes much smaller than the scale set by the mass.\footnote{Interestingly, the authors argue that one also needs magnetic charges in the UV to get the log-coefficient to match the conformal anomaly.} This is a very different mechanism for the interpretation of the anomaly-coefficient  than the one based on symmetry, that nevertheless crucially relies on the imposition of Gauss law. In particular, the analysis of \cite{Casini:2019nmu} does not require the existence of singularly smeared operators. This is in contrast to \cite{Donnelly:2014fua, David:2022jfd, Ball:2024hqe}, where the corner modes exist as a classical stochastic field with divergent covariance. This covariance needs regularisation and is hence scheme-dependent. However, this regularised definition of corner mode is sufficient to compute the universal log-coefficient in entanglement entropy which is scheme-independent. 


Our goal in this paper is to  revisit the question of the existence of spatial corner mode operators in a continuum QFT from a new angle that requires matter, taking scalar QED as a case study. Our contributions come in two flavours.

\begin{enumerate}
    \item The first is a proposal for a {new} centre mode operator in scalar QED. Namely, we argue that a natural choice is given by the Coulombic (so longitudinal) electric field measured inside some region, sourced by the matter \emph{outside} the region, via the Gauss constraint. The very definition of this mode is \emph{non-local}, in that it relies on a specific choice of in/out split of the Cauchy slice and cannot be written as a local operator at a point, in contrast with the usual perpendicular electric field at the corner. From an operational perspective, our mode can be accessed in the \emph{bulk} of our subregion, simply as a particular mode of the total electric field. It is not a choice whether we measure it or not. 

In contrast with the usual corner-supported charges \cite{Donnelly:2016auv,Geiller:2019bti,Carrozza:2021gju,Araujo-Regado:2024dpr,Ball:2024hqe,Gomes:2016mwl,Riello:2021lfl,Gomes:2018dxs,Gomes:2019xto,Hoehn:2025pmx,Shimizu:2026zug,Shimizu:2026wjz,Ball:2024gti,Borsboom:2026xrd}, our mode is not a joint centre for the two complementary regions.\footnote{The complement region will have its own different centre operator given by the Coulombic electric field measured outside, sourced by matter inside the original region.} This means that it will have genuine \emph{quantum} fluctuations in the vacuum, coming from the existence of a myriad of complement operators which  do not commute with it. In some sense, we can think of it as a left-over degree of freedom arising after ``tracing out'' the complement. In scalar QFT, this would lead to a factor algebra. Here, by contrast, the Gauss constraint induces a quantum centre. In addition, it is really a consequence of \emph{interactions}, arising as a first-order correction to the free algebra of QED. At zero coupling, the Gauss constraint reduces to $\nabla\cdot\, E=0$, eliminating the possibility for the non-local sourcing of a subregional mode that is independent from local radiation. 

Finally, we note that our centre mode is in fact an \emph{edge} mode because it is determined entirely by the boundary value of a harmonic function inside the region. 

\item Though our new centre mode may seem natural in many ways, it still does not lead straightforwardly to a well-defined quantum operator in the continuum. Our second main contribution is then a technical one, namely, to explicitly perform the necessary  correlation function renormalisation to leading-order in the coupling. As will turn out, the required renormalisation is purely \emph{corner-supported}, even though our mode is smeared in the bulk. At the end of the day, we will obtain a renormalised vacuum that supports our centre mode, thereby completing its quantisation. 
\end{enumerate}

More in detail, we start with the classical theory of a massive charged scalar minimally coupled to the electromagnetic field and consider the Poisson $\star$-algebra of purely local gauge-invariant functions on a Cauchy slice $\Sigma$ in Minkowski spacetime. This algebra is generated by $\langle\rho(f), J(V), E(V), B(V), |\psi|^2(f)\rangle$, where $\rho(f),J(V)$ are the smeared charge and current, respectively, and $|\psi|^2(f)$ is the smeared local number operator, while $E(V),B(V)$ are the smeared electric and magnetic fields.\footnote{These are all the purely \emph{local} gauge-invariant operators in the theory. This algebra is closed under Poisson brackets. We could also add (suitably smeared versions of) gauge-invariant bi-local operators of the form $W(x,y)=\psi^\dagger(x)P\exp \{ie\int_\gamma A\}\psi(y)$, but we choose not to do so in this paper because we do not study carefully its two-point function. Nevertheless, conceptually at least, they could be included without fundamentally changing any of our results for the new longitudinal electric corner mode and will make a reappearance in the discussion section.} An important aspect to note for later is that gauge invariance forces us to work with quadratic composite operators in the matter sector. In the Hamiltonian formulation, the Gauss constraint, $\nabla\cdot E=e\rho$, can be imposed on this algebra as it forms a two-sided ideal, since gauge-invariant operators commute with the constraint. The quotient by the Gauss ideal leads to a global Cauchy slice algebra generated by $\langle\rho(f), J(V), E_{T}(V), B(V), |\psi|^2(f)\rangle$, where $E_{L}\coloneqq E-E_T$ is the longitudinal electric field sourced by $\rho$, which now becomes a dependent operator under the identification, while the transversal mode satisfies $\nabla\cdot E_T=0$.

The story becomes more interesting when we consider algebras associated to subregions of $\Sigma$. Given a spatial open set ${\cal O}\subset\Sigma$ with compact closure, we start with the algebra of gauge-invariant operators with support strictly inside $\cal O$. After quotienting by the Gauss constraint, the resulting Poisson $\star$-algebra is \pah{${\langle\rho(f), J(V_{1}), E(V_{2}), B(V_{3}), |\psi|^2(g)\rangle}$, where $f, g\in C^\infty_c(\cal O)$, $V_{i}\, \in\, C^{\infty}_{c}(T{\cal O})$ are smooth functions and vector fields, respectively, compactly supported inside $\cal O$. Note that $E=E_T+E_L$ now contains the longitudinal mode because that also depends on the charge density outside $\mathcal{O}$ and is thus not fully determined from data inside the region. 

At this stage, we perform one additional step: we split the longitudinal field inside $\mathcal{O}$, $E_L(V)=E_L^{\rm in}(V)+E_L^{\rm out}(V)$, into the parts that are sourced by matter inside and outside of $\mathcal{O}$, respectively. As we will see, the outer mode $E_L^{\rm out}$ is independent of the other subregional operators and will become the main player in our work after we add it to the regional algebra (which also adds $E_L^{\rm in}$), which then becomes
\begin{align}
    {\cal A}_{\text{C}}({\cal O})\coloneqq \big\langle\rho(f), J(V_{1}), E(V_{2}), B(V_{3}), |\psi|^2(g), E_{L}^{\textrm{\text{out}}}(V_{4})\big\rangle\,.\label{A_C}
\end{align}}
  This classical algebra has several interesting properties. Most crucially for us, for any given subregion ${\cal O}$, the algebra has a centre generated by $E_{L}^{\textrm{\text{out}}}(V)$, since all operators inside $\cal O$ commute with the charge density in the complement.\footnote{This is why we label the algebra by the subscript C for centre.} 
  
Our primary goal in this paper is to investigate the existence of a GNS state in which this algebra ${\cal A}_{\textrm{C}}({\cal O})$ is quantised. We show that such a state exists and is obtained by pulling back the perturbative Fock vacuum of scalar QED to the Cauchy slice and performing a renormalisation of the correlators for the composite operators including the new centre mode on $\Sigma$. This is the main technical result of the paper. 

Throughout the paper, it is important to distinguish two conceptually \emph{distinct} types of renormalisation that we are required to perform. 
\begin{itemize} 
\item[(i)] The first is a \emph{bulk} renormalisation of the two-point functions of all \emph{composite} operators arising from the fact that we choose to work with operators that are sharply smeared in time. Although strictly speaking, in a renormalisable Lorentz-invariant QFT, such as scalar QED, the algebra generated by $\langle\rho, J, E, B, |\psi|^2\rangle$  does not exist unless we do an infinitesimal time smearing, we show that there exists an effective field theory (EFT) with higher (spatial) derivative corrections to the Hamiltonian that can be used to quantise the algebra of quadratic operators on a Cauchy slice.\footnote{Typically higher-derivative terms in the Lagrangian generate ghosts in the theory and hence lead to negative norm states in the Hilbert space of the quantum theory. However, if the interactions only contain higher spatial derivatives than it does not generate negative norm states.}\footnote{Note that for the linear electromagnetic fields $E$ and $B$ no such renormalisation is needed and their equal-time two-point functions are finite.} Because this EFT differs from scalar QED by the addition of irrelevant operators, it agrees with our theory of interest in the IR.\footnote{Its UV behaviour is, however, modified.} This means that we can use this EFT to extract lessons about our new longitudinal electric corner mode in scalar QED, provided we only consider operators smeared below some UV cut-off scale.  

Our motivation of using such an EFT is then simply to analyse the constraints that the existence of such corner mode operators places on the continuum QFT, while working with Cauchy slice algebras.\footnote{It is for this reason that the title does not directly refer to corner modes in scalar QED.} If we were to work instead with open spacetime regions (like double-cones), this bulk renormalisation step would not be necessary \cite{Haag:1992hx,jaffe1966wick,Yazdi_2023,streater2016pct}, and we would be able to work directly with scalar QED without high-derivative corrections. That being said, the other type of renormalisation described below would still be necessary for the   centre mode to be well-defined.

\item[(ii)] The second type is a \emph{boundary} renormalisation of the two-point function of the charge density operator at the location of the boundary $\partial\cal O$ of some chosen region $\mathcal{O} \subset \Sigma$.  This is necessary because the definition of the would-be centre mode smeared with a smooth, compactly supported vector field inside $\cal O$, still involves a step-function type of smearing induced by the Gauss law. Concretely, it reads 
\begin{align}
    E_L^{\text{out}}(V)\coloneqq-\frac{e}{4\pi}\int_{\cal O} d^3x \int_{\mathcal{O}^c} d^3y \frac{\nabla\cdot V(x)}{|x-y|} \rho (y)\,, \label{E_L^out}
\end{align}
where $G(|x-y|)=-\frac{1}{4\pi|x-y|}$ is the free Green's function of the Laplace operator in 3D. The integral over $\cal O$ has a smoothly decaying smearing, but the integral over $\mathcal{O}^c$ is truncated sharply at $\partial\cal O$. Thus, the two-point function of $E_L^{\text{out}}(V)$ is not a priori well-defined, unless one renormalises the operator product. As we explain in detail in the paper, the required renormalisation is purely supported on $\partial\cal O$. This is consistent with our centre mode being an actual \emph{edge} mode.
\end{itemize}

The renormalisation strategy we employ in this paper is the Epstein-Glaser method \cite{Epstein:1973gw}, or more specifically, the closely related method of \emph{differential renormalisation} \cite{Freedman:1991tk, Haagensen:1992am, Latorre:1993xh, Gracia-Bondia:2014zwa, Gracia-Bondia:2015nda}. This is a particular scheme for renormalising the coincidence limit of operator products via the notion of distributional extension of singular functions, as explained in appendix \ref{app:EG}, without the need to introduce any UV regulator. Put simply, we replace the singular functions by well-defined distributions constructed from derivatives of other well-defined distributions. As always, any renormalisation scheme leads to a set of undetermined finite counter-terms, to be fixed by actually measuring the corresponding correlators in an experiment. 

However, generically there is a tension between renormalisation and \emph{positivity} of the state, which is the requirement that
\begin{align}
    \omega(a^*a)\geq0\,,\qquad\forall a\in\cal A\,,
\end{align}
for $\omega$ some pertinent algebraic state $\omega:\cal A \to \mathbb{C}$ (e.g.\ the vacuum). While the bare, unrenormalised state is guaranteed to be positive, but composite operators do not have well-defined correlators in them, the renormalised state supports well-defined,  composite operators (with finite correlators), but is not guaranteed to be positive. In particular, positivity can be lost for operators smeared with functions that have support at large momenta. To restore positivity, new finite counter-terms need to be added to the operator product. Whenever these counter-terms correspond to the addition of new \emph{irrelevant} operators to the Hamiltonian, a modification of the UV physics of the theory is taking place.\footnote{No such irrelevant counter-terms are needed if we instead take the route of restricting our space of smearing functions to those with support below some momentum scale. In other words, the state is already positive provided we introduce a UV cut-off in the theory. Either way, we end up in the setup of EFTs.}  

It will turn out that for both the bulk and boundary renormalisations described above, counter-terms need to be added for positivity. However, as previously mentioned, the bulk renormalisation is really a feature of our restriction to sharp Cauchy slice operators and disappears the moment we allow for time smearings \cite{Haag:1992hx,jaffe1966wick,Yazdi_2023,streater2016pct}. For time-smeared operators, the Fock vacuum of scalar QED is already positive for the standard local algebras. In contrast, the need for a boundary renormalisation is really a feature of the inclusion of the centre mode in \eqref{A_C}, as well as the existence of a subregional boundary, and survives the lifting to spacetime algebras.\footnote{We plan to show this explicitly in upcoming work.} The conclusion is then that a modification of the UV structure of the Fock vacuum of QED is necessary to support well-defined centre modes. This is one of the main conceptual results of our work.

Although we will carry out our explicit construction in Minkowski spacetime, this simplification is not strictly necessary. This is because the renormalisation always has to do with the UV behaviour of the two-point functions, at scales much smaller than any curvature scale of the ambient spacetime. Similarly, at no point in our analysis does the existence of an asymptotic boundary play any role. Thus, our conclusions should apply equally well to a closed universe. 

\subsection{Summary of the main results}

In this section, we summarise the important points and messages of this paper, for the benefit of the reader. 

\begin{enumerate}

\item We start with the abstract net of Cauchy slice $\star$-algebras for open sets $\mathcal{O}$ on a flat slice $\Sigma$ in Minkowski spacetime, generated by purely local gauge-invariant operators\footnote{$\langle A,B\rangle$ denotes the $\star$-algebra generated by operators of the form $A,B$.}\pah{
\begin{align}
    \mathcal{A}_\text{kin}(\mathcal{O})\coloneqq \big\langle E(V), B(W), \rho(f), J(U), |\psi|^2(g)\,\big\vert\,f,g\in C_c^\infty(\mathcal{O}), U,V,W\in C_c^\infty(T\mathcal{O})\big\rangle\,\nonumber
\end{align}
with appropriate compactly-supported smearings. Here, $E=E_T+E_L$ denotes the full electric field containing both transversal and longitudinal modes. In other words, this gauge-invariant algebra does not yet obey the Gauss law (though it commutes with it) and we call it the gauge-invariant kinematical algebra. The commutators of the generators are defined by the classical canonical commutation relations (CCR for short). We are essentially working with the Cauchy slice pull-back of the standard ``universal $C^\star$-algebra'' of scalar QED \cite{Buchholz:2015epa}, except that we work at the $\star$-algebra level. The resulting $\star$-algebra net (on the Cauchy slice) admits no corner modes and it satisfies all the algebraic QFT axioms in a spatial version, including isotony. }

\item \pah{When extending this algebra to the full Cauchy slice $\Sigma$, $\bar{\mathcal{A}}_{\rm kin}(\Sigma)$, crucially, we include non-compact smearing in the form $f,g\in C_{r^{-1}}^\infty(\Sigma)$ and $U,V,W\in C_{r^{-2}}^\infty(T\Sigma)$, for otherwise it will turn out that the global algebra does not allow for longitudinal electric fields inside any subregion $\mathcal{O}$ that is sourced by matter in its complement. }

\item \pah{On this extended kinematical gauge-invariant Cauchy slice algebra $\bar{\mathcal{A}}_{\rm kin}(\Sigma)$, we then impose the Gauss law,
\begin{align}
{\cal G}(f)\,\coloneqq\, \int_{\Sigma}\dt^3x\, ( \nabla \cdot E  - e \rho\, )(x)\, f(x) , \qquad f \in C_{r^{-1}}^\infty (\Sigma)\,,
\end{align}
for all smearing fields featuring a $1/r$ fall-off, by quotienting it ($\mathcal{G}(f)$ generates a two-sided ideal). This results in our global physical Cauchy slice algebra
\begin{equation}
    {\cal A}_{\rm phys}(\Sigma)\simeq\, \Big\langle\, E_T(V), \, B(V)\,, \rho (f ), \, J(V), \, |\psi|^2 (f)\, \Big|\  \forall\, V\in C^\infty_{r^{-2}}(T\Sigma),f\in C^\infty_{r^{-1}}(\Sigma) \Big\rangle\,,\nonumber
\end{equation}
in which the longitudinal electric field $E_L$ is no longer independent, but determined by $\rho$-smearings. We note that this algebra is exact, i.e.\ not perturbative in $e$ despite the coupling appearing.\footnote{As we are working on a Cauchy slice, there is no dynamics to be imposed yet. By contrast, for spacetime algebras, turning on the coupling would have to be incorporated algebraically at the level of constraints and dynamics together for consistency.}}

\item We then pull-back the standard covariant Fock vacuum state $\omega_{F}$ to the Cauchy slice, in order to build a GNS representation for  $\mathcal{A}_{\rm phys}(\Sigma)$ with  finite two-point functions on the resulting Hilbert space. This requires renormalisation of the original Fock vacuum two-point functions due to the presence of composite matter operators together with singular time smearings. We use differential renormalisation to define the new pulled-back renormalised state $\omega_{F}^{R}$. As we show,  $\omega_{F}^{R}$ can be interpreted as the vacuum of a Hamiltonian obtained by perturbing the scalar QED Hamiltonian with higher spatial derivative terms. Thus, there exists a continuum theory for which the physical Cauchy slice algebra of local gauge-invariant operators can be well-defined.

\item \pah{Next, we \emph{fix} a region $\mathcal{O}\subset\Sigma$ and consider the algebra
\begin{equation}
    \mathcal{A}_{\rm phys}(\mathcal{O})\coloneqq\Big\langle\, E_T(V),E_L(V), B(V), \rho (f ),  J(V), \, |\psi|^2 (f)\, \Big|\  V\in C^\infty_c(T\mathcal{O}),f\in C^\infty_c(\mathcal{O})\Big\rangle ,\nonumber
\end{equation} 
which encodes all independent operators from the global Cauchy slice algebra that are in principle measurable in the region. Note that the longitudinal field now reappears as an independent operator since it also depends on the matter outside.

To isolate the contribution from the outside, we split the longitudinal field $E=E_L^{\rm in}+E_L^{\rm out}$ in $\mathcal{O}$ into the inner and outer part, which are sourced by the scalar field in- and outside of $\mathcal{O}$, respectively. Due to the non-smooth smearing in eq.~\eqref{E_L^out}, the outer (and also inner) mode is independent of the other operators in $\mathcal{A}_{\rm phys}(\Sigma)$ and we augment the algebra by it, to obtain a new regional algebra
\begin{align}
    {\cal A}_\text{C}({\cal O})=\mathcal{A} _{\text{phys}}({\cal O})\vee \big\langle E_L ^{\text{out}}(V)\,\big|\,V\in C_c^{\infty}(T\mathcal{O})\big\rangle\,.
\end{align}
Since $E_L^{\rm out}$ commutes with all elements in $\mathcal{A}_{\rm phys}(\mathcal{O})$, we have that $\mathcal{A}_C$ features a centre with the outer longitudinal electric mode. In particular, this mode can be associated with the corner $\p\mathcal{O}$ since it is associated with a harmonic smearing of $\rho$ in the complement (which thus is entirely determined by the configuration on the corner). By contrast, $E_L^{\rm in}$ is sourced inside $\mathcal{O}$ and not an edge mode.
}

\item The pulled-back state $\omega^R_F$ does \emph{not} yield finite correlators involving the longitudinal electric centre mode $E_L^{\rm out}$. As a consequence of the non-locality of the Gauss law, a further \emph{corner} renormalisation on $\partial\cal O$ is required. We show that, given a \emph{fixed} subregion ${\cal O}$, there exists a quasi-free GNS state $\omega_0$ which can be used to quantise the centrally extended algebra ${\cal A}_\text{C}({\cal O})$ in the following (schematic) way:
\begin{align*}
\omega_{0}({\cal A}_{\textrm{\text{phys}}}(\mathcal{O})) &\coloneqq \omega_{F}^{R}({\cal A}_{\textrm{\text{phys}}}({\cal O})) \\
\omega_{0}(E_{L}^{\textrm{\text{out}}}(V)\, A) &\,=\ \textrm{is finite,}\qquad \forall\, A\, \in {\cal A}_\text{C}({\cal O})\,.
\end{align*}

Crucially $\omega_0$ is \emph{unitarily inequivalent} to $\omega_F^{R}$ and, in fact, breaks translation invariance on the Cauchy slice
\begin{align*}
\omega_{0}(E_{L}^{\text{out}}(V)\, E_{L}^{\text{out}}(W))\, \neq\, \omega_0 (E_{L}^{\text{out}}(\alpha_{a} V)\, E_{L}^{\text{out}}(\alpha_{a} W))\,,
\end{align*}
where $\alpha_{a}$ stands for the action of translations by a vector $a$ on the support of the smearing fields.
This tells us that the centre modes in this setup can only be defined for a \emph{unique} fixed subregion.\footnote{Actually, they can be defined for disconnected regions $\cal O$ with multiple simply-connected components also.} The addition of the longitudinal electric corner mode has thus physically distinguished the region $\mathcal{O}$.
Furthermore, correlation functions in ${\cal O}$ in the new vacuum $\omega_{0}$ can be realised as correlators in $\omega_{F}^{R}$ shifted by a superposition of classical Coulombic fields sourced by the matter in the complement. The centre mode provides a stochastic background for the local algebra.

\item The central operators are parametrised by functions on $\partial {\cal O}$, which identifies them as new edge modes for our subregion algebra. We fully characterise the allowed space of such functions, by demanding that they lead to finite norm under the renormalised boundary kernel, which we extract from the two-point correlator of $E_L^{\rm out}$ on $\p\mathcal{O}$. This kernel is (here in a planar limit approximation) given by 
\begin{align}
    C_{\p \p } ^{\rm ren}(r_\parallel )&\coloneqq \frac{1}{48m^3} \triangle_{R^2}^2 \ln ^2 (mr_\parallel ) + \frac{2\gamma - \ln 4 + 1}{8m}\triangle_{R^2} \ln^2 (mr_\parallel)  + \frac{1}{24m} \triangle_{R^2} \ln ^3 (m r_\parallel)\nonumber\\
    &\qquad\qquad\qquad +c_0+ c_1\Delta \delta^2(r_\parallel)\,,\nonumber
\end{align}
where $r_\parallel$ and $\Delta_{R^2}$ denote the position vector and Laplacian operator on the co-dimension-two boundary (of radius $R$), respectively. This is one of our primary results. We also use this result to describe the completion of the corner smearing functions of the centre mode, which yields a generalised Sobolev space that is a strict subspace of the Hilbert space of square integrable functions on the corner. 
\end{enumerate}

We finally end this section with a few remarks.
\begin{itemize}
    \item We are working in the framework that the algebra of operators is exact, and it is the GNS state that is expanded in perturbation theory in coupling $e$. At leading-order, the vacuum is the free Fock vacuum, modulo the renormalisation required for singular time smearings
    \begin{align}
        \omega[e]=\omega^R_F + \sum_{n=1}^\infty e^{2n} \omega_n^R\,.
    \end{align}
    The regional algebra, as well as its commutation relations, are taken to be exact, independent of the coupling. This is the opposite procedure to what is usually done in perturbative algebraic quantum field theory \cite{Hollands:2001nf, Brunetti:2009qc, Fredenhagen:2015iia,Brunetti:2025ocr,Rejzner:2016hdj}, where the algebra is formally expanded in a power-series in the coupling constant, whereas the state is taken to be exact in that algebra. Note that the two procedures are equivalent.

    \item $E_L^\text{out}\sim O(e)$ and so its correlation functions in $\omega[e]$ are of a higher perturbative order than those of the other operators. But we also take the central mode as an exact operator. The fact that it scales with $e$ is simply telling us that its emergence is an effect due to constraints, not present in free theory. Given this, we emphasise that our renormalisation analysis throughout is at the first non-vanishing order of the coupling, at the level of the leading-order vacuum state $\omega^R_F$. We leave the incorporation of loop-corrections for elsewhere.
\end{itemize}

Let us now turn to the detailed exposition.

\section{Motivating the centre mode classically}

The goal of this section is to present the intuition behind our definition of the centre mode for the subregional algebra, which comes from the classical theory.

\subsection{Physical subregion phase space}\label{subsec:subregion phase space}

In this section we present the analysis of the canonical phase space of a subregion in the theory of electromagnetism coupled to a complex massive scalar field $\psi$. We use the formalism of dynamical reference frames for abelian gauge theory developed in \cite{Carrozza:2021gju,Araujo-Regado:2024dpr,Hoehn:2025pmx}. 

The kinematical pre-symplectic form of a spacetime subregion with a spatial Cauchy surface $\mathcal{O} \subset \Sigma $ is:
\begin{equation}
    \Omega = \int _{\mathcal{O}} \sqrt{g} \dt^3 x \ \l  \d A_i \curlywedge \d F^{ti}  + \d \psi \curlywedge \d D^t  \psi ^\dagger  - \d  \psi ^\dagger  \curlywedge \d D^t  \psi \r\,,
\end{equation}
where $D_\mu  = \p _\mu - i e A_\mu $, $g$ is the determinant of the induced metric on $\Sigma$, $\curlywedge$ denotes the field space wedge product. This symplectic form transforms non-trivially under gauge transformations of the global theory, with parameter that does not vanish on the boundary $\p \mathcal{O}$. To restore gauge invariance one needs to add to the symplectic form a corner term, that depends on a field $\Phi ^b$ which lives on the boundary \cite{Donnelly:2016auv,Geiller:2019bti,Carrozza:2021gju,Araujo-Regado:2024dpr,Hoehn:2025pmx}:
\begin{equation}
    \Omega ^\text{phys} \coloneqq \Omega - \int _{\p \mathcal{O}}  \d \Phi^b \curlywedge \d \star F\,,
\end{equation}
where $\star$ denotes the Hodge star operator on spacetime.

Such fields need to transform covariantly under gauge transformations on the boundary, and therefore can be realised via dynamical reference frames \cite{Carrozza:2021gju,Araujo-Regado:2024dpr,Hoehn:2025pmx}. An ideal dynamical reference frame  is a possibly non-local functional $U[A,\psi](x)$ of the connection $A_\mu (x)$ and the matter field $\psi (x)$, that locally takes value in $\rm{U}(1)$ and satisfies $U(x)\to g(x)U(x)$ under gauge transformations $g(x)\in\rm{U}(1)$. In other words, its phase $\Phi[A,\psi](x)\coloneqq-i\ln U[A,\psi](x)$ obeys infinitesimally:
\begin{equation}
    A _\mu (x) \mapsto A_\mu (x) + \p _\mu \lambda (x)  \quad \psi (x)\mapsto e^{i e \lambda (x)} \psi (x), \quad \Rightarrow \quad \Phi ^b (x) \mapsto \Phi ^b  (x) + \lambda(x)\,.
\end{equation}
In what follows, we shall for brevity often refer to the phase $\Phi^b$ of the frame as the frame.

The boundary reference frame $\Phi^{b}$ can then be extended in the bulk of the subregion via some linear differential equation that depends on the bare fields $A$ and $\psi$. A common example is the Coulomb frame:
\begin{equation}\label{Phi example}
    \triangle \Phi  = -\nabla ^i A_i , \qquad \Phi \big|_{\p \mathcal{O}} = \Phi ^b\,.\end{equation}
With a bulk extension, the symplectic form can be written as:
\begin{equation}\label{eq:omegaphys}
    \Omega^{\rm phys} = \int _{\mathcal{O}} \sqrt{g}\dt^3 x \,\l  \d A_i ^\dr \curlywedge \d F^{ti}  + \d \psi ^\dr  \curlywedge \d  D ^t \psi ^{\dagger \, \dr } - \d  \psi ^{\dagger \, \dr } \curlywedge \d  D^t  \psi ^\dr \r \,,
\end{equation}
where
\begin{equation}
    A^\dr \coloneqq A_\mu - \p _\mu  \Phi, \quad \psi ^\dr \coloneqq e^{ - ie \Phi } \psi, \quad  D = \p _\mu - i e A^\dr _\mu 
\end{equation}
are gauge-invariant and called ``dressed'' fields. They satisfy a dressing condition \cite{Araujo-Regado:2024dpr}, which in the case of the Coulomb frame is $\nabla ^i A_i ^\dr = 0$.

In \cite{Araujo-Regado:2024dpr} it was understood that from the point of view of the subregion there are two types of reference frames -- intrinsic $\tilde{\Phi}$ and extrinsic $\Phi$ (see \cite{Araujo-Regado:2025ejs} for their lattice realisation). The former is built from degrees of freedom purely from the subregion including its boundary, while the latter also includes degrees of freedom from the complement, which implies we have extended the subregion phase space. This definition implies that we can shift the value of the extrinsic frame while keeping $A$ and $\psi$ in the subregion fixed. Such transformations in the extended subregion phase space are called frame reorientations and have the following effect on the bare and dressed fields:
\begin{equation}
    \begin{split}
        &A _\mu  \mapsto A _\mu, \quad \psi \mapsto \psi, \quad  \Phi  \mapsto \Phi + \alpha \\
        \Rightarrow \quad & A _\mu ^\dr  \mapsto A _\mu ^\dr  - \dt \alpha , \quad \psi ^\dr \mapsto e^{-ie\alpha}\psi ^\dr\,, 
    \end{split}
\end{equation}
i.e.\ they shift the gauge-invariant fields. Physically, these reorientations can be realised by large gauge transformations, as well as by the addition or subtraction of hard or soft radiation in the complement of $\mathcal{O}$ \cite{Araujo-Regado:2024dpr}.

The existence of this direction in the symplectic phase space allows us to define a new gauge-invariant variable, called subregion Goldstone mode \cite{Araujo-Regado:2024dpr}, by dressing the intrinsic with the extrinsic frame,\footnote{It is called the regional Goldstone mode because it parametrises the physical degeneracy of $F^{\rm dr}\equiv F$.}
\begin{equation}
    \varphi \coloneqq  \Phi -\tilde{\Phi}\,,
\end{equation}
and thereby to re-express the physical symplectic form in eq.~\eqref{eq:omegaphys}\footnote{We assume that eq.~\eqref{eq:omegaphys} is written in terms of an extrinsic frame dressing.} as 
\begin{equation}\label{eq:Omega phys}
\Omega ^\text{phys} =   \int _{\mathcal{O}} \sqrt{g} \,\dt^3 x \,\left( \d \tilde A_i ^\dr \curlywedge \d F^{ti}  + \d \tilde \psi ^\dr  \curlywedge \d D ^t \tilde{\psi }^{\dagger \, \dr } - \d  \tilde{\psi }^{\dagger \, \dr } \curlywedge \d D ^t  \tilde \psi ^\dr\right)  + \int _{\p \mathcal{O}} \d \varphi \curlywedge \d \star F\,,
\end{equation}
where the over-script `` $\tilde{\cdot}$ '' indicates dressing with respect to the intrinsic frame. One way to understand the appearance of the symplectic pair of corner fields is to think of them as the data needed to solve two independent elliptic equations -- the local Gauss's law for the electric field, and the dressing condition for the dressed potential $A^\dr$. The role of both corner fields is interesting in its own right, but for now we focus on the corner charges: 

\begin{equation}
    Q[\alpha] = \int _{\p \mathcal{O}} \alpha \star F\,.\end{equation}

There are infinitely many of them for each smooth function $\alpha$ on $\p \mathcal{O}$. These charges generate the \emph{physical} symmetries of the subregion phase space, extended by an extrinsic reference frame \pah{\cite{Donnelly:2016auv,Geiller:2019bti,Carrozza:2021gju,Araujo-Regado:2024dpr,Shimizu:2026zug}}. They are analogous to the large gauge charges living at $\scri ^+ _-$.

Furthermore, they Poisson commute with all gauge-invariant operators built purely from degrees of freedom in the interior of the subregion. Therefore, in the quantum theory they are viable candidates for centre operators for the subregion algebra. While quantum operators living on the boundary of a subregion are perfectly well-defined on the lattice \cite{Casini:2013rba,Ghosh:2015iwa,Delcamp:2016eya,Araujo-Regado:2025ejs,Soni:2015yga,VanAcoleyen:2015ccp}, their construction in the continuum remains an open problem when embedding the regional data into the global theory.\footnote{For a finite region in isolation, progress was made recently in \cite{Fewster:2025ijg}. There are also ways to quantise edge or Goldstone modes in the presence of non-local boundary conditions \cite{Hoehn:2025pmx}.} The main goal of this paper is to revisit the challenge of constructing well-defined corner-supported centre operators in a continuum quantum field theory. However, rather than considering the above types of normal electric centre modes (which we only do in appendix~\ref{app:empm}), we will propose a new longitudinal electric centre mode that requires the presence of matter. In the next subsection, we present the classical intuition motivating our construction.

\subsection{Corner modes}\label{subsec:corner modes}

As it is well known, quantising a field at a given point in space $\hat\phi (x)$, does not give a well-defined operator on the Hilbert space, but rather an operator-valued distribution e.g.\ see \cite{Haag:1992hx,streater2016pct,halvorson2006algebraic}. To get an (unbounded) operator one typically smears in spacetime with a  $C_c^\infty$\footnote{The subscript $c$ stands for compactly supported, and the superscript $\infty$ for infinitely differentiable.} function $f(x)$ (also called a test function) with support in some open set $m \subset M$, where $M$ is the global spacetime:
\begin{equation}
    \hat{\phi} (f) \coloneqq \int _{ m } \sqrt{|g|} \dt^4 x \, f(x) \hat{\phi} (x)\,.
\end{equation}
Lower-dimensional smearings can, under certain conditions, also result in well-defined operators.\footnote{We saw several examples of this in the introduction.}  For example, smearing $\hat\phi(x)$ along timelike curves was shown to achieve this in \cite{Borchers1964FieldOA}, while smearing it along a spatial Cauchy slice yields well-defined operators for free theories \cite{Haag:1992hx,streater2016pct}. However, for operators in interacting field theories, it is typically expected that purely spatial smearings are not sufficient to obtain well-defined quantum operators \cite{Haag:1992hx,Witten:2023qsv}. Furthermore, even in free theories, composite, i.e.\ non-linear field operators remain ill-defined when smearing them only spatially, e.g.\ see \cite{jaffe1966wick,Yazdi_2023,streater2016pct}, a feature we will also encounter in section~\ref{sec:GNS A_loc}. 
On the other hand, smearing linear fields on spatial structures of co-dimension $> 1$ will not completely remove their distributional nature even in free theories, yielding  operators with divergent correlation functions. 

We can already see the tension between algebraic quantum field theory and the quantisation of gauge theories. The former deals almost exclusively with local operators with four-dimensional spacetime support, while the latter introduces non-locality through the constraints and the principle of gauge invariance, which naturally give rise to degrees of freedom that ``live'' on a co-dimension-two surface. Therefore, the existence of the corner modes requires smearing along the boundary, and this does not yield well-defined quantum operators. (See, however, \cite{Fewster:2025ijg} for an interesting perspective on the corner modes in pure Maxwell theory for isolated regions.)

For the global theory, these boundary fields are the large gauge charges that live on $\scri ^+ _-$ and their symplectic conjugate -- the global Goldstone mode \cite{He:2014cra, Strominger:2017zoo, He:2019pll}. Even though the Goldstone mode is completely independent of the radiative data for the propagating photon and matter fields, it is crucial for the existence of finite non-trivial scattering amplitudes and QED. Significant progress towards the derivation of their two-point function form the action was made in \cite{He:2024ddb}.

In the case of a subregion, the corner charges are propagating degrees of freedom whose dynamics is dictated by the global theory. They are simply the pull-back of the radial electric field to a co-dimension two closed surface. However, from the point of view of the global theory, there should not be (and there is not) anything special about this surface. This issue is quite evident in the case of free Maxwell theory, which on the reduced phase space, in a  fixed large gauge charge sector, is classically to a theory of two free massless fields. It seems that we do not have reasons to have well defined corner operators for electromagnetism, any more than we have in an ordinary massless field theory. 

Indeed, to find a way forward, we need to exploit the features which make QED different from other quantum field theories. A crucial characteristic, related to the gauge symmetry, is the constraint on initial data on some spatial Cauchy slice $\Sigma$ given by Gauss's law:
\begin{equation}
    \l \nabla \cdot E  - e \rho \r (x) = 0 \,,
\end{equation}
where $e$ is the coupling constant and $\rho (x) \coloneqq i \left(  \psi (x) \pi (x) -  \psi ^\dagger (x) \pi ^\dagger  (x)   \right)$ with ${\pi(x)\coloneqq (D_0\psi(x))^\dag}$. This equation implies that the longitudinal electric field $E_L$ on a Cauchy slice  $\Sigma$ is a non-local functional of the charge distribution:
\begin{equation}
    E_L (x) = -\frac{e}{4\pi} \int _{ \Sigma }  \dt^3 y  \, \nabla ^x \frac{1}{|x-y|} \rho (y)\,.
\end{equation}
We can now split this observable into two parts -- one with support inside the subregion, and one with support outside. We will refer to these fields as  $E_L ^{\text{in}}$ and $E_L ^{\text{out}}$, respectively:
\begin{equation}
\begin{split}\label{eq:E-l-out-def}
 E_L  ^{\text{in}} (x) & \coloneqq -\frac{e}{4\pi} \int _{ \mathcal{O} }  \dt^3 y  \, \nabla ^x \frac{1}{|x-y|} \rho (y)\,,\\
  E_L ^{\text{out}} (x) & \coloneqq -\frac{e}{4\pi} \int _{\mathcal{O}^c}  \dt^ 3 y  \, \nabla ^x \frac{1}{|x-y|} \rho (y),
\end{split}
\end{equation}
where $\mathcal{O}^c$ is the complement of $\mathcal{O}$ in $\Sigma$. It is straightforward to see that $E_L ^{\text{out}}$ commutes with every local gauge-invariant observable in the interior $\mathring {\mathcal{O}}$ of $\mathcal{O}$. Furthermore, $\forall\, x \, \in \, \mathring{\mathcal{O}}$ we have, 
\begin{equation}
    \nabla \cdot  E_L ^{\text{out}} (x) = e  \int _{\mathcal{O}^c}  \dt^3 y  \, \triangle ^x\left( -\frac{1}{4\pi|x-y|}\right) \rho (y)=  e 
    \int _{\mathcal{O}^c}  \dt^3 y  \,  \d^3(x-y) \rho (y) =0\,.
\end{equation}
This means that inside $\mathcal{O}$,  $E_L ^{\text{out}}$ is (minus) the gradient of the \emph{harmonic} function 
\begin{equation}
    \Psi_{{\cal O}^{c}}(x)\coloneqq \int_{\mathcal{O} ^{c}}\, \dt^3 y \,  \frac{e}{4\pi|x-y|}\rho(y)\,.
\end{equation}
As such, inside $\mathcal{O} $, $E_L ^{\text{out}}$ is completely specified by its value on the corner $\p \mathcal{O}$, even though in global theory it is \emph{not} a corner field. In particular, since we are not imposing any boundary conditions on the corner (meaning that both $\Psi_{{\cal O}^{c}}$ and its normal derivative are free to take any value), Green's third identity gives the following bulk reconstruction formula  
\begin{align}\label{eq:eloutsurface}
E_{L}^{\rm out} (x) & =\, \frac{1}{4\pi} \nabla ^x \int_{\partial{\cal O}}\,\sqrt{\gamma} \dt ^{2} x_0 \, \l \Psi_{{\cal O}^{c}}(x_0)\, \nabla _\perp ^{x_0}\frac{1}{|x_0-x|} -\nabla_{\perp} ^{x_0 }\Psi_{{\cal O}^{c}}(x_0)\, \frac{1}{|x_0-x|} \r  \,,
\end{align}
where $\nabla_\perp$ indicates the outward normal derivative on the subregion boundary $\p \mathcal{O}$ and $\{x_0\}$ and $\gamma$ are the coordinates and induced metric on $\partial\cal O$, respectively.

In this sense, this externally sourced Coulombic mode is really an \emph{edge} mode, even though it lives in the bulk of the region.\footnote{Note that the internally sourced Coulombic mode $E_L^{\rm in}$ \emph{cannot} be determined by boundary data, since it is not divergence-free inside $\cal O$. Therefore, it is not an edge mode.} To re-iterate, this is a mode of the total Coulombic field that can be accessed and measured inside $\cal O$. We are not adding anything by hand. This is simply a consequence of the fact that imposing the (interacting) Gauss law identifies locally accessible electric field modes with particular smearings of the charge density in the complement.

This is precisely how we will attempt to define the central operators of our regional algebra in scalar QED -- they are not simply the pull-back of the fields on the corner, but rather the degrees of freedom that affect non-locally\footnote{By ``non-locally'' we are not referring to any violation of microcausality.} the physics inside the subregion via the constraint.

Notice that this definition works only in the interacting theory, which means that our corner modes are not the ones usually appearing in the discussion of edge mode. Specifically, they are not the ones contributing to the entanglement entropy of the free theory \cite{Donnelly:2014fua,Donnelly:2015hxa,Ball:2024hqe}, and they are thus also distinct from the edge modes \cite{Donnelly:2016auv,Geiller:2019bti,Carrozza:2021gju,Araujo-Regado:2024dpr,Ball:2024hqe,Riello:2021lfl,Shimizu:2026zug} alluded to in section~\ref{subsec:subregion phase space}, which likewise aleady appear in the free theory.

\section{Algebras on a Cauchy slice}
After our classical motivation, let us now construct $\star$-algebras of gauge-invariant operators in  scalar QED on a flat Cauchy slice $\Sigma$ in Minkowski spacetime, as well as on open subsets of it. 

Our strategy is to implement an algebraic incarnation of Dirac's ``constraint quantisation'' programme. Rather than imposing constraints on states, we shall start with an abstract kinematical $\star$-algebra of gauge-invariant operators on which we then impose the Gauss law by means of quotienting by the ideal generated by it. This yields a global physical $\star$-algebra $\mathcal{A}_{\rm phys}(\Sigma)$. Next, we also explore several options for studying physical subalgebras that one may associate with open subregions of the slice. Here, we also consider an extension of the physical algebra by adding the outer longitudinal electric field as a new centre mode to a regional algebra, just as in the classical case.  In the next sections~\ref{sec:GNS A_loc} and~\ref{sec:GNS A_C}, we then build a GNS representation of these various algebras. Given that they ``solve'' the Gauss law constraints, the GNS Hilbert space constitutes the physical Hilbert space of the theory.

Since we are working with the exact Gauss law, the resulting $\star$-algebras we obtain are likewise comprised of exact operators. The perturbative nature of the theory, by contrast, will only reflect in the perturbative expansion of the GNS state, of which we only consider the lowest order in the next section~\ref{sec:GNS A_loc}.

\subsection{Gauge-invariant kinematical Cauchy slice algebras}

Given a flat Cauchy slice $\Sigma$, we construct an abstract kinematical $\star$-algebra by smearing gauge-invariant local operator-valued distributions in a suitable manner across the slice.\footnote{The reason we do not start from a non-invariant kinematical algebra obtained by smearing the elementary linear fields $\psi(x), \pi(x),E(x),\vec{A}(x)$ is that, for the scalar sector, the resulting algebra would not contain our algebra as an invariant subalgebra due to the presence of composite operators. Furthermore, for simplicity of the later renormalisation, we keep multi-local operators, such as those obtained by smearing $\psi(x)W(x,y)\psi^\dag(y)$ with $W(x,y)$ a Wilson line between $x,y\in\Sigma$, out for the time being.} More precisely, for any open subset $\mathcal{O}\subset\Sigma$ our starting point is the gauge-invariant operators:
\begin{equation}
\begin{split}
     E (V)&\coloneqq\int _{\mathcal{O} \subset \Sigma }  \dt^3 x \,   V_{i } (x)E  ^{ i} (x)\,, \quad  B(V)\coloneqq\int _{\mathcal{O} \subset \Sigma }  \dt^3 x \,   V_{i } (x )B ^i (x)\,, \\
    \rho (f )& \coloneqq\int _{\mathcal{O} \subset \Sigma }  \dt^3 x \, f (x) :\rho (x):\,, \quad  J (V) \coloneqq\int _{\mathcal{O} \subset \Sigma }  \dt^3 x \,   V^{i } (x) :J _i (x):\,,  \\
     |\psi|^2 (f )& \coloneqq\int _{\mathcal{O} \subset \Sigma } \dt^3 x \, f (x) :|\psi|^2 (x):\,,
\end{split}
\end{equation}
which we take as our generators, where 
\begin{align}
    \rho (x) &\coloneqq i \left(  \psi (x) \pi (x) -  \psi ^\dagger (x) \pi ^\dagger  (x)   \right)\,,\\
    |\psi|^2 (x) &\coloneqq \psi^\dagger (x) \psi (x)\,, \\
    J_i(x) &\coloneqq i \l \psi  (x) (D _i \psi)^\dagger (x) - \psi ^\dagger (x) D_i \psi (x) \r\,,
\end{align}
and $V^i \in C^\infty _c \l T \mathcal{O} \r$ and $f \in  C^\infty _c \l \mathcal{O} \r $ are compactly supported smooth vector fields and functions inside $\mathcal{O}$, respectively. We shall see shortly why we include smearings of $|\psi|^2$ among the generators. As our generators in the matter sector are quadratic in terms of the Fourier coefficients, we always assume that these operators are normal ordered (as denoted by $:\cdot:$). This gives us a canonical basis with respect to the free field vacuum such that the CCR remain unchanged. Henceforth, we will drop the normal ordering symbol for notational simplicity unless explicitly required.

The  magnetic operators obey the Bianchi identity in the sense that:\footnote{For example, this follows from defining $B=\nabla\times{A}$ in terms of the elementary fields.} 
\begin{equation}\label{eq:0 GL Bi}
    B(\nabla f) = 0\qquad\forall\,f\in C^\infty_c(\cal O)\,.
\end{equation}
By contrast, the Gauss law 
\begin{align}
{\cal G}(f)\,\coloneqq\, \int_{\Sigma}\dt^3x\, ( \nabla \cdot E  - e \rho\, )(x)\, f(x) , \qquad f \in C_c^\infty (\cal O)
\end{align}
is not yet obeyed as an identity, so $E,\rho$ are independent and $E=E_L+E_T$ contains both transversal and longitudinal modes. However,  all operators commute with $\mathcal{G}(f)$, and so are gauge-invariant; hence, $\mathcal{G}(f)$ is central.

The non-vanishing commutators between the generators are
\begin{align}\label{eq:comm rel 0}
[E(V), B(W)]&=\, i \int_{O} \dt^3 x \, (\nabla \times V(x))\cdot W(x)\, ,\nonumber\\
[\rho(f),\, J(V)\, ]&=  -2 i  |\psi| ^2 (f \nabla \cdot V)\,,\nonumber\\
[E(V), J(W)] &=\, -2 i \,\vert\psi\vert^2\left(V\cdot W\right)\, .
\end{align}
The last two relations are the reason why we included smearings of $|\psi|^2$ among the generators. Note that $|\psi|^2(f)$ commutes with all other generators and is therefore central too.

Accordingly, we shall call the $\star$-algebra generated by these operators as the kinematical gauge-invariant algebra ${\cal A}_{\textrm{\text{kin}}}({\cal O})$ associated with the open set $\mathcal{O}\subset\Sigma$: 
\begin{align}
{\cal A} _{\text{\text{kin}}}({\cal O})\, \coloneqq\, \Big\langle\, E(V), \, B(V)\, ,\rho (f ), \, J(V), \, |\psi|^2 (f)\, \Big|\  \forall\, V\in C^\infty_c(T\mathcal{O}),f\in C^\infty_c(\mathcal{O}) \Big\rangle\,.
\end{align}
Here, $\langle A,B\rangle$ denotes the $\star$-algebra generated by $A,B$, i.e.\ the algebra closed under the $\star$-involution, linear combinations, and the operator product (and therefore the commutator).

These algebras are local: given two disjoint open sets $\mathcal{O}_{1}$ and $\mathcal{O}_2$,  ${\cal A}_{\textrm{\text{kin}}}(\mathcal{O} _1)$ and ${\cal A}_{\textrm{\text{kin}}}(\mathcal{O}_2)$ commute with each other. Thus, they constitute a net of local gauge-invariant kinematical algebras that we may combine into a global $\star$-algebra as follows:
\begin{align}
{\cal A}_{\text{kin}}(\Sigma)\, \coloneqq\, \bigcup_{\mathcal{O}\subset \Sigma} {\cal A}_{\textrm{\text{kin}}}(\mathcal{O})\,
 .\label{eq:Akin}
\end{align}

However, this is not quite the global Cauchy slice algebra we shall henceforth work with. Recall that all the local algebras are built from compact smearings. In particular, $\mathcal{A}_{\rm kin}(\Sigma)$ as currently defined contains only compactly smeared Gauss law operators as part of its centre. As we shall see shortly, this will preclude longitudinal electric field modes in subregions that are sourced by matter outside it.  However, physically, the longitudinal modes as Coulombic fields are sourced by the full matter content on the slice, and the algebra must account for that. 

It turns out that a slight extension of the \emph{global} algebra will remedy this issue. Rather than working with $\mathcal{A}_{\rm kin}(\Sigma)$, we shall invoke the following $\star$-algebra
\begin{align}
\bar{{\cal A}} _{\text{\text{kin}}}({\Sigma})\, \coloneqq\, \Big\langle\, E(V), \, B(V)\, ,\rho (f ), \, J(V), \, |\psi|^2 (f)\, \Big|\  \forall\, V\in C^\infty_{r^{-2}}(T\Sigma),f\in C^\infty_{r^{-1}}(\Sigma) \Big\rangle\,,\label{eq:slightextkin}
\end{align}
where $C^\infty_{r^{-1}}(\Sigma)$ and $C^\infty_{r^{-2}}(T\Sigma)$ denote the spaces of smooth functions and vector fields on $\Sigma$ such that $f\sim O(r^{-1})$ and $V^i\sim O(r^{-2})$ as $r\to\infty$, respectively.  Clearly, $\cal{A}_{\rm kin}\subsetneq\bar{\cal{A}}_{\rm kin}$, and now we have $\mathcal{G}(f)\in\bar{\cal{A}}_{\rm kin}$ for $f\in C^\infty_{r^{-1}}(\Sigma)$. Note that, while $f$ is allowed to have non-compact support, it still decays fast enough such that all $\mathcal{G}(f)$ remain small gauge constraint operators. Furthermore, notice that also our extended algebra does not contain any large-gauge charged observables, which in principle could be added.

\subsection{Global physical Cauchy slice algebra}\label{subsec:imposing the contraint}

Next, let us impose the Gauss law on the algebra. Being central elements, the Gauss law operators generate a two-sided ideal
\begin{equation}
    \mathcal{I}\coloneqq\big\langle \mathcal{G}(f)\,\big\vert\,f\in C^\infty_{r^{-1}}(\Sigma)\big\rangle\subset\bar{\cal{A}}_{\rm kin}\,.
\end{equation}
We will thus impose the Gauss constraints by means of the quotient
\begin{equation}
    \mathcal{A}_{\rm phys}(\Sigma)\coloneqq \bar{\cal{A}}_{\rm kin}/\mathcal{I}\,,\label{eq:quotient}
\end{equation}
yielding the global \emph{physical} algebra, which we will henceforth adopt. Each of its elements $[a]\in\mathcal{A}_{\rm phys}(\Sigma)$ is a co-set $[a]\coloneqq\{a+I\,|\,a\in\bar{\cal A}_{\rm kin},\forall\,I\in\mathcal{I}\}$. 

In physical terms, we identify the charge distribution $e\rho(f)$ with the electric field $E(\nabla f)$ for all $f\in C^\infty_{r^{-1}}(\Sigma)$. More precisely, since the transversal part of the field obeys $E_T(\nabla f)=0$, we identify the longitudinal component $E_L(\nabla f)$ with $e\rho(f)$, and so may eliminate either operator in favour of the other. In what follows, we choose to eliminate the former, so that our description of the global physical $\star$-algebra reads
\begin{equation}
    {\cal A}_{\rm phys}(\Sigma)\simeq\, \Big\langle\, E_T(V), \, B(V)\,, \rho (f ), \, J(V), \, |\psi|^2 (f)\, \Big|\  \forall\, V\in C^\infty_{r^{-2}}(T\Sigma),f\in C^\infty_{r^{-1}}(\Sigma) \Big\rangle\,.\label{eq:physalgebra}
\end{equation}
This algebra still features a centre generated by $|\psi|^2(f)$. The only non-trivial commutator of its generators that differs from eq.~\eqref{eq:comm rel 0} is the last one, which now becomes
\begin{equation}
    [E_T(V), J(W)] =\, -2 i \,\vert\psi\vert^2\left(V_T\cdot W_T\right)\, .
\end{equation}

Even though we eliminated the longitudinal electric field as an independent observable, ${\cal A}_{\rm phys}$ still includes it as a \emph{dependent} operator. To this end, note that an arbitrary ${V\in C^\infty_{r^{-2}}(T\Sigma)}$ can be uniquely written via the Hodge decomposition as
$V=V_T+V_L$
provided we require $\lim_{r\to\infty}V_{T,L}=0$. In this case, we have
\begin{equation}
    V_L = \nabla \Psi _V \,,\qquad\qquad V_T=\nabla \times {U} \,,
\end{equation}
where
\begin{equation}\label{eq:psi_v def}
    \Psi _V (x) = -\frac{1}{4 \pi } \int _{\Sigma } \dt^3 y \, \frac{\nabla \cdot V (y)}{|x-y|} , \quad \quad  {U}  = \frac{1}{4 \pi } \int _{\Sigma } \dt^3 y \, \frac{\nabla \times V (y)}{|x-y|} \,.
\end{equation}
In particular, this implies that $\Psi_V\in C^\infty_{r^{-1}}(\Sigma)$ and ${U}\in C^\infty_{r^{-1}}(T\Sigma)$. Now ${E_L (\nabla \times {U}) = 0}$ for any such $\vec{U}$, so that we may write 
\begin{equation}\label{eq:El as rho}
    E_L (V) = E_ L \l \nabla \Psi _V \r  \coloneqq e \rho (\Psi _V)\,\qquad\forall\,V\in C^\infty_{r^{-2}}(T\Sigma)\,.
\end{equation}
Since $\Psi_V\in C^\infty_{r^{-1}}(\Sigma)$, the right hand side resides in $\mathcal{A}_{\rm phys}(\Sigma)$. Thus, we may use eq.~\eqref{eq:El as rho} to define $E_L$ within our description of the physical algebra in eq.~\eqref{eq:physalgebra}, as indicated by the colon. This definition also entitles us to think of $E_L$ as being smeared by the same  functions $V\in C^\infty_{r^{-2}}(T\Sigma)$ as the other generators in $\mathcal{A}_{\rm phys}$ that correspond to vector fields.

\subsection{Regional physical Cauchy slice algebras}\label{ssec:regionalphysalg}

We noted above that $\mathcal{A}_{\rm kin}$ is a global net of local kinematical gauge-invariant $\star$-algebras on $\Sigma$ embedded in the slightly extended global algebra $\bar{\cal A}_{\rm kin}$. Can one similarly think of $\mathcal{A}_{\rm phys}(\Sigma)$ as a global net of local \emph{physical} algebras $\mathcal{A}_{\rm loc}(\mathcal{O})$ ascribed to any open subset $\mathcal{O}\subset\Sigma$? There is an obvious sense in which this can be made to hold. Namely, if we define
\begin{equation}\label{eq:AlocO}
   \mathcal{A}_{\rm loc}(\mathcal{O})\coloneqq\Big\langle\, E_T(V), \, B(V)\,, \rho (f ), \, J(V), \, |\psi|^2 (f)\, \Big|\  \forall\, V\in C^\infty_c(T\mathcal{O}),f\in C^\infty_c(\mathcal{O})\Big\rangle \,,
\end{equation} 
i.e.\ in terms of compact regional smearing, then $\mathcal{A}_{\rm loc}(\mathcal{O}_1)$ and $\mathcal{A}_{\rm loc}(\mathcal{O}_2)$ would commute for any disjoint $\mathcal{O}_1,\mathcal{O}_2$. In this case, $\mathcal{A}_{\rm phys}(\Sigma)$ would be a slight extension of the union of this local net, in analogy to their kinematical counterparts in eqs.~\eqref{eq:Akin} and~\eqref{eq:slightextkin}.

However, if we want our regional physical algebras to encompass all the observables (built from the generators of the global algebra) that could in principle be locally measured in that region, then we have to proceed differently. Indeed, also the longitudinal electric field is a mode that is in principle measurable, and so $E_L(V)$ for $V\in C^\infty_c(T\mathcal{O})$ should be included in the definition of regional physical algebras. 

This observable has also previously been part of the local kinematical gauge-invariant algebras $\mathcal{A}_{\rm kin}(\mathcal{O})$; upon the quotient in eq.~\eqref{eq:quotient} it still resides in $\mathcal{O}$, though now as a dependent observable, depending through eq.~\eqref{eq:El as rho} non-locally on the matter distribution anywhere on $\Sigma$. This is because $\Psi_V$, defined in eq.~\eqref{eq:psi_v def}, is not necessarily compactly supported in $\mathcal{O}$, even when $V\in C^\infty_c(T\mathcal{O})$. 
Since $\mathrm{supp}(\nabla\cdot V)\subset\mathcal{O}$, we have that $\Psi_V$ is harmonic in the complement $\mathcal{O}^c\coloneqq \Sigma\setminus\mathcal{O}$. As such, it is entirely defined by its boundary value on $\partial\mathcal{O}$. If it vanishes there, $\Psi_V$ is compactly supported in $\mathcal{O}$, and $E_L(V)=e\rho(\Psi_V)$ is already contained in $\mathcal{A}_{\rm loc}(\mathcal{O})$ above. Otherwise, $\Psi_V$ does not have compact support even on $\Sigma$, in which case $E_L(V)=e\rho(\Psi_V)$ does not reside in $\mathcal{A}_{\rm loc}(\mathcal{O})$. Thus, if we want our longitudinal electric field to be sourced by the matter content on the entire slice, as it physically should, then we have to include $E_L(V)=e\rho(\Psi_V)$ for all $V\in C^\infty_c(T\mathcal{O})$ in our regional algebras, not only for those $V$ such that $\Psi_V$ is compact. This is the reason for the extension in eq.~\eqref{eq:slightextkin}. 

As the basic regional physical algebra, we therefore take
\begin{equation}
    \mathcal{A}_{\rm phys}(\mathcal{O})\coloneqq\Big\langle\, E_T(V),E_L(V), \, B(V)\,, \rho (f ), \, J(V), \, |\psi|^2 (f)\, \Big|\  \forall\, V\in C^\infty_c(T\mathcal{O}),f\in C^\infty_c(\mathcal{O})\Big\rangle \,,
\end{equation} 
with $E_L(V)$ defined in eq.~\eqref{eq:El as rho}. Note that $E_L(V)$, unlike all the other generators, is linear in $e$ and so of higher order in the perturbative expansion. Owing to the implemented Gauss law, these regional algebras are no longer local: $\mathcal{A}_{\rm phys}(\mathcal{O}_1)$ and $\mathcal{A}_{\rm phys}(\mathcal{O}_2)$ do not commute even when $\mathcal{O}_1,\mathcal{O}_2$ are disjoint. Indeed, for any $V\in C^\infty_c(T\mathcal{O}_1)$ and $W\in C^\infty_c(T\mathcal{O}_2)$, we have 
\begin{equation}
    [ E_L (V), \, J(W)] =e\,[ \rho (\Psi _V ), \, J(W)]=  -2i e |\psi|^2( \nabla\cdot{ W} \Psi _V )\,.\label{eq:newcommutator}
\end{equation}
Hence, whenever ${\rm supp}(\Psi_V)\cap {\rm supp}(\nabla\cdot W)\neq\emptyset$, the commutator does not vanish and belongs to $\mathcal{A}_{\rm phys}(\mathcal{O}_2)$ because ${\rm supp}(\nabla\cdot{ W} \Psi _V)\subset{\rm supp}(W)\subset\mathcal{O}_2$. In particular, setting $\mathcal{O}_1=\mathcal{O}_2=\mathcal{O}$, we see that the commutator already belongs to $\mathcal{A}_{\rm phys}(\mathcal{O})$. 

Notice that $\mathcal{A}_{\rm phys}(\mathcal{O})$ is an algebra with non-trivial matter centre; as before, $|\psi|^2(f)$ commutes with all other generators and so generates this centre. In the next sections, we will see that the standard Fock vacuum suffices to give rise to a GNS representation of these regional algebras (and thus of $\mathcal{A}_{\rm loc}(\mathcal{O})$ too).

\subsection{Splitting longitudinal fields into inner and outer modes}

We have not yet considered a notion of corner modes by which we might want to augment our regional algebras; all fields are compactly supported in $\mathcal{O}$. This includes the full longitudinal electric field $E_L(V)$ if viewed as an independent field, though via the Gauss law it is expressed in terms of the matter content on all of $\Sigma$. Thus, as a next step, and just as in the classical case, we split this field into an inner and outer matter contribution. This will result in an independent \emph{central} generator  which may be viewed as a corner mode and that we will add to the algebra associated with a fixed region $\mathcal{O}$. 

This step is  more consequential  than any of our previous ones: it distinguishes the corner $\partial\mathcal{O}$ across which we make this split because the new central generator is not yet contained even in the global algebra $\mathcal{A}_{\rm phys}(\Sigma)$. In other words, for any corner $\partial\mathcal{O}$ across which one wishes to make this split, one has to extend the global physical algebra by hand.

In line with this, we will see in section~\ref{sec:GNS A_C} that extending the Fock vacuum based GNS representation of the regional algebras to this new centre mode necessitates an additional corner renormalisation. The resulting representation is unitarily inequivalent and corresponds to an effective corner field theory, rendering the cut physically significant. Specifically, one would have to carry out such a corner renormalisation for any region for which one wishes to include this split, each time  at the cost of modifying the physical theory. Our main point is, however, that a consistent representation of the centrally extended regional algebra does exist.\footnote{Our step of splitting $E_L(V)$ into an inner and an outer piece is somewhat akin to splitting a modular boost Hamiltonian $H=H_L+H_R$ into a contribution to the left and right of a Killing horizon, which results in highly UV-divergent objects $H_L,H_R$. Even bounded functions of these cannot simply be added to the regional  algebras on the left or right of the horizon in a given representation. The crux in our case is that an additional corner renormalisation turns the inner and outer longitudinal modes well-defined, cf.~section~\ref{sec:GNS A_C}. } This observation is broadly consistent with recent developments in the standard covariant formulation of algebraic quantum field theory, where it was shown that for a \emph{fixed} region and its complement one may include edge modes in the theory \cite{Fewster:2025ijg}.

Let us now consider a \emph{fixed} $\mathcal{O}\subset\Sigma$ and perform the split similarly to section \ref{subsec:corner modes}:
\begin{equation}\label{eq:totalEL}
    E_L(V) = E_L ^{\text{out}} (V) +  E_L ^{\text{in}} (V)\,,\qquad V\in C^\infty_c(T\mathcal{O}) \,,
\end{equation}
where
\begin{equation}\begin{split}
E_L ^{\text{out}} (V) & \coloneqq e \rho ( 1_{\mathcal{O}^c}\Psi _V )  =  e \int _{\mathcal{O}^c} \,  \dt ^3 y \, \Psi _V (x) \rho (x)\,, \\
E_L ^{\text{in}} (V) & \coloneqq e \rho ( 1_{\mathcal{O}}\,\Psi _V )  \,= e \int _{\mathcal{O}} \,  \dt ^3 y \, \Psi _V (x) \rho (x) \,,
    \end{split}\label{eq:starELout}
\end{equation}
and we used the characteristic function
\begin{equation}
\mathbf{1}_{\mathcal{S} }(x)\coloneqq \begin{cases}
    1 & \quad \text{if} \quad x \in \mathcal{S} \\
    0 & \quad \text{otherwise}\,.
\end{cases}
\end{equation}
Clearly, due to the discontinuous smearing, neither $E_L^{\rm out}(V)$ nor $E_L^{\rm in}(V)$ are contained in $\mathcal{A}_{\rm phys}(\Sigma)$.  Inserting the definition of $\Psi_V$ in eq.~\eqref{eq:psi_v def} for compactly supported $V$ into eq.~\eqref{eq:starELout} gives $E_L^{\rm out}(V)$, $V\in C_c^\infty(T\mathcal{O})$, in the transparent form of eq.~\eqref{E_L^out}.

Since their sum resides in $\mathcal{A}_{\rm phys}(\mathcal{O})$, it suffices to add only one of them, say the outer longitudinal  field $E^{\rm out}_L(V)$, as a new generator. We noted in section~\ref{ssec:regionalphysalg} that $\Psi_V$ is harmonic in $\mathcal{O}^c$ and so is entirely determined by its value on the corner $\partial\mathcal{O}$. It is in this sense that we may regard the outer longitudinal electric field as a corner mode also in the quantum theory. (By contrast, the inner mode $E_L^{\rm in}$ is sourced by matter inside $\mathcal{O}$, smeared by a non-harmonic function, and thus not a corner mode.) Its addition results in an augmented regional $\star$-algebra associated with $\mathcal{O}$: 
\begin{equation}
      \mc{A} _C (\mathcal{O}) \coloneqq \mc{A} _{\text{phys}} (\mathcal{O}) \vee  \mc{A} _{\rm out} (\mathcal{O}), \qquad  \mc{A} _{\rm out} (\mathcal{O}) 
     \coloneqq \langle E_L^{\rm out}(V)\,|\,V\in C^\infty_c(T\mathcal{O})\rangle \,.\label{eq:Aout}
\end{equation}

The added outer algebra is Abelian, and commutes with all other elements in $\mathcal{A}_C(\mathcal{O})$. Indeed, eq.~\eqref{eq:newcommutator} implies for $V,W\in C^\infty_c(T\mathcal{O})$
\begin{equation}
   [ \rho ( \Psi _V ), \, J(W)] =[ \rho ( \mathbf{1} _\mathcal{O} \Psi _V ), \, J(W)]=  -2i |\psi|^2( \nabla\cdot\, W \Psi _V )  \quad \text{and} \quad  [ \rho ( \mathbf {1}_{\mathcal{O}^c}\Psi _V ), \, J(W)] =0 \,.
\end{equation}
$\mathcal{A}_C(\mathcal{O})$ thus has an expanded centre, which is the reason for the $C$ subscript. In fact, for later purpose, we may also consider the global Cauchy slice algebra extended by the outer longitudinal electric field for a \emph{fixed} subregion $\mathcal{O}$, i.e.
\begin{equation}
    \mathcal{A}_C(\Sigma)\coloneqq\mathcal{A}_{\rm phys}(\Sigma)\vee \mathcal{A}_{\rm out}(\mathcal{O})\,.
\end{equation} 

We now have all the ingredients to build representations for both $\mathcal{A}_{\rm phys}(\Sigma)$ and $\mathcal{A}_C(\Sigma)$. This is where the perturbative nature of the theory will come to the fore, namely in the perturbative expansion of the states. Note that in our abstract discussion above, by contrast, we have worked with the full Gauss law. Accordingly, all our algebras are exact.

\section{Building a perturbative GNS state for the Cauchy slice algebra}
\label{sec:GNS A_loc}

In this section, we begin quantising the $\star$-algebra ${\cal A}_{\textrm{phys}}(\Sigma)$ perturbatively using the GNS construction. Here, we will  demonstrate that correlation functions of this algebra are well-defined in the Fock vacuum, provided we impose certain restrictions on the overlap of smearing functions. In the next section, we shall deal with lifting these restrictions by renormalising all the correlation functions. This will then define a new vacuum state for our algebra. Once correlation functions are well-defined, the corresponding GNS construction exists, though we will not explicitly build the Hilbert space in this paper.

In an interacting field theory,  one route to GNS quantisation is to start with the $\star$-algebra and specify a positive linear functional $\omega[[e]]$ which is a formal power series in the  coupling\footnote{Thanks to Furry's theorem, only even powers of $e$ appear in correlators and so here in $\omega$.} 
\begin{align}
\omega[[e]]\, \coloneqq \, \omega_{F}^{R}\, +\, \sum_{n=1}^{\infty}\, e^{2n}\, \omega_{n}\,,
\end{align} 
where the ``leading order'' term in this expansion is denoted as $\omega_{F}^{R}$ for reasons that will become clear shortly. We will  refer to $\omega_{F}^{R}$ as the quasi-free GNS state and $\omega[[e]]$ as the perturbative GNS functional.\footnote{We remind the reader that $\omega[[e]]$ is simply a formal power series and does not converge in the interacting theory. It is thus not a GNS state in the true sense.}

Let us now show here and in the next section that we can use the free field Fock vacuum state $\omega_F$ on the covariant free field algebra generated by the complex scalar and Maxwell fields to define the quasi-free GNS state $\omega^R_F$ on the physical Cauchy slice algebra given in eq.~\eqref{eq:physalgebra}. We achieve this by pulling back the Fock vacuum $\omega_F$ to this canonical algebra; $\omega^R_F$ is this pullback, i.e.\ $\omega^R_F(a)\coloneqq\omega_F(a)$ for $a$ in which the matter operators have non-overlapping support. 
 As $\mathcal{A}_{\rm phys}(\Sigma)$ contains both $\mathcal{A}_{\rm loc}(\mathcal{O})$ and $\mathcal{A}_{\rm phys}(\mathcal{O})$ for any region $\mathcal{O}$, $\omega^R_F$ also defines a GNS state for our regional canonical algebras.

We expect that the sub-leading terms in the perturbative expansion can be constructed by adapting the methods of perturbative algebraic quantum field theory \cite{Hollands:2001nf, Brunetti:2009qc, Fredenhagen:2015iia,Brunetti:2025ocr,Rejzner:2016hdj}. However, such a perturbative expansion will not be analysed in this paper.

\subsection{Pulling back the Fock vacuum to a slice}\label{subsec:omeonalocononover}

For notational compactness, we denote by $\Psi_{{I}}(F_{I})$ the  generators  of ${\cal A}_{\textrm{phys}}({\Sigma})$ in eq.~\eqref{eq:physalgebra}, where $F_{I}$ are the corresponding test fields dual to $\Psi_{I}$ and $\Psi_1=E_T, \Psi_2=B$, while $I=3,4,5$ label the gauge-invariant matter operators (which are quadratic in $\psi$ and its derivatives).

We now establish that the covariant Fock vacuum $\omega_{F}$ can be pulled back to a linear functional on the subspace generated by all the elements of ${\cal A}_{\textrm{phys}}(\Sigma)$, which are linear combinations of monomials $\prod_{i=1}^{n}\Psi_{I_{i}}(F_{I_{i}})$, such that 
for all the matter operators in this monomial, their corresponding smearing fields have pairwise trivial overlap. In appendix  \ref{sec_positivity} we discuss what needs to be proven to establish the positivity of this state.

First, on any linear polynomial in ${\cal A}_{\textrm{phys}}(\Sigma)$, the pullback of the Fock vacuum is trivial thanks to normal ordering 
\begin{align}
\omega_{F}^{R}\l \sum_{I=1}^{5}\, a_{I}\, \Psi_{I}(F_{I})\r \, \coloneqq\,  \sum_{I=1}^{5} \, a_{I}\, \omega_{F}(\Psi_{I}(F_{I}))\, =\, 0\quad \, \forall \,a_I\, \in \mathbb{C}\,.
\end{align}

Let us thus determine $\omega_{F}^{R}$ on (the above specified subset of) quadratic monomials.
We begin by considering $\omega_{F}^{R}(E_{T}(V)\, E_{T}(W))$ for all $V,\, W\, \in\, C_{r^{-2}}^{\infty}(T{\Sigma})$. The Fock vacuum $\omega_{F}$ evaluates the un-smeared quadratic monomials to a well defined distribution,
\begin{align}\label{eixejyff}
\omega_{F} ^R ((E_T )_{i}(x)\, ( E_T ) _{j}(y))\, =\, \frac{1}{8}\, (\partial_{i}\partial_{j} - \delta_{ij}\triangle)\, \frac{1}{r^{2}}\, =\, \omega_F ^R(B_{i}(x) B_{j}(y))\,,
\end{align}
where $r = \vert x - y\vert$. 
Hence, using partial integration, we get
\begin{align}
\omega_{F}^{R}(E_T(V)\, E_T(W))\, & =\, \frac{1}{8}\, \int_{{\Sigma}^{2}}\dt^3{x}\,\dt^3{y}\, V^{i}(x) W^{j}(y) (\partial_{i}\partial_{j} - \delta_{ij}\triangle)\, \frac{1}{r^{2}}\nonumber\\
& =\, \frac{1}{8}\int_{{\Sigma}^{2}}\dt^3{x}\,\dt^3{y}\, V(x) \cdot (\, \nabla \times\, (\, \nabla \times\, W)\, )(y)\, \frac{1}{\vert x - y \vert^{2}}\,.
\end{align}
We also note that the corresponding momentum space kernel is simply
\begin{align}
C^{ij}_{E_{T}E_{T}}(k)\, &=\, C^{ij}_{BB}(k)\, =\, \vert k\vert\, \l \, \delta_{ij} - \frac{k_{i}k _{j}}{|k|^2}\, \r \, \geq\, 0\quad  \forall\, k\,.\nonumber
\end{align}
This kernel behaves as $|k|$ for large $|k|$. The norm for test vector fields induced by $\omega_{F}^{R}$ on $C_{r^{-2}}^{\infty}(T{\Sigma})$ is given via the Parseval-Plancherel theorem,\footnote{This is the norm of the Sobolev space $H^{\frac{1}{2}}({\bf R}^{3})$.}
\begin{align}
\vert\vert\, V \vert \vert^{2}\, =\, \int \frac{\dt ^{3}k}{(2\pi)^{3}}\, V(k) \cdot V(-k) \,C_{E_{T}E_{T}}(k)\,.
\end{align}
The same result also holds for the magnetic field. The two-point functions for the (transversal) electric and magnetic fields are thus well-defined.

Similarly, as long as the smearing fields do not overlap, the covariant $\omega_{F}$ can be pulled back to the quadratic polynomials in the matter sector. We leave the details of this computation to the reader, as they simply involve Wick contractions of free fields in the Fock vacuum. To leading order in perturbative expansion the result is, 
\begin{align}\label{eq:tpfomdop}
\omega^R_{F}(\rho(f) \, \rho(g))\, &= 2 \, 
\, \int _{\Sigma^2}\dt^3{x}\,\dt^3{y}\, f(x) g (y)\, G_{+}(r)\, G_{-}(r)\,,\nonumber\\
\omega^R_{F}(J(V)\, J(W))\, &=\, \frac{1}{2}\, \int_{{\Sigma}^{2}}\dt^3{x}\,\dt^3{y}\, V^{i}(x) \,W^{j}(y)\, \big(-\partial_{i}G_+(r)\partial_{j}G_+(r)+ G_+\partial_{i}\partial_{j}G_+ (r)\, \big)\,\nonumber\\
\omega^R _F \l |\psi| ^2 (f ) \, |\psi| ^2 (g )  \r & =  \frac{1}{4} \int_{{\Sigma}^{2}}\dt^3{x}\,\dt^3{y}\, f(x) \,g (y)\, G_+ ^2 (r)  \,,
\end{align}
 where 
\begin{align}\label{eq:Gpm def}
G_{+}(x-y)\, &=\, \omega_{F}(\psi(x)\psi^{\dagger}(y))\, = \int \frac{\dt ^{3}p}{(2\pi)^{3}}\frac{1}{\omega_{p}}\, e^{i p \cdot (x - y)} \, =\,  \frac{m}{2\pi^{2} \vert x - y \vert}\, K_{1}(m \vert x - y \vert)\,,\nonumber\ \nonumber\\
G_{-}(x - y)&=\, \omega_{F}(\pi_{\psi}(x)\, \pi_{\psi}^{\dagger}(y))\, =\, \int \frac{\dt ^{3}p}{(2\pi)^{3}}\omega_{p}\, e^{i p \cdot (x - y)} =\,  - \frac{m^{2}}{2 \pi \vert x - y \vert^{2}}\, K_{2}(m \vert x - y \vert)\,,
\end{align}
and $K_{i}$ are the modified Bessel functions of the second kind. When the smearing fields do not overlap, the singular $r\to0$ limit of these Bessel functions is avoided and the two-point functions remain finite. The case with non-trivial overlap is considered in the next section.

Let us now consider how the covariant Fock vacuum $\omega_{F}$ evaluates higher order monomials in ${\cal A}_{\textrm{phys}}(\Sigma)$ of the form $\Psi_{I_{1}}(F_{I_{1}})\, \dots\, \Psi_{I_{n}}(F_{I_{n}})$. Without loss of generality, we focus on the scalar field quantities, i.e.\ we assume
\begin{align}
3\, \leq\, I_{i}\, \leq\, 5\,, \quad  \forall\, 1\, \leq\, i\, \leq\, n\,,
\end{align}
since the $n$-point function of linear fields, ($E_T(V)$ and $B(V)$) on a Cauchy slice in a free field vacuum is simply the product of one- and two-point functions. Let us further assume here that the involved generators do not overlap, i.e.\ that
\begin{align}
\textrm{supp}(F_{I_{i}})\, \cap\, \textrm{supp}(F_{I_{j}})\, =\, \emptyset \, \quad  \forall\, (i,j) \,.
\end{align}
In this case, once again $\omega_{F}^{R}$ can be defined as simply the pullback of $\omega_{F}$.  We demonstrate this only for a specific example of such a $n$-point monomial, namely the scalar field density
\begin{align}
\omega_{F}^{R}\left(\prod_{i}\, \rho_{i}(f_{i})\right)\, \coloneqq\, \omega_{F}\left(\prod_{i}\, \rho_{i}(f_{i})\right)\,.\label{eq:omegarhof}
\end{align}
The right hand side can be evaluated using Wick's theorem and we find that (momentarily reinstating the normal ordering symbol)
\begin{align}
\omega_{F}^{R}\left(\prod_{i}\, :\rho_{i}(f_{i}):\right)\, =\, \sum_{P}\, \prod_{(i,j) \in\, P} \omega_{F}^{R}\left(:\rho_{i}(f_{i}):\, :\rho_{j}(f_{j}):\right)\, +\, G_{c}^{F}(f_{1}, \dots\, f_{n})\,,\label{eq:Wick}
\end{align}
where $P$ is the set of pairs $(i,j)$ with 
$i < j$ mod $n$. $G_{c}^{F}(f_{1}, \dots,\, f_{n})$ is the connected smeared $n$-point function of the density operator in a quasi-free Fock vacuum. Recall from above that $\omega_{F}^{R}\left(:\rho_{i}(f_{i}):\, :\rho_{j}(f_{j}):\right)$ is well-defined when $f_i,f_j$ are non-overlapping as assumed.

We now argue that $G_{c}^{F}(f_{1}, \dots\, f_{n})\, \in\, {\mathbb{C}}$, i.e.\ that also the connected piece does not diverge. Indeed, every term in $G_{c}^{F}$ is of the form
\begin{align}
 \int_{{\Sigma}^{n}}\, \prod_{i=1}^{n}\, \dt^{3}x_{i}\, f_{i}(x_{i})\, G_{\alpha_{1}}(x_{i_{1}} - x_{i_{2}})\, G_{\alpha_{2}}(x_{i_{2}} - x_{i_{3}})\, \dots\, G_{\alpha_{n}}(x_{i_{n}} - x_{i_{1}})\,,
\end{align}
where   $\alpha_i\, \in\, \{+, -\}$ and $i_{k}\, \in\, \{1,\, \dots\, n\}$; $G_c^F$ is simply the sum over all terms obtained (up to sign) by permuting the arguments $x_{i_k}$ in the two-point correlators $G_{\alpha_i}$. Since all $f_i$ are non-overlapping, we have
\begin{align}
\Big\vert\,\prod_{i=1}^n f_i(x_i) \,G_{\alpha_{1}}(x_{i_{1}} - x_{i_{2}})\, G_{\alpha_{2}}(x_{i_{2}} - x_{i_{3}})\, \dots\, G_{\alpha_{n}}(x_{i_{n}} - x_{i_{1}})\, \Big\vert\, <\, M\Big\vert\,\prod_{i=1}^n f_i(x_i) \Big\vert\qquad\forall \,\alpha_i\,,
\end{align}
for some finite positive constant $M$, which may be chosen the same for all the permutations. Consequently, taking all permutations into account,
\begin{align}
\vert\, G_{c}^{F}(f_{1}, \dots,\, f_{n})\vert\, <\, (n-1)!\,M\, \prod_{i}\, \vert\vert f_{i} \vert\vert_{L^{1}}\,,
\end{align}
which is finite. Thus, the expression in eq.~\eqref{eq:omegarhof} is well-defined.

This observation extends straightforwardly to higher order monomials in other matter operators, completing the specification of $\omega_{F}^{R}$ on all elements of ${\cal A}_{\textrm{phys}}(\Sigma)$ which are given by combinations of generators corresponding to sets of pairwise non-overlapping test fields.

As a next step, we turn to the case of overlapping smearing fields in the matter sector, which will require renormalisation.

\subsection{Renormalising the canonical vacuum}
\label{sec:extending domain}

In this section, we will show how to extend the domain of $\omega_{F}^{R}$ to all of ${\cal A}_{\textrm{phys}}(\Sigma)$, i.e.\ including for correlators with overlapping smearing fields. The primary obstruction to such an extension comes from the well-known result in QFT that equal time correlation functions of operators (that are not linear in free fields) which are smeared only in spatial directions are ill-defined \cite{jaffe1966wick,Yazdi_2023,streater2016pct}. Specifically, in the present context this means that $\omega_{F}\l \Psi_{I}(F_{I}) \Psi_{J}(F_{J}) \r $ diverges $\forall\, I\, \in\, \{3,4,5\}$ when $\textrm{supp}(F_{I})\, \cap\, \textrm{supp}(F_{J})\, \neq\, \emptyset$. We will now prove that all of the divergences present in two-point functions can be renormalised. This will subsequently allow us to define a positive linear functional $\omega_{F}^{R}$ on all of ${\cal A}_{\textrm{phys}}(\Sigma)$.

We first summarise the main results of this section as it will assist the reader in navigating several technical details that follow.
\begin{itemize}
\item In section \ref{subsec:time_smear}, we revisit the known result that a purely spatial smearing of composite operators yields UV divergences by explicitly computing $\omega_{F}(\rho(f, \epsilon)\, \rho(g, \epsilon))$ in $\Sigma \times [-\epsilon, \epsilon]$ and examining the precise nature of the divergence as $\epsilon\to0$. This example also illustrates why renormalisation is not a free lunch, in the sense that the existence of a renormalised vacuum, locality of the Hamiltonian, and absence of negative norm states in the Hilbert space built from such a vacuum, require a careful choice of the subtraction scheme.
\item In section \ref{subsec:ren}, we show that we can renormalise the equal time matter correlators on $\Sigma$ using differential renormalisation \cite{Freedman:1991tk}. This then leads to a well-defined linear functional $\omega_{F}^{R}$ on ${\cal A}_{\textrm{phys}}(\Sigma)$.

\item In section \ref{subsec:restoring_positivity}, we show that a suitable choice of local counter-terms can ensure that $\omega_{F}^{R}$ is positive-definite on ${\cal A}_{\textrm{phys}}(\Sigma)$. In other words, it can be used as a starting point for a perturbative GNS quantisation of ${\cal A}_{\textrm{phys}}(\Sigma)$.
\item In section \ref{subsec:higherptren}, we finally show that the higher-point functions are rendered UV-finite in $\omega_{F}^{R}$.
\end{itemize}

A quasi-free positive GNS state on $\Sigma$ that generates finite two-point functions for quadratic operators such as $\rho$ cannot be the vacuum state of a free theory, but belongs to an effective field theory (EFT) with higher-derivative spatial interactions. This will then be our continuum model in the next section to explore the existence of the longitudinal electric corner modes associated to a spatial subregion in $\Sigma$.

\subsubsection{Absence of time smearing as the origin of divergences} \label{subsec:time_smear}
Even in the theory of a free (charged) scalar field, the GNS state associated to the Fock vacuum can not be pulled back onto the Cauchy slice algebra generated by the composite operators by $\rho, J$. The simplest way to see this is to consider the pullback of the Fock vacuum state $\omega_F$ on a time-thickened Cauchy slice $\Sigma \times [-\epsilon,\, \epsilon]$, where the thickness is parametrised by $\epsilon$. The two-point functions diverge in the limit $\epsilon\, \rightarrow\, 0$ if the spatial support of the smearing functions overlap, as we now demonstrate.

Let $F,\, G$ be two smearing functions in spacetime which have temporal support inside the thin strip $[-\epsilon,\, \epsilon]$. We take the set of functions which are of the form
\begin{align}
F(X)\, =\, \kappa_{\epsilon}(t_{X})\, f(x)\,,
\end{align}
where $X$ denotes a point in spacetime, $\kappa_{\epsilon}(t_{X})$ is the bump function defined as 
\begin{align}
\kappa_{\epsilon}(t)\, =\begin{cases}
    \frac{{\cal N}}{\epsilon}\, e^{\frac{1}{(\frac{t}{\epsilon})^{2} - 1}}\, &\textrm{for}\, \vert t \vert < \epsilon\,,\nonumber\\
    0\,& \textrm{for}\, \vert t \vert \geq \epsilon\,,
\end{cases}
\end{align}
and ${\cal N}$ is a normalisation constant. 
Note that
\begin{align}
\int_{-\infty}^{\infty} \kappa_{\epsilon}(t) = 1\nonumber\,, \quad  \quad \lim_{\epsilon\, \rightarrow\, 0}\, \kappa_{\epsilon}(t) = \delta(t)\,.
\end{align}

The (covariant) Wightmann two-point function of the charge density at different spacetime points in momentum representation is: 
\begin{align}
\omega_{F}(\rho(X)\, \rho(Y))\, =\,  \int\, \frac{\dt ^{3}p}{(2\pi)^{3}2\omega_{p}}\, \frac{\dt ^{3}{q}}{(2\pi)^{3} 2\omega_{q}}\, (\omega_{p} - \omega_{q})^{2}\, e^{-i (p + q)\, \cdot\, (X - Y)}\,.
\end{align}
For $F(X)=\kappa_{\epsilon}(t_X)f(x)$ and $G(Y)=\kappa_{\epsilon}(t_Y)g(y)$, this formula leads to: 
\begin{align}
\omega_{F}(\rho(F)\, \rho(G))\, =\, \int\, \frac{\dt ^{3}p}{(2\pi)^{3}2\omega_{p}}\, \frac{\dt ^{3}{q}}{(2\pi)^{3} 2\omega_{q}}\, (\omega_{p} - \omega_{q})^{2}\, \tilde{f}(p+ {q})\, \tilde{g}(-p-{q})\, \vert\, \tilde{\kappa}_{\epsilon}(\omega_{p} + \omega_{q}) \vert^{2}\,.
\end{align}
We can simplify this expression by choosing $k = p + {q}$. Then,
\begin{align}
\omega_{F}(\rho(F)\, \rho(G))\, =\, \int\, \frac{\dt ^{3}k}{(2\pi)^{3}}\, \tilde{f}(k)\, C_{\rho\rho}^{\epsilon}(k)\, \tilde{g}(-k)\,,
\end{align}
where the kernel $C_{\rho\rho}^{\epsilon}$ is defined as 
\begin{align}
C_{\rho\rho}^{\epsilon}(k)\, \coloneqq\, \frac{e^{2}}{4}\,  \int\, \frac{\dt ^{3}{q}}{(2\pi)^{3}}\, \frac{1}{\omega_{q}\, \omega_{k - q}}\, (\omega_{k - q} - \omega_{q})^{2}\, \vert \tilde{\kappa}_{\epsilon}(\omega_{q} + \omega_{k - q}) \vert^{2}\,.
\end{align}
This kernel is manifestly non-negative and finite $\forall\, \epsilon > 0$, because the Fourier transform $\tilde{\kappa}_{\epsilon}$ of the smooth, compactly supported bump function decays faster than any inverse power of frequency at fixed $\epsilon$. 

In order to derive a finite spatial kernel $C_{\rho\rho}(k)$, we need to analyse the $\epsilon\, \rightarrow\, 0$ limit of $C^{\epsilon}_{\rho\rho}(k)$. The simplest way to do this is to express $C_{\rho\rho}^{\epsilon}(k)$ in terms of the ``spectral density'':
\begin{align}
C_{\rho\rho}^{\epsilon}(k)&=\frac{e^{2}}{4}\, \int \dt x\, \int\, \frac{\dt ^{3}{q}}{(2\pi)^{3}}\, \frac{1}{\omega_{q}\, \omega_{k - q}}\, (\omega_{k - q} - \omega_{q})^{2}\, \vert \tilde{\kappa}_{\epsilon}(\omega_{q} + \omega_{k - q}) \vert^{2}\, \delta( x - (\, \omega_{q} + \omega_{k - q}\, )\, )\nonumber\\
&=\, \frac{e^{2}}{4 (2\pi)^{3}}\, \int \dt x\, \vert \tilde{\kappa}_{\epsilon}(x) \vert^{2} \int_{m}^{\infty}\, \vert q \vert^{2}\, \dt \omega_{q} \frac{\omega_{q}}{\vert {q} \vert}\, \frac{(2\omega_{q}\, -\, x)^{2}}{\omega_{q}\, (x - \omega_{q})}\, \int\, \dt ^{2}\hat{q}\, \delta(x - (\omega_{q} + \omega_{k - q})) \,.
\end{align}
For $\vert k\vert>0$, the angular integral gives 
\begin{align}
\int \dt ^{2}\hat{q}\, \delta(x - (\omega_{q} + \omega_{k - q}))  = 2 \pi\, \frac{(x-\omega_{q})}{\vert q \vert \vert k \vert}\,,
\end{align}
provided the delta function has a solution with $\vert\cos\theta\vert\leq 1$ and $x-\omega_q>0$, and is zero otherwise. Here $\theta$ is the angle between $k$ and ${q}$. These conditions give the following limits on the $x$- and $\omega_q$-integrals:
\begin{align}
\vert \cos\theta \vert \leq\, 1\, \implies\, \omega_{-}\, \leq\, \omega_{q}\, \leq\, \omega_{+}\,,\nonumber\\
\textrm{where}\, \,\,\omega_{\pm}\, =\, \frac{x}{2}\, \pm\, \frac{\vert k \vert}{2}\sqrt{1-\frac{4m^{2}}{x^{2} - k^{2}}}\,,
\end{align}
Since $x=\omega_q+\omega_{k-q}>0$, the spectral threshold is 
\begin{align}
x\, \geq\, \sqrt{\vert k\vert^{2}+4m^{2}} \,.
\end{align}
Finally, combining all of these ingredients, we see that
\begin{align}\label{ceprhorhop1}
C_{\rho\rho}^{\epsilon}(k)&=
\frac{e^{2}}{4(2\pi)^{3}}\, \frac{2\pi}{\vert k \vert}\, \int_{\sqrt{k^{2} + 4 m^{2}}}^{\infty}\, \dt x\,\vert \tilde{\kappa}_{\epsilon}(x) \vert^{2}\, \int_{\omega_{-}}^{\omega_{+}}\, \dt  \omega_{q}\, (2\omega_{q} - x)^{2}\nonumber\\
&=\, \frac{e^{2}}{48 \pi^{2}}\, \vert k \vert^{2} \int_{\sqrt{k^{2} + 4 m^{2}}}^{\infty}\, \dt x\,\vert \tilde{\kappa}_{\epsilon}(x) \vert^{2}\, \left(1\, -\, \frac{4m^{2}}{x^{2} - \vert k \vert^{2}}\right)^{\frac{3}{2}}\,. 
\end{align} 
This is the final formula for the kernel for the density-density two-point function in the quasi-free Fock vacuum on the time-thickened slice. At $k=0$, the kernel vanishes, as follows directly from its defining integral. 

We can analyse this kernel at small $\epsilon$ as follows. We set
\begin{align}
\tilde{\kappa_{\epsilon}}(x)\, =\, {\cal N}\, \int_{-1}^{1}\dt s\, e^{\frac{1}{(s^{2} - 1)}}e^{i\epsilon x s}\, =:\, \tilde{\kappa}(\epsilon x)\,,
\end{align}
and in eq.~(\ref{ceprhorhop1}) change variables to $y=\epsilon x$ to find
\begin{align}
C_{\rho\rho}^{\epsilon}(k)\, =\, \frac{e^{2}}{48 \pi^{2}}\, \vert k \vert^{2} \frac{1}{\epsilon} \int_{\epsilon\sqrt{k^{2} + 4 m^{2}}}^{\infty}\, \dt y\,\vert \tilde{\kappa}(y)\vert^{2}\, \left(1\, -\, \frac{4m^{2}}{(\frac{y}{\epsilon})^{2} - \vert k \vert^{2}}\right)^{\frac{3}{2}}\,. 
\end{align}
 Next, setting
\begin{align}
S_{\epsilon}(y,k)\, \coloneqq\, \left(1\, -\, \frac{4m^{2}}{(\frac{y}{\epsilon})^{2} - \vert k \vert^{2}}\right)^{\frac{3}{2}} \,,
\end{align}
 we have 
\begin{align}
C_{\rho\rho}^{\epsilon}(k)&=\frac{e^{2}}{48 \pi^{2}}\, \vert k \vert^{2} \frac{1}{\epsilon} \left[\, \int_{0}^{\infty}\, \dt y \vert \tilde{\kappa}(y) \vert^{2} - \int_{0}^{\epsilon\sqrt{k^{2} + 4 m^{2}}}\, \dt y \vert \tilde{\kappa}(y) \vert^{2}\,\right.\nonumber\\
&\qquad\qquad\qquad\qquad
\left.+  \int_{\epsilon\sqrt{k^{2} + 4 m^{2}}}^{\infty}\, \dt y\,\vert \tilde{\kappa}(y)\vert^{2}\, \left( S_{\epsilon}(y,k)\, -\, 1 \right)\right] .
\end{align}
The behaviour of the first two terms as $\epsilon\, \rightarrow\, 0$ is 
\begin{align}
\textrm{1st term}\, &\sim\ \frac{1}{\epsilon} \vert k \vert^{2}\,,\nonumber\\
\textrm{2nd term}\, &\sim\, -\vert k \vert^{2} \sqrt{k^{2} + 4 m^{2}}\,.\nonumber
\end{align}
It can also be verified that in the $\epsilon\, \rightarrow\, 0$ limit, the third term is finite.\footnote{This can be checked by changing the variables in that integral back to $x$ and estimating the integral at small $\epsilon$. We will leave these details for a diligent reader.}

Thus, it is the first term which diverges on the Cauchy slice and which requires a renormalisation when we evaluate this two-point function in the strict $\epsilon\, \rightarrow\, 0$ limit. However, notice that the second term is universal (that is, scheme-independent) and grows as $-\vert {k} \vert^{3}$ at large ${k}$ which spoils the positivity of the kernel $C_{\rho\rho}^{\epsilon=0}({k})$ on the Cauchy slice. 

Finally, we note that similar computations can be done for the $C_{JJ}$ correlator to illustrate the same point.

\subsubsection{Renormalisation of two-point functions}\label{subsec:ren}

In this subsection, we extend the domain of $\omega^{{R}}_{F}$ to all the monomials in ${\cal A}_{\textrm{phys}}({\Sigma})$. We start by specifying  the leading order (in $e$) two-point functions of the composite matter operators ($\rho$, $J$ and $|\psi|^2$) using a differential renormalisation scheme. In the two subsections thereafter, we give evidence that the resulting functional is positive. We also argue that it can be extended to higher monomials generated by the matter operators. Thus, $\omega_{F}^R$ can be used to define a perturbative GNS representation of the algebra generated by all the local operators in ${\cal A}_{\textrm{phys}}({\Sigma})$.

Recall from \eqref{eq:tpfomdop} and \eqref{eq:Gpm def} that
\begin{align}
C_{\rho\rho}(r)\, & =\, -\,  \frac{m^{3}}{4\pi^{3}}\, \frac{K_{1}(mr)\, K_{2}(mr)}{r^{3}} \nonumber\\
C_{J_i J_j}(r)\, & =\,\frac{m^2}{4 \pi ^4 } \l - \p _i \frac{K_1(mr)}{r} \p _j \frac{K_1(mr)}{r}  +  \frac{K_1(mr)}{r} \p _i \p _j \frac{K_1(mr)}{r} \r \\
C_{\psi ^2  \psi ^2 }(r)\, & =\, \frac{m^2}{16 \pi ^2}\frac{K_1 ^2 (mr)}{ r^2}\,. \nonumber
\end{align}
The trouble resides in the coincidence limit. As $r\, \rightarrow\, 0$, we have 
\begin{align}
C_{\rho\rho}(r) \, & \sim\, -\, \frac{m^{3}}{4\pi^{3}}\, \l  \frac{2}{m^{3} r^{6}}\, +\, \frac{\ln mr}{m r^{4}}\, +\, \frac{(\gamma - \ln 2 - 1)}{m r^{4}}\, +\, O\l \frac{\ln mr}{r^{2}}\r  \r  = : C^\infty_{\rho \rho }(r) \\
C_{J_i J_j}(r)\, & \sim\, \,\frac{m^2}{4 \pi ^4 }\left( \l -2 \d _{ij} +  4 \hat{r}_i \hat{r}_j  \r \frac{1}{m^2 r^6 } +  \l (1-\gamma +\ln 2 )\d _{ij} +(-1+4\gamma -4\ln 2) \hat{r}_i \hat{r}_j  \r \frac{1}{r^4 } \right. \nonumber\\
&\qquad \qquad \qquad \qquad \qquad \left.- \l \d _{ij} - 4 \hat{r}_i \hat{r}_j \r\frac{\ln mr}{r^4} \right) = : C^\infty_{J_i J_j}(r)   \nonumber \\
C_{\psi ^2  \psi ^2 }(r)\, & \sim \, \frac{1}{16 \pi ^2 r^4} = : C ^\infty  _{ \psi^2 \psi ^2 }(r) \,,\nonumber
\end{align}
where $\gamma$ is the Euler-Mascheroni constant.

Such distributions are UV divergent when integrated with a three-dimensional test function. We therefore renormalise the kernels using the differential renormalisation which is summarised in appendix \ref{app:EG}. Essentially, this boils down to replacing each divergent term in the kernel with the Laplacian of some well-defined distribution:
\begin{align}\label{eq:ren_kernels}
 C_{\rho\rho} ^{\text{ren}} (r)\coloneqq\,\,&  C_{\rho\rho} (r) -  C_{\rho\rho} ^\infty (r) -\,  \frac{m^{3}}{4\pi^{3}}\left[\, \frac{1}{12m^{3}} \triangle^{2}\left(\frac{1}{r^{2}}\right)\, + \frac{1}{2m}\, \triangle\left(\frac{\ln mr}{r^{2}}\right)\, \right.\nonumber\\
&\qquad\qquad\qquad\qquad\qquad\left.+\,\frac{1}{2m} \left(\gamma - \ln 2 + \frac{1}{2}\right)\, \triangle\left(\frac{1}{r^{2}}\right)\, +\, c_{0}\delta^{3}(x)\, +\, c_{1} \triangle \delta^{3}(x)\, \right]\nonumber \\
    C_{J_iJ_j} ^{\text{ren}} (r)\coloneqq \,\,&    C_{J_iJ_j}  (r) -  C^\infty _{J_i J_j}(r)   +  e^2 \left[ \frac{m^2}{4 \pi ^4 }   \l -\frac{1}{9}\d_{ij} \triangle +  \frac{1}{6} \p _i \p _j \r \triangle \l \frac{1}{2m^2 r^2 }\r \right.\\
   &  + \left. \l \frac{1}{4} \d_{ij} \triangle +\frac{2 +4 \gamma - 4 \ln 2 }{8} \p _i  \p _j \r  \frac{1}{r^2}  +  \frac{1}{2} \p _i \p _j \l \frac{\ln mr}{r^2} \r  \right] + d_0 \d^3 (x) + d_1 \triangle \d^3 (x) \nonumber \\
   C^{\text{ren}} _{\psi ^2  \psi ^2 }(r) \coloneqq \,\,&  C_{\psi ^2  \psi ^2 }(r)- C ^\infty  _{\psi ^2  \psi ^2 }(r) + \triangle \left(\frac{1}{32 \pi ^2 r^2}\right) +f_0 \d^3 (x) +f_1 \triangle \d^3 (x)\,, \nonumber 
\end{align}
where the terms that involve partial derivatives or the Laplacian operator are defined to act on test functions as:
\begin{align*}
 \int \dt^3 x \, f(x) \triangle \mathcal{K} (r) & \coloneqq 
    \int \dt^3 x \, \triangle   f(x) \,\mathcal{K} (r)  \\
     \int \dt^3 x \, f(x) \p _i \p _j \mathcal{K}(r) & \coloneqq 
    \int \dt^3 x \, \p _i \p _j   f(x) \mathcal{K} (r)  \,.
\end{align*}
This procedure yields ultra-violet finite kernels $C_{\dots}(r)$. That is, differential renormalisation allows us to smoothly extend the distributions to the coincident point $r = 0$ while keeping their profiles the same away from this point. 

It is now not difficult to check that  ${\int f(x)  C_{\rho\rho}^{\text{ren}} (x-y) g(y)}$, ${\int f(x)  C_{J_i J_j} ^{\text{ren}} (x-y) g(y)}$, and  ${\int f(x) C_{\psi^2\psi^2}^{\rm ren}(x-y) g(y)}$ are finite. From now on, we will set $c_0=d_0=f_0=0$ because we require the kernels to vanish at $|k|=0$. 

To check whether they are also positive we need to look at their large $k$ behaviour:\footnote{As shown in appendix \ref{app:fourier}, this result can be derived by integrating $C_{\rho\rho}^{\textrm{ren}}$ in a ball of radius $\frac{1}{p}$ when $p>>1$.}
\begin{align}
C_{\rho\rho}^{\text{ren}} (k) \, & \sim \, -\, \frac{\pi}{3} \vert k \vert^{3} \, < \, 0 \\
C_{J_i J_j}^{\text{ren}} (k) \,  & \, \sim \frac{e^2}{16 \pi ^2} \l \frac{1}{9} |k|^3 - \frac{1}{6} |k| k _i k _j  \r \, < \, 0 \\
C_{\psi^2 \psi^2 }^{\text{ren}} (k) \,  & \, \sim -\frac{|k|}{16}  \, < \, 0 \,.
\end{align}
We see that all kernels are negative. The negativity of $C^\text{ren} _{\psi^2 \psi ^2}$ can be fixed by choosing an appropriate $f_1{<}0$ constant for the $\Delta \d^3 (x)$ counter-term which is just $-\vert k\vert^2$ in momentum space, see appendix~\ref{app:fourier}. Unfortunately, such a simple solution does not work for $C_{\rho\rho}^{\text{ren}} $ and $C_{J_i J_j}^{\text{ren}}$. Even after adding the counter-terms of the form $c_1 \triangle\, \delta^{3}(x )$ and $d_1 \triangle\, \delta^{3}(x)$ for $c_1, d_1{<}0$ the kernel still becomes negative for large enough $\vert k \vert $. The resulting GNS state is therefore no longer positive and requires addition of counter-terms associated to the introduction of irrelevant couplings in the theory.

\subsection{Restoring positivity of the renormalised state}\label{subsec:restoring_positivity}
In order to illustrate this subtle relation between renormalisation and violation of positivity of the GNS state, let us examine the following simple example. 
Consider a two-point function defined as 
\begin{align}
W(f,g) = \int \frac{d^{3}k}{(2\pi)^{3}} \tilde{f}(k) \tilde{g}(-k)\, {\cal W}(k)\,,
\end{align}
where 
\begin{align}
{\cal W}(k)\, =\, \textrm{{\cal FT}} \left[\, \frac{K_{2}(mr)}{r}\, \right]
\end{align}
and $\text{FT}$ stands for Fourier transform in $D = 3$.

${\cal W}(k)$ is thus divergent in the UV. We can understand the divergent term by evaluating the integrand in the range $\frac{1}{\Lambda}<r<\frac{1}{k}$, 
\begin{align}
{\cal W}_{\Lambda}(k)\, =\, -\frac{8\pi}{m^{2}}\ln\left(\frac{\vert k\vert}{\Lambda}\right)\, +\, \textrm{finite}\,,
\end{align}
where $\Lambda$ is a UV cut-off. Such a kernel is always positive because by definition $k$ is always smaller than $\Lambda$. We can then define the renormalised two-point function as follows:
\begin{align}
{\cal W}^{\text{ren}}(k)\, &=\, \lim_{\Lambda \rightarrow \infty}\, \left(\, {\cal W}_{\Lambda}(k)\, + \frac{8\pi}{m^{2}}\ln\left(\frac{\mu}{\Lambda}\right)\, \right)\nonumber\\
&=\, -\frac{8\pi}{m^{2}}\ln\left(\frac{\vert k\vert}{\mu}\right)\, +\, \textrm{finite}
\end{align}
with $\mu$  the renormalisation scale. We now notice that the renormalised function is not positive once $\vert k \vert$ is larger than $\mu$. This is a simple example of how renormalisation can lead to violation of positivity. However, in this case the local counter-term  that restores positivity is simply given by\footnote{Since the mass dimension of $\beta$ is $\frac{1}{M^{4}}$ we have expressed $\beta$ in terms of a dimensionless parameter $\tilde{\beta}$ as
$
\beta\, :=\, {\tilde{\beta}}/{\mu^{4}}$.}
\begin{align}
{\cal W}^{\textrm{ren}}_{\textrm{pos}}(k)\, =\, {\cal W}^{\text{ren}}(k)\, +\, \frac{\tilde{\beta}}{\mu^{4}}\, k^{2}
\end{align}
for a sufficiently large $\tilde{\beta}$.

Similarly, in our case, to maintain the positivity of the kernel for all $k$ and preserve locality, we need to add the following local counter-terms
\begin{align}
C_{\rho\rho} ^{\text{ren}} & (\vert x - y \vert)\, \rightarrow\, C_{\rho\rho}^{\text{ren}}(\vert x - y\vert)\, + c_{2} \triangle^{2}\delta^{3}(x - y) \nonumber \\
C_{J_i J_j} ^{\text{ren}} & (\vert x - y \vert)\, \rightarrow\, C_{J_i J_j}^{\text{ren}}(\vert x - y \vert)\, + d_{2} \triangle^{2}\delta^{3}(x - y)\,,\label{eq:positiverenorm}
\end{align}
where $c_{2}> 0$ and $d_2>0$.  At large $\vert k \vert$ these counter-terms scale as $\vert k \vert^{4}$, which will dominate the negative contribution. However, restoring positivity comes at a price. The scaling dimension of the counter-term is $[c_2]=[d_2]= -1$ and corresponds to adding  an irrelevant operator to the effective classical Hamiltonian generated by $\rho(x), \vec{J}(x)$. In other words, the  contribution of the matter sector to the resulting effective interaction Hamiltonian contains terms of the form\footnote{As the fundamental variables in the matter sector are $\rho, J, \vert \Psi \vert^{2}$, it is plausible that this  effective field theory may be interpreted as one of a relativistic charged plasma interacting with electro-magnetic field \cite{Morrison:1980zz}. However, a deeper understanding of the dynamical aspects of our effective theory is outside the scope of the present paper.}
\begin{align}
H_{\textrm{int}}\, =\, \int\, \rho(k)\, ( C_{\rho\rho}^{\text{ren}}(k) ) ^{-1}\, \rho(-k) \, +  \, J^i (k)\, ( C_{J_i J
_j}^{\text{ren}}(k) ) ^{-1}\, J^j(-k)\, +\, \cdots\,.
\end{align}

\subsection{Extending the canonical vacuum to higher-order monomials}\label{subsec:higherptren}

In subsection \ref{subsec:omeonalocononover}, we saw that $\omega_{F}^{R}$ is well defined on any polynomial in ${\cal A}_{\textrm{phys}}(\Sigma)$ in which the set of smearing fields dual to matter operators are pairwise non-overlapping. In subsection~\ref{sec:extending domain}, we then showed that we can renormalise the state to encompass also two-point functions with overlapping smearing functions. Here, we will now argue that this renormalised $\omega_F^R$ can, in fact, be extended to all elements of the Cauchy slice algebra $\mathcal{A}_{\rm phys}(\Sigma)$.

Rather than proving the result for generic monomials, we will compute $\omega_{F}^{R}(\, \prod_{i=1}^{n}\rho(f_{i})\, )$ and demonstrate that it is finite $\forall\, n$. The generalisation of this computation to any element of ${\cal A}_{\textrm{phys}}(\Sigma)$ should be straightforward, albeit rather tedious, and we will simply comment on {it} at the end of this subsection.   

Thanks to Wick's theorem (cf.~eq.~\eqref{eq:Wick}), the $n$-point correlator can be written in terms of products of two-point correlators $\omega_{F}^{R}\left(:\!\rho_{i}(f_{i})\!:\, :\!\rho_{j}(f_{j})\!:\right)$, as well as connected $n$-point functions. In the previous subsections, we showed how to renormalise $\omega_F^R$ in such a way that the two-point correlators are finite. Thus, in what follows we only need to concern ourselves with connected $n$-point functions. 

Let us first note that the connected component obeys
\begin{align}
G_{c}^{F}(f_{1},\, \dots,\, f_{n})\, =\, 0
\end{align}
for odd $n$. This is because the state $\o _F ^R$ is invariant under charge conjugation, whereas the charge density operator satisfies $\hat{C}:\!\rho(f)\!:  \hat{C}^{\dagger}\, =\, - :\!\rho(f)\!:$.

To compute the connected correlator for even $n$, it is convenient to express $\rho(f)$ as follows. The complex scalar field on $\Sigma$ can be written in terms of Fourier modes as
\begin{align}
\psi(x)\, =\, \int \frac{\dt ^{3}p}{(2\pi)^{3}2\omega_{p}}\, (\, d_{p} \cdot (1,1)\, ) e^{- i p \cdot x}\,,
\end{align}
where the doublet $d_{p}$ and it's ``adjoint'' are  defined as 
\begin{align}
d_{p} &\coloneqq\, (a_{p},\, b_{-p}^{\dagger})^{T}\nonumber\\
d^{\dagger}_{p}&\coloneqq\, (a^{\dagger}_{p},\, b_{-p})^{T}\,.
\end{align}
In terms of these operators, $\rho(f)$ takes the form 
\begin{align}
\rho(f)\, =\, \int\, \frac{\dt ^{3}k}{(2\pi)^{3}2\omega_{k}}\, \tilde{f}(-k)\, \int\, \frac{\dt ^{3}p}{(2\pi)^{3}2\omega_{p}}\, :\! d_{p}^{\dagger T}\, {\cal Q}(p, k)\, d_{p + k}\! :\,.
\end{align}
Here, ${\cal Q}$ is the $2\, \times\, 2$-matrix defined as follows 
\begin{align}
{\cal Q}\, :=\, \begin{pmatrix} {\cal Q}_{+} & {\cal Q}_{-} \\ -{\cal Q}_{-} & -{\cal Q}_{+} \end{pmatrix}\,,
\end{align}
where
\begin{align}
{\cal Q}_{\pm}(k, p)\, =\, 2\, \frac{(\omega_{p} \pm \omega_{p + k})}{\sqrt{\omega_{p}\, \omega_{p + k}}}\,.
\end{align}
To leading order in $e$, the only non-trivial two-point functions  involving $d_{p}, d^{\dagger}_p$ are 
\begin{align}\label{wickdddag}
\omega_{F}(d^{\dagger}_{p}\, d_{p^{\prime}} )\, &=\, (2\pi)^{3}\, \Pi_{L}\, \delta^{3}(p - p^{\prime})\nonumber\\
\omega_{F}(d_{p}\, d^{\dagger}_{p^{\prime}})\,& =\, (2\pi)^{3}\, \Pi_{R}\, \delta^{3}(p -p^{\prime})
\end{align}
with
\begin{align}
\Pi_{L}\, =\, \begin{pmatrix} 0 & 0\\ 0 & 1 \end{pmatrix}, \quad \quad 
\Pi_{R}\, =\, \begin{pmatrix} 1 & 0\\ 0 & 0 \end{pmatrix}\,.
\end{align}
Using eq.~(\ref{wickdddag}), it can be easily verified that
\begin{align}\label{npttrace}
G_{c}^{F}\left(\prod_{i=1}^{n}\, \rho_{i}(f_{i})\right)=\int_{k_{1}, \dots,\, k_{n}}\, \delta^{3}\left(\sum_{j=1}^{n} k_{j}\right)\, \prod_{j=1}^{n}\, \tilde{f}_{j}(-k_{j})\int_{p}\, \sum_{\sigma \in S_{n}/Z_{n}}\, {\cal I}(\sigma(1), \dots,\, \sigma(n))\,,
\end{align}
where 
\begin{align}\label{calIijkl}
{\cal I}(\sigma(1),\, \dots,\, \sigma(n))&\coloneqq\nonumber\\
&\hspace*{-0.9in}\textrm{Tr}\big(\, {\cal Q}(k_{\sigma(1)}, p_{\sigma(1)})\, \Pi_{\sigma(1), \sigma(2)}\, {\cal Q}(k_{\sigma(2)}, p_{\sigma(2)})\, \Pi_{\sigma(2),\sigma(3)}\, \dots\, {\cal Q}(k_{\sigma(n)}, p_{\sigma(n)})\, \Pi_{\sigma(n), \sigma(1)}\, \big),
\end{align}
and $p_{i}\, =\, p + \sum_{j=1}^{i-1}\, k_{j}$, $S_{n}/Z_{n}$ is the $n$-element permutation group modulo cyclic permutations, and finally
\begin{align}\label{piijlom}
\Pi_{i,j}\, &\coloneqq \begin{cases}
    \Pi_{R}\, &\textrm{if }\, i < j\\
 \Pi_{L}\, &\textrm{if }\, i > j.
\end{cases} 
\end{align}

Although the RHS of eq.~\eqref{npttrace} looks rather complicated, we will argue that for any even $n\, \geq\, 4$, the integral over virtual momenta is always convergent. In the interest of pedagogy, we consider $n = 4$ as the generalisation to higher points is simply a tedious exercise in algebra. Ignoring the integrals for a moment, we have
\begin{align}
{G}_{c}^{F}\, \sim\,\, \sum_{\sigma\, \in\, S_{6}/Z_{6}}\, {\cal I}(\sigma(1)\, \dots\, \sigma(4))\,.
\end{align}
There are 6 terms in this sum. A direct computation reveals that these terms are divided into two classes. The first class consists of the terms that are quartic monomials in ${\cal Q}_{-}$. These terms are convergent after the integral over virtual momenta, since at large $\vert p \vert$ 
\begin{align}
{\cal Q}_{-}\, \sim\, O\l \frac{1}{\vert p \vert}\r\,.
\end{align}
This class consists of the permutations that satisfy the condition
\begin{align}
\sigma(1) < \sigma(2) > \sigma (3)\, <\, \sigma(4)\,.
\end{align}
There are precisely two such $\sigma$s (up to cyclic permutations), namely 
\begin{align}
\sigma : (1,2,3,4)\, &\rightarrow\, (1,3,2,4)\nonumber\\
\sigma : (1,2,3,4)\, &\rightarrow\, (1,4,2,3)\,.
\end{align}
Hence,
\begin{align}
{\cal I}(1,3,2,4)\, \sim\, {\cal I}(1,4,2,3)\, =\, O\left(\frac{1}{\vert p \vert^{4}}\right)\,.
\end{align}

On the other hand, for the remaining permutations, in which the increasing or decreasing sequence is of length at least two,\footnote{The remaining permutations (modulo cyclic permutations) map $(1,2,3,4)$ to $(1,2,3,4), (1,2,4,3), (1,3,4,2), (1,4,3,2)$. The first three have increasing sequence of length at least two and the last one has a decreasing sequence $(4,3,2)$ of length two.} ${\cal I}$ is quadratic in both ${\cal Q}_{-},\, {\cal Q}_{+}$. There are four such terms, and it can be verified that since $\sum_{i=1}^{4} k_{i} = 0$, the sum of these terms is convergent. We leave the details to the reader and summarise the main steps involved in the verification. 
Using eq.~(\ref{calIijkl}), we find that 
\begin{align}
{\cal I}(1,2,3,4)&=\, \textrm{Tr}\big(\, {\cal Q}(k_{1}, p_{1})\, \Pi_{R}\, {\cal Q}(k_{2}, p_{2})\, \Pi_{R}\, {\cal Q}(k_{3}, p_{3})\, \Pi_{R}\, {\cal Q}(k_{4}, p_{4})\, \Pi_{L} \big)\nonumber\\
&=\, - {\cal Q}_{-}(k_{1}, p_{1})\, {\cal Q}_{+}(k_{2}, p_{2})\, {\cal Q}_{+}(k_{3}, p_{3})\, {\cal Q}_{-}(k_{4}, p_{4})\,.
\end{align} 
Similarly, it can be checked that 
\begin{align}
{\cal I}(1,2,4,3)\,& =\, {\cal Q}_{-}(k_{1}, p_{1})\, {\cal Q}_{+}(k_{2}, p_{2})\, {\cal Q}_{-}(k_{3}, p_{3})\, {\cal Q}_{+}(k_{4}, p_{4})\nonumber\\
{\cal I}(1,3,4,2)\,& =\, {\cal Q}_{-}(k_{1}, p_{1})\, {\cal Q}_{+}(k_{3}, p_{3})\, {\cal Q}_{+}(k_{4}, p_{4})\, {\cal Q}_{-}(k_{2}, p_{2})\nonumber\\
{\cal I}(1,4,3,2)\, &=\, -\, {\cal Q}_{-}(k_{1}, p_{1})\, {\cal Q}_{-}(k_{4}, p_{4})\, {\cal Q}_{+}(k_{3}, p_{3})\, {\cal Q}_{+}(k_{2}, p_{2})\,.
\end{align}
For large $\vert p_{i}\vert $, we have
\begin{align}
{\cal Q}_{+}(k, p)\, =\, 1 + O\left(\frac{1}{\vert p \vert^{2}}\right)\,.
\end{align}
Thus, it can now be immediately seen that the leading order contribution vanishes at large $p$ and the most dominant contribution is 
\begin{align}\label{bbaa}
{\cal I}(1,2,3,4)\, +\, {\cal I}(1,2,4,3)\, + {\cal I}(1,3,4,2)\, +\, {\cal I}(1,4,3,2) =\, O\left(\frac{1}{\vert p \vert^{4}}\right)\,.
\end{align}
Hence, in the large $p$ limit 
\begin{align}
 \sum_{\sigma_{4}}\, {\cal I}(\sigma(1),\, \dots\, \sigma(4))\, \rightarrow\, \frac{1}{\vert p \vert^{4}}\,,
\end{align}
whose integral over $p$ converges.

The proof can be extended to arbitrary orders and requires one to classify all the cycles which contribute to monomials with different powers of ${\cal Q}_{+}$.

Although the correlation function $\omega_{F}^{R}(\rho_{1}(f_{1})\, \dots\, \rho_{n}(f_{n}))$ of the scalar field charge density is finite for $n > 2$,  the generic $n$-point connected correlation functions requires further renormalisation of the GNS state. For example, the reader can verify that the finiteness of the current correlators $G_{c}^{F}(\prod_{i=1}^{4} J_{i}(V_{i}))\, =:\, \omega_{F}^{R}(\prod_{m=1}^{4} J_{m}(V_{m})\, )$ requires renormalisation. Unlike $\omega_{F}^{R}(\prod_{i=1}^{4} \rho_{i}(f_{i}))$ where cancellations between particle and anti-particle contributions lead to a convergent expression, in this case it can be shown that 
\begin{align}
\omega_{F}^{R}(J_{1}(V_{1})\, \dots\, J_{4}(V_{4}))\, \sim\, O(\Lambda^{3})\,.
\end{align}
Hence, the GNS state needs to be defined in such a way that all connected correlators are finite to leading order in $e$. We will not provide explicit details of this renormalisation; however, it can be accomplished using differential renormalisation in precisely the same way as we carried it out for two-point correlation functions.

Let us finally summarise the main result of this section. We have carried out the main steps in building a GNS state for the Cauchy slice algebra $\mc{A} _{\text{phys}} (\Sigma)$. By adding counter-terms to the scalar QED Fock vacuum $\o _F$, we have constructed a new state $\o _F ^R$ that is positive on the linear elements of the algebra. More in detail, we have shown that differential renormalisation leads to a $\omega_{F}^{R}$ which satisfies 
\begin{align}
\omega_{F}^{R}(a^{\dagger}\, a)\, >\, 0\qquad\forall\, a \coloneqq \sum_{I=1}^{5}\, c_{I}\Psi_{I}(F_{I}) \,. 
\end{align}
While the proof of  positivity of $\omega_{F}^{R}$ for the entire algebra ${\cal A}_{\textrm{phys}}(\Sigma)$ is necessary for completing the GNS construction, it is beyond the scope of this paper.\footnote{The complication in proving positivity arises from the connected part of $n$-point functions.} In what follows, we will simply assume this to be the case. Our assumption is based on the renormalisation scheme which leads to positive kernels for all generators and manifest finiteness of connected $n$-point functions. 

Finally, we saw that the origin of the divergence of the two-point function of $\rho$, as well as the lack of positivity for the Fock vacuum state, is due to the fact that we are constraining the operators to live on a Cauchy slice. In this paper, we have chosen to work with Cauchy slice algebras as this enables one to define the longitudinal electric corner modes as additional regional centre modes which will be added to a regional quantum algebra in the next section. 

The question of how to reconcile positivity and finiteness of the state with the existence of a regional centre mode in a covariant formulation of local algebras is left for future work.

In the next section we extend $\omega_{F}^{R}$ to  $\mc{A} _{C} (\mathcal{O})$ for a fixed open subset $\mathcal{O}\subset\Sigma$.

\section{Quantisation of the longitudinal electric corner mode} \label{sec:GNS A_C}
In this section, we extend the definition of the GNS functional $\omega _F ^R$ to a new state $\o _0$, which has a well-defined action on the entire algebra  ${\cal A}_{C}(\Sigma) = {\cal A}_{\textrm{phys}}(\Sigma) \vee \mathcal{A} _{\rm out} ({\cal O}) $ for some \emph{fixed} open subset $\mathcal{O}\subset\Sigma$ with $\mathcal{A}_{\rm out}(\mathcal{O})$ defined in eq.~\eqref{eq:Aout}. For operators in ${\cal A}_{\textrm{phys}}(\Sigma)$, we define $\omega_{0}\,\equiv\, \omega_{F}^{R}$, i.e.
\begin{align}
\omega_{0}(A)\, \coloneqq\, \omega_{F}^{R}(A)\, \qquad  \forall\, A \, \in \mathcal{A} _{\rm phys} (\Sigma) \,.
\end{align}
Then what is left is to fix the correlation of the longitudinal centre modes $E_L ^{\rm out} (V)$. This is achieved by using the definition of $E_{L}^{\textrm{out}}$ in eq.~\eqref{eq:starELout} and then renormalising the kernel so that the resulting two-point function is finite. This renders higher $n$-point functions also finite, as the connected component of the  $n$-point function of $\rho$ is UV finite as shown in section~\ref{subsec:higherptren}.

\subsection{General form of the longitudinal two-point function}

In this and the following subsections, we will show that the new state $\o _0$ yields the following form for the two-point functions of the longitudinal electric corner modes of the fixed $\mathcal{O}$:
\begin{align}\label{eq:omega02pt}
\omega_{0}({E}_{L}^{\rm out}(V) {E}_{L}^{\rm out}(W))&=e^2\int_{({\cal O}^{c})^{2}}\dt^3x\,\dt^3y\, F_{V}(x)\, C_{E^{\rm out}_{L}E^{\rm out}_{L}}(\vert x\, -\, y\vert)\, F_{W}(y)\nonumber \\
&\quad+\frac{m^{3}{e^2}}{4\pi^{3}} \int_{(\partial{\cal O})^{2}}\dt^2\hat x\,\dt^2\hat y\Big( F_{V}(x)\, \p _{n_x} \p _{n_y} C^{\rm ren}_{\partial\partial}(\vec{r})\, F_{W}(y)  \\
&\qquad\quad  - \p _{n_x} F_{V}(x)\,  \p _{n_y} C^{\rm ren}_{\partial\partial}(\vec{r})\, F_{W}(y)  - F_{V}(x)\, \p _{n_x}  C^{\rm ren}_{\partial\partial}(\vec{r})\, \p _{n_y} F_{W}(y)\nonumber \\
&\qquad\qquad\qquad+ \p _{n_x} F_{V}(x)\,   C^{\rm ren}_{\partial\partial}(\vec{r})\, \p _{n_y} F_{W}(y)\Big)\,.\nonumber
\end{align}
Here, $\p_{n_x},\p_{n_y}$ denote normal derivatives, $V, W\, \in\, C^{\infty}_{c}(T{\cal O})$, and as a shorthand we define
\begin{align}
F_{V}(x)\, =\, \mathbf{1}_{{\cal O}^{c}}\, \Psi_V(x)\,, \qquad \Rightarrow\qquad\triangle F_V(x) = 0, \quad \forall\, x \in {\overset{\circ}{\mathcal{O}^c}}\,
\end{align}
for the relevant smearing of the longitudinal field in eq.~\eqref{eq:starELout}. Furthermore,
$C_{E^{\rm out}_{L}E^{\rm out}_{L}}( \vec{r} )$ is a renormalised kernel with support in the regional complement ${\cal O}^{c}$, and $C^{\rm ren}_{\partial\partial}$ is a kernel with support on the corner $\partial{\cal O}$.

To arrive at the above expression, our starting point is to note that $E_L^{\rm out}$ is defined in terms of the scalar charge density in eq.~\eqref{eq:starELout} (minding the $e$ prefactor), so that we may use the \emph{positive} renormalised kernel $C_{\rho \rho} ^{\textrm{ren}}$ to examine the UV divergences in the na\''ive two-point correlator
\begin{align}
\omega_{F}^{R}(E_{L}^{\rm out}(V)\, E_{L}^{\rm out}(W))\, =\, e^2\int_{(\Sigma - \overline{{\cal O}})^{2}}\dt^3x\,\dt^3y\, F_{V}(x) F_{W}(y)\, C^{\rm ren}_{\rho\rho}(\vert x - y\vert)\,
\end{align}
that subsequently we need to renormalise to define $\omega_0$.
Using the expressions from eqs.~\eqref{eq:ren_kernels} and \eqref{eq:positiverenorm}, we find:
\begin{align}\label{eq:omega0eloutveloutw}
\omega_{F}^{R}(E_{L}^{\rm out}(V)\, E_{L}^{\rm out}(W))&=e^2\times\,\textrm{finite bulk term}\, \\
&\,- \frac{m^{3}e^2}{4\pi^{3}} \int_{(\Sigma - \overline{{\cal O}})^{2}}\!\!\!\!\!\!\!\!\dt^3x\,\dt^3y\, F_{V}(x) F_{W}(y) \!\left( \frac{1}{12m^{3}} \triangle^{2}\left(\frac{1}{r^{2}}\right) + \frac{1}{2m}\, \triangle\left(\frac{\ln mr}{r^{2}}\right)  \right. \nonumber\\ 
&\qquad \qquad  \left. +\,\frac{1}{2m} \left(\gamma - \ln 2 + \frac{1}{2}\right) \triangle\left(\frac{1}{r^{2}}\right)+c_{1} \triangle \delta^{3}(\vec{r}) + c_{2}\, \triangle^{2} \delta^{3}(\vec{r}) \right).\nonumber
\end{align}
The terms proportional to $c_{1}, c_{2}$  are zero since $F_{V}, F_{W}$ are harmonic in ${\cal O}^{c}$.\footnote{Thus, we see that the bulk counter-term ambiguity parametrised by $c_{1},\, c_{2} $ does not affect the boundary kernel; also, we  have used the global constraint $C_{\rho\rho}(k=0) = 0$ to set $c_{0}$ to zero.} Now we will analyse the remaining terms in the second and third line in the above equation. Defining
\begin{align}\label{eq:tildeC}
\tilde{C}({r})\, \coloneqq\, - \frac{1}{  12m^3 r^{2} } - \frac{-1+2\gamma -\ln 4}{4m}\ln r \, -\, \frac{1}{4m} \ln^{2}(mr)\,,
\end{align}
it can be readily checked that 
\begin{align}\label{bosstildecrdef}
    \triangle ^2 \, \tilde{C} (r)=  - \frac{1}{12m^{3}} \triangle^{2}\left(\frac{1}{r^{2}}\right)\, - \frac{1}{2m}\, \triangle\left(\frac{\ln mr}{r^{2}}\right)\, -\,\frac{1}{2m} \left(\gamma - \ln 2 + \frac{1}{2}\right)\, \triangle\left(\frac{1}{r^{2}}\right)\,,
\end{align}
which is precisely what we have across the second and third line of \eqref{eq:omega0eloutveloutw}. Integrating by parts and again using the fact that $F_V$ and $F_W$ are both harmonic, we arrive at:
\begin{align}\label{eq:w(el)_intial}
\omega_{F}^{R}(E_{L}^{\rm out}(V)\, E_{L}^{\rm out}(W))&=\,e^2\times\,\textrm{finite bulk term}\,\nonumber \\
&\quad+\frac{m^{3}{e^2}}{4\pi^{3}}\!\int_{(\partial{\cal O})^{2}}\!\!\!\!\!\!\dt^{2}\hat x\, \dt^{2}\hat y\, \Big(\, F_{V}(x)\, F_{W}(y)\, \partial_{n_{x}}\, \partial_{n_{y}}\tilde{C}({r}) \nonumber\\
&\qquad\qquad\qquad - F_{V} (x)\partial_{n_{y}}F_{W} (y)\partial_{n_{x}} \tilde{C} ({r})- F_{W} (x)\partial_{n_{x}}F_{V} (y) \partial_{n_{y}}\tilde{C} ({r})\nonumber\\
&\qquad\qquad\qquad+ \partial_{n_{x}}F_{V}\, \partial_{n_{y}}F_{W}\, \tilde{C} ({r})\, \Big)\,.
\end{align} 
This is now almost of the shape given in eq.~\eqref{eq:omega02pt}, except that the corner contributions are divergent, as we shall see shortly, and therefore still need to be renormalised to define $\omega_0$.

\subsection{Local approximation on the boundary}

Let us now analyse the boundary contribution in eq.~\eqref{eq:w(el)_intial} under the assumption that $\partial{\cal O}$ is $S^{2}_{R}$ and choose its centre as the origin of the co-ordinate system. Although the actual form of the boundary two-point function does depend on the geometry of $\partial{\cal O}$, the UV structure and consequent renormalisation is fixed by the local topology of the boundary, which is $\mathbb{R}^{2}$ for any $\p \mathcal{O}$. Therefore by analysing this structure for $S_R^2$ we can obtain general results. 

The radius of the sphere is denoted $R$, while it will also be convenient to label points on it in an infinitesimal neighbourhood, where we approximate $S^2_R$ by $\mathbb{R}^2$, by $x_\parallel$ and $y_\parallel$, so that
\begin{equation}
    |x - y|^2 \Big|_{x, y \in \p \mathcal{O}} = \big|x _\parallel - y _\parallel\big| ^2= r^2 _\parallel  = R^2 \sin ^2 \frac{\theta}{2}\,.
\end{equation}

Straightforward computations show that several of the terms in the kernel in eq.~\eqref{eq:w(el)_intial} behave as $\theta^{-n}$ for $n\geq 2$, when $x_\parallel \to y_\parallel$. Such terms are not integrable on $S_R^2$, which shows that the current boundary kernel is divergent. Because this is a colinear divergence, we can analyse the renormalisation of the kernel in a local patch, where the surface looks flat up to $1/R$ corrections. Since all distances in this local region are much smaller than $R$, we can treat $1/R$ as a small parameter in the local expansion and drop all terms proportional to it. It allows us to make several crucial simplifications. Firstly, we have
\begin{equation}
    \begin{split}
        \p_{n_x} \tilde{C} \l| x - y | \r \Big|_{\p \mathcal{O}} &= \p_{n_y} \tilde{C} \l| x - y | \r = \tilde{C} ' \l r_\parallel \r \frac{r_\parallel}{2R} \\ 
         \p_{n_x} \p _{n_y}\tilde{C} \l| x - y | \r \Big|_{\p \mathcal{O}} &= \tilde{C} '' \l r_\parallel \r \frac{r_\parallel}{2R} + \tilde{C} ' \l r_\parallel \r \l -\frac{1}{r_\parallel}    +  \frac{r_\parallel}{4R^2} \r \,.
    \end{split}\label{eq:planarapprox}
\end{equation}
The local limit allows us to drop  all terms above apart from $- \tilde{C} ' \l  r_\parallel \r  \frac{1}{r_\parallel}  $. Furthermore, because the functions $F_V $ are harmonic in $\mathcal{O}^c$, they can be written in terms of spherical harmonics as:
\begin{equation}
\begin{split}
     F_V (r, \theta, \phi ) = \sum  _{l,m} \, f_V ^ {l,m} \l \frac{R}{r} \r^{l+1} Y_{l,m} \l \theta, \phi \r \,.
\end{split}
\end{equation}
This immediately implies that the spherical mode coefficients of $\p _r F_V\big|_{\p \mathcal{O}}$ and $\p _r ^2  F_V\big|_{\p \mathcal{O}}$ are those of $ F_V \big|_{\p \mathcal{O}}$  multiplied by $\frac{-(l+1)}{R}$ and $\frac{l(l+1)}{R^2}$, respectively. This means that only large $l>l_\ast$ contributions are relevant for the local approximation. This is called the short chordal limit, in which $l \to \infty$ and $R \to \infty$, while $\frac{l}{R}$ remains finite. In this limit, we can define a parameter $k  \coloneqq \frac{l+1}{R}$, which we treat as continuous, and we also have the following relations:
\begin{align}
    \p _r ^2 F_V \big|_{\p \mathcal{O}} &\sim \frac{1}{R^2} \Delta _{R^2} F_V \big|_{\p \mathcal{O}} = - \Delta _{\p \mathcal{O} } F_V \big|_{\p \mathcal{O}}\,, \nonumber\\\text{FT}\l  \p _r  F_V \big|_{\p \mathcal{O}}  \r (k) &\sim  -k   \text{FT} \l F_V \big|_{\p \mathcal{O}}  \r (k)\,.
\end{align}

In conclusion, the nature of the boundary kernel divergence allows us to focus only on its short distance behaviour in the limit where $r_\parallel \to  0$. In this limit, the surface looks planar, the Fourier space is continuous, and all terms proportional to $1/R$ are irrelevant.

\subsection{Renormalisation on the boundary}\label{subsec:boundary renorm}

In this local planar limit, we can rewrite the first and fourth corner contributions in eq.~\eqref{eq:w(el)_intial} (which thanks to eq.~\eqref{eq:planarapprox} are the only ones to survive in this approximation) as
\begin{equation}\label{omega0eloutVeloutW}
\begin{split}
    & \int_{(\p\mathcal{O})^2 }  \, \dt^{2}x\, \dt ^{2}y\l  F_{V}(x)\, F_{W}(y)\, \partial_{n_{x}}\, \partial_{n_{y}}\tilde{C}({r})\, +\ \partial_{n_{x}}F_{V}\, \partial_{n_{y}}F_{W}\, \tilde{C} ({r})\r \\
    &\qquad
    \qquad\qquad\qquad =  \int \, \frac{\dt ^{2}k}{(2\pi)^{2}}\,  \tilde{F} _V  (k)  \tilde{F} _W ( -k) \l  \text{FT}\l -\frac{\tilde{C}'(r)}{r} \r (k) + k^2 \text{FT} (\tilde{C}) (k)  \r \\
    &\qquad\qquad\qquad\qquad =:  \int \, \frac{\dt ^{2}k}{(2\pi)^{2}}\,  \tilde{F} _V  (k)  \tilde{F} _W ( -k) \l  \text{FT}\l -\frac{\tilde{C}'(r)}{r}  - \triangle _{R^2} \tilde{C} \r (k)   \r \,.\nonumber
\end{split}
\end{equation}
In this expression, the Fourier transform of $\tilde{C}$ should be thought of as the Fourier transform of whatever the renormalised kernel is. Once again we can use the identities from the local approximation to establish that:
\begin{align}
   \l  - \frac{\tilde{C}'(r)}{r}  - \triangle _{R^2} \tilde{C} \r \Bigg|_{\p \mathcal{O}} &=   - \triangle _{R^3} \tilde{C}  \Big|_{\p \mathcal{O}}  = \l
    \frac{1}{6 m^3 r^{4}}\, +\, \frac{2\gamma - \ln 4 + 1}{4mr^2}\, +\, \frac{\ln(m\, r)}{2mr^{2}} \r   \Bigg|_{\p \mathcal{O}} \nonumber\\
    &= :C_{\p \p } (r_\parallel)\,.
\end{align}

This boundary kernel $C_{\p \p } (r_\parallel)$ is divergent, so we once again appeal to the Epstein-Glaser renormalisation procedure. In particular, we can now define the renormalised kernel via differential renormalisation as:
\begin{equation}
    \begin{split}
        C_{\p \p } ^{\rm ren}(r_\parallel )\coloneqq & \frac{1}{48m^3} \triangle _{R^2}^2 \ln ^2 (mr_\parallel ) + \frac{2\gamma - \ln 4 + 1}{8m}\triangle _{R^2} \ln^2 (m r_\parallel)  + \frac{1}{24m} \triangle _{R^2} \ln ^3 (m r_\parallel) \\
        &\qquad \qquad \qquad +c_0+ c_1\Delta \d^2(r_\parallel)\,.
    \end{split}
\end{equation}
To check for positivity, we need to evaluate $\text{FT}(  C_{\p \p } ^{\rm ren}) (k)$, using the formulas given in appendix~\ref{app:fourier}, finding: 
\begin{equation} 
     \text{FT}(  C_{\p \p } ^{\rm ren}) (k) =  \frac{k^2}{48m^3 \pi} \ln \frac{k}{2m} + \frac{3}{4m \pi} \ln ^2 \frac{k}{2m} +  \frac{-1+10\gamma +\ln 4}{8 m \pi }\ln \frac{k}{2m} + c_0  + k^2 c_1\,. \label{bdy kernel}
 \end{equation}
It can be verified (using Mathematica) that
\begin{equation} \label{eq: positive corner 2pt}
  \text{FT}(  C_{\p \p } ^{\rm ren}) (k) +k^2 > 0 , \quad \forall\, k >0\,.
\end{equation}
Using the last line to define our renormalised \emph{positive} boundary kernel, we obtain a new positive state $\o _0$ for which the two-point function of $E_L ^{\rm out}(V)$ is well-defined, via the expression in eq.~\eqref{eq:omega02pt}. Although our derivation has focussed on UV divergence and the subsequent renormalisation of the boundary kernel in the planar limit approximation, the complete derivation of the renormalised kernel can be explicitly written down without approximation in the spherical basis in the case where $\partial{\cal O}$ is a sphere of radius $R$, as we summarise in appendix~\ref{app:S2kernel}. Crucially, there we found that the same counter-term is necessary for positivity. In appendix \ref{sec_positivity}, we will provide a preliminary analysis for the positivity of $\omega_0$ to extend to all observables in $\mathcal{A}_{C}(\Sigma)$ perturbatively,  including mixed observables involving both $E_L^{\rm out}$ and operators from $\mathcal{A}_{\rm phys}(\Sigma)$.

Therefore, $\o _0 $ can provide us with a well-defined GNS representation of the extended algebra $\mathcal{A}_C (\Sigma)$.

Several comments are in order at this stage.
\begin{itemize}
\item Although the renormalised kernel $\omega_{0}(E_{L}^{\textrm{out}}(V)\, E_{L}^{\textrm{out}}(W))$ in eq.~\eqref{eq:omega02pt} is the sum of a bulk and a boundary term, a moment of meditation reveals that it can be expressed purely as a boundary kernel. This can be seen as follows: the functions $F_V,\, F_{W}$ are harmonic in ${\cal O}^{c}$, and thus they are completely fixed by their boundary values $F_V(x) =  \int _{\p \mathcal{O}}\dt^2x_0 G(x,x_0) F_V(x_0)$, where $G(x,x_0)$ is the appropriate Dirichlet Green's function in $\mathcal{O}^c$. We can then rewrite the bulk contribution of $\o _0 (E_L^{\rm out} (V)E_L^{\rm out} (W))$ as: 
\begin{equation}
    \begin{split}
        \o _0 (E_L^{\rm out} (V)\,E_L^{\rm out} (W)) _{\text{bulk}}  &=  e^2\int _{(\mathcal{O}^c)^2}\dt^3x\,\dt^3y \, F_V(x) C_{E_L^{\rm out}E_L^{\rm out}} (\vert x-y\vert) \,F_W(y) \\
        &= e^2\int _{(\p \mathcal{O}^c)^2}\dt^2x_0\,\dt^2 y_0 \,   F_V(x_0)  \tilde{C}_{E_L^{\rm out}E_L^{\rm out}} (\vert x_0-y_0\vert)  \,F_W(y_0)  \,,
    \end{split}\label{eq:purebdy}
\end{equation}
where $\tilde{C}_{E_L^{\rm out}E_L^{\rm out}} (\vert x_0-y_0\vert)$ is $ C_{E_L^{\rm out}E_L^{\rm out}} (\vert x-y\vert)$ smeared with $G(x,x_0)$ and $G(y,y_0)$. Because this contribution is finite, it contributes at subleading orders in the momentum $k$ and so does not change the analysis we did above. 

\item An immediate consequence of extending $\omega_{0}$ to ${\cal A}_{C}({\Sigma})$ is that the resulting state is no longer translation-invariant. It can be shown via direct computation that 
\begin{align}\label{transnoninv}
\omega_{0}(E_{L}^{\rm out}\left(\alpha_{a}V)\, E_{L}^{\rm out}(\alpha_{a}W)\right)\, \neq\, \omega_{0}\left(E_{L}^{\rm out}(V)\, E_{L}^{\rm out}(W)\right)\,,
\end{align}
where $\alpha_a V(x)=V(x+a)$ denotes translating the support of $V$ by some vector $a$.
We will leave it to a diligent reader to verify eq.~\eqref{transnoninv}. However, it should be clear that the extension $\omega_{F}^{R}\, \rightarrow\, \omega_{0}$ is obtained by renormalising the boundary kernel which renormalises the modes supported on the fixed co-dimension-one boundary $\p\mathcal{O}$. Thus, the extension  $\omega_F^R\to\omega_{0}$ from $\mathcal{A}_{\rm phys}(\Sigma)$ to incorporating the outer longitudinal electric modes on multiple corners can only be carried out for a countable collection of mutually disjoint open sets $\{{\cal O}_{I}\}_{I\in\mathbb{N}}$, with ${\cal O}_{I}\cap {\cal O}_{J}=\emptyset$ for all $I\neq J$. 

\item Let us place our discussion in the broader context of edge modes and electric corner observables that reside in the centre of a regional algebra. There are different notions of edge modes in the literature \cite{Araujo-Regado:2024dpr}, but what they agree on in Maxwell theory (and  gauge theory more generally) is that, classically, the momentum degrees of freedom of these edge modes are given by the normal component of the electric field at the boundary $\p\mathcal{O}$ \cite{Donnelly:2016auv,Geiller:2019bti,Carrozza:2021gju,Araujo-Regado:2024dpr,Ball:2024hqe,Riello:2021lfl,Gomes:2018dxs,Shimizu:2026zug}. At the gauge-invariant level, the conjugate to the normal electric field is a Goldstone mode, which has cross-boundary support \cite{Araujo-Regado:2024dpr}. Thus, if one restricts to gauge-invariant observables which have support exclusively in the region $\mathcal{O}$ (including on its boundary $\partial\mathcal{O}$) one is left with a normal electric field that commutes with all other observables and so resides in the centre of the purely regional algebra. 

Owing to the distributional nature of this corner mode, it is non-trivial to realise  in quantum field theory. While such a regional electric centre has been standardly discussed in lattice gauge theories \cite{Casini:2013rba,Soni:2015yga,Delcamp:2016eya,Araujo-Regado:2025ejs,VanAcoleyen:2015ccp} (arising in particular when dropping the Goldstone mode \cite{Araujo-Regado:2025ejs}), its realisation in the continuum is significantly more subtle and tightly related to entanglement entropy computations \cite{Donnelly:2014fua,Donnelly:2015hxa,Soni:2016ogt,Casini:2015dsg,Casini:2019nmu,David:2022jfd,Fewster:2025ijg}. However, for example in \cite{David:2022jfd}, it was shown how to regularise the two-point function of the normal component of the electric field on $\p\mathcal{O}$ in the continuum $\rm{U}(1)$ quantum gauge field theory and that it is proportional to the conformal anomaly at this entangling surface.\footnote{The conformal anomaly at the entangling surface is intimately tied to the universal logarithmic correction to the entanglement entropy \cite{Casini:2019nmu,David:2022jfd}.}

It should by now be clear that the longitudinal electric centre for $\mathcal{A}_C(\mathcal{O})$, generated by $E_L^{\rm out}$ and discussed in the present manuscript, is qualitatively distinct from the normal electric flux centre alluded to above. There are two key differences: 
\begin{enumerate}
    \item The normal electric flux centre modes are of order $O(e^0)$ and so also arise in pure Maxwell theory, while $E_L^{\rm out}=O(e)$, which means that our longitudinal electric centre mode disappears as $e\to0$ in the free theory.
    \item The normal electric centre modes are codimension-one fields in $\Sigma$ even classically. By contrast, $E_L^{\rm out}$ is a codimension-zero field in $\Sigma\setminus\mathcal{O}$ and only via constraint imposition associated with $\mathcal{O}$. Thanks to the harmonic smearing it may be associated with the corner $\p\mathcal{O}$, which is also why $\omega_0$ required a boundary renormalisation and the addition of counter-terms localised on $\p\mathcal{O}$ to ensure positivity.
    \item The normal electric centre comprises the centre for the gauge-invariant algebras associated with both $\mathcal{O}$, as well as its complement $\mathcal{O}^c$. In contrast to that, $E_L^{\rm out}$ only resides in the centre of $\mathcal{A}_C(\mathcal{O})$, however, does not commute with other operators localised in $\mathcal{O}^c$.
\end{enumerate}
Nevertheless, one may ask whether the approach taken in the present paper can also be used to explore the realisation of the distinct normal electric centre in pure Maxwell theory. That is, given a quasi-free GNS state $\omega_0$ on the local $\star$-algebra $\big\langle E_T(V),B(V)\,\big|\,V\in C_c^\infty(T\mathcal{O})\big\rangle$ associated with $\mathcal{O}$, can we extend this state to boundary operators $E_\perp(V_\p)$, $V_\p\in C^\infty(T\p\mathcal{O})$, measuring the normal electric flux through the corner and which commute with all other regional operators. In appendix~\ref{app:empm}, we show that this is indeed possible, demonstrating that the GNS kernel $C_{E_TE_T}(k)$ from pure Maxwell theory can be renormalised and extended to define a positive linear functional on the centrally extended algebra for $\mathcal{O}\cup\p\mathcal{O}$.
This, however, requires the extension of the admissible regional smearing fields to encompass, in a suitable limit, a smearing purely on $\p\mathcal{O}$. Like eq.~\eqref{transnoninv}, the price to pay for accommodating the normal electric centre in the continuum theory in this way is that the extended GNS state no longer is translation-invariant.

This construction makes the distinction in point~2 above explicit, highlighting how the normal electric centre is obtained from a codimension-one distributional smearing, while the longitudinal electric centre arises a priori from a regional codimension-zero step function smearing (which, however, yields a pure codimension-one two-point function via eq.~\eqref{eq:purebdy}). It also highlights that the broad perspective adopted here is more generally applicable than to the longitudinal centre modes.

\item The extension of $\omega_{0}$ to the enlarged algebra $\mathcal{A}_C(\Sigma)$ (and thereby ${\cal A}_{C}({\cal O})$) is scheme-dependent and more explicitly should be written as $\omega_{0}(\mu)$ where $\mu$ is the renormalisation scale.  More in detail, 
\begin{align}
\omega_{0}(\mu)\big({E}_{L}^{\rm out}(V) {E}_{L}^{\rm out}(W)\big)&\coloneqq\, \int_{\Sigma^{2}}\dt^3x\,\dt^3y\, (\nabla \cdot V)(x)\, (\nabla \cdot W)(y)\, K_{\rm ren}(x,y; {c_{0}, c_{1}};  \mu) \nonumber\\
\omega_{0}(\mu)\big(A\, \cdot B\big)&=\, \omega_{F}^{R}(A\, \cdot B)\,\qquad
\forall\, A, B\, \notin\mathcal{A}_{\rm out}(\mathcal{O})\,,
\end{align} 
where $c_{0}, c_{1}$ were defined in eq.~(\ref{bdy kernel}), and  $K_{\rm ren}$ denotes the renormalised kernel at scale $\mu$. However, this scheme dependence in the renormalisation of the two-point function involving $E_L^{\rm out}$ should be understood as arising from the choice of local surface counter-terms in the Hamiltonian which determines the vacuum state. The physical (scheme-independent) vacuum should be understood in terms of renormalised couplings such that the kernel obeys the Callan-Symanzik equation
\begin{align}\label{cseqn}
\mu\, \p _{\mu}\, K_{\textrm{ren}}(k;\, e, c_{0}(\mu),\,  c_{1}(\mu); \mu)\, =\, 0\,.
\end{align}
We note that to leading order in perturbation theory $\frac{d e}{d \mu} = O(e^{3}) = 0$ and hence  at this order $e(\mu)=e$. We also note that the left hand side of eq.~\eqref{cseqn} generically contains the anomalous dimension $(\frac{d}{d \mu} + 2\,  \gamma_{E_{L}^{\textrm{out}}})$. It is defined as ${d Z_{E_{L}^{\textrm{out}}}^{\frac{1}{2}}}/{d \mu}$, where $Z_{E_{L}^{\textrm{out}}}$ is the multiplicative renormalisation of the corner mode operator. Once again, to leading order in $e$, $Z_{E_{L}^{\textrm{out}}}\, =\, 1$ and so the anomalous dimension vanishes.
\item Lastly, thanks to eq.~\eqref{eq:totalEL}, $\omega_{0}$ is also a positive linear functional for the inner longitudinal electric mode defined in eq.~\eqref{eq:starELout}, which already resides in $\mathcal{A}_C(\mathcal{O})$,\footnote{That is, once $E_L^{\rm out}(V)$ is added to $\mathcal{A}_{\rm phys}(\Sigma)$ (and $\mathcal{A}_{\rm phys}(\mathcal{O})$), so is, of course, $E_L^{\rm in}(V) = E_L(V)-E_L^{\rm out}(V)$. However, if instead one defined another regional algebra with longitudinal centre as $\mathcal{A}_C'(\mathcal{O})\coloneqq \mathcal{A}_{\rm loc}(\mathcal{O})\vee\mathcal{A}_{\rm out}(\mathcal{O})$, with $\mathcal{A}_{\rm loc}(\mathcal{O})$ given in eq.~\eqref{eq:AlocO}, then adjoinging $E_L^{\rm in}(V)$ by hand would be equivalent to simply invoking $\mathcal{A}_{C}(\mathcal{O})$ as done above.}
\begin{equation}
    E_L ^{\rm in} (V)\coloneqq e \int _{\mathcal{O}} \dt^3 x\, \Psi_V(x) \rho (x)\,,
\end{equation}
where $\Psi_V(x)$ does not vanish on $\p\mathcal{O}$ and satisfies $\triangle \Psi_V  = \nabla \cdot V $. Once again, even though we have smeared with a compact vector field $V \in C_c^\infty (T\mathcal{O})$, because of the constraints, we end up with an operator with a non-compact smearing. Because the function $\Psi_V$ is not harmonic in $\mathcal{O}$, the two-point function of $  E_L ^{\rm in} (V)$ has three types of terms: bulk-bulk, bulk-boundary and boundary-boundary. The first ones, like  for $E_L^{\rm out}$ are finite. The second ones are also finite, because the bulk part involves $\triangle \Psi_V = \nabla \cdot V $, which has vanishing support on the boundary, and therefore there are no coincidence point divergences. Finally, the boundary-boundary part is exactly the same as for $E_L ^{\rm out}$ and is finite when evaluated using $\o _0$.
\end{itemize}


\subsection{Completion of corner smearings}\label{ssec_completecornersmearing}

In this section, we use the renormalised GNS state $\omega_{0}$ to Cauchy-complete the space of all allowed two-dimensional smearing functions for the corner mode operator. The main result of this section is the statement that the space of all such corner smearing  functions is strictly smaller than the space of square integrable functions on $\partial{{\cal O}}$. 

As we have seen earlier, the kernel of the two-point function of $E_L^\text{out}(V)$ is corner-supported, despite $V\in C^\infty_c(T\cal{O})$ being a bulk smearing. Its leading behaviour in the UV is given by (see eq.~\eqref{bdy kernel})
\begin{align}
    \tilde{C} _{\p \p } ^{\rm ren} (k) \sim  \frac{k^2}{48m^3 \pi} \ln \frac{k}{2m}\,,
\end{align}
where $k$ is the wave number on a planar patch of the corner.\footnote{$C^{\rm ren}_{\p\p}$ only encodes the pure corner term of the two-point function.} We leave it for the reader to convince themselves that the full two-point function kernel, including the ``bulk'' finite pieces in eqs.~\eqref{eq:omega02pt} and~\eqref{eq:w(el)_intial}, defines a positive norm (provided we include the counter-term in eq.~\eqref{eq: positive corner 2pt}). This essentially follows from the discussion of $C_{\rho\rho}$ in section~\ref{sec:GNS A_loc}. This provides a convergence requirement on the corner values of $F_V$. Namely, we need that 
\begin{align}
    ||F_V||_{\tilde{C} _{\p \p } ^{\rm ren}}\coloneqq\int\dt^2k \;|\Tilde{F}_V(k)|^2\; k^2\ln(k) < \infty  \,.
\end{align}
Crucially, convergence under this norm is \emph{stronger} than convergence in the $L^2$-norm, thereby not allowing any square integrable function as a valid smearing.  Assuming the corner is topologically a sphere, $\p\mathcal{O}\simeq S^2$,\footnote{In fact, as only the UV information matters, it should suffice to assume that the corner is simply connected to obtain the same result.} the space of allowed smearings is in fact a \emph{generalised logarithmic Sobolev space}, denoted by $H^{1,1/2}({S}^2)$, see appendix \ref{app:sobolev} for details. 

The space $H^{1,1/2}({S}^2)$ is a Hilbert \emph{subspace} of $L^2({S}^2)$ defined by
\begin{align}
    H^{1,1/2}({S}^2)&\coloneqq\Big\{f\in L^2({S}^2)\,\Big|\;||f||_{H^{1,1/2}({S}^2)}<\infty\Big\}
\end{align}
with norm induced by the inner product\footnote{There are many inner products one could write here leading to unitarily equivalent Hilbert spaces. All that is required is that the kernel $W(l)$ in $\langle f,g\rangle_{H^{1,1/2}({S}^2)}= \sum_{l=0}^\infty\sum_{m=-l}^l W(l)f_{l,m}\overline{g_{l,m}}$ has the same growth rate at $l\gg1$ and is a smooth, strictly positive function of $l$, bounded away from $0$. For example, a kernel of the form $W(l)=\alpha+l(l+\beta)\ln(\gamma+l)$ for $\alpha,\beta,\gamma>0$ would define the same Hilbert space.}
\begin{align}
    \langle f,g\rangle_{H^{1,1/2}({S}^2)}&\coloneqq\sum_{l=0}^\infty\sum_{m=-l}^l\left(1+l(l+1)\ln (1+ l)\right)f_{l,m}\overline{g_{l,m}}\,,\\
    \text{where }f_{l,m}&=\int_{{S}^2} d\Omega\; f(\theta,\phi)\overline{Y_{l,m}(\theta,\phi)}\,,
\end{align}
for $f,g\in L^2({S}^2)$ and $\{Y_{l,m}\}$ the standard spherical harmonic basis. In particular, notice the behaviour of the norm at large $l$
\begin{align}
    ||f||_{H^{1,1/2}({S}^2)}\underset{{l\gg1}}{\sim}\sum_l l^2\ln (l) \left(\sum_{m=-l}^l|f_{l,m}|^2\right)\,,
\end{align}
whose convergence imposes a stronger decay rate as $l\gg 1$ on the function, compared to that imposed by the $L^2({S}^2)$-norm. The space $H^{1,1/2}({S}^2)$ includes plenty of non-smooth, non-continuous and even singular functions on the sphere. But they are still much ``smoother'' than the square integrable functions. 

We have the dense inclusions
\begin{align}
    C^\infty({S}^2)\subsetneq H^{1,1/2}({S}^2)\subsetneq L^2({S}^2)\,.
\end{align}
The valid corner mode smearings in our renormalised GNS state lie in the middle space. This is a quantitative characterisation of our longitudinal electric centre modes in the quantum theory.

\section{Discussion}

In this paper, we have proposed a new longitudinal electric corner mode for spatial subregions in a continuum  quantum field theory in which a complex scalar is coupled to an Abelian gauge field. Our definition follows from the classical intuition that whenever we separate space into inside and outside, the charged matter outside sources a longitudinal electric field that can be measured inside the region. In addition, the profile of this field within the region can be deduced uniquely from harmonic data on the corner, allowing us to think of it as ``living on the boundary''.  

This is different from the standard electric corner mode considered in the literature, given by the local electric flux on the corner.  Although this corner flux can be defined classically \cite{Donnelly:2016auv,Geiller:2019bti,Carrozza:2021gju,Araujo-Regado:2024dpr,Ball:2024hqe,Riello:2021lfl,Gomes:2018dxs,Shimizu:2026zug}, it is too singular to lead to a well-defined quantum operator in continuum Maxwell theory. Attempts to incorporate it, therefore, usually proceed via some choice of regularisation scheme or treat it as a stochastic background \cite{Donnelly:2014fua,Donnelly:2015hxa,Casini:2015dsg,Casini:2019kex,Soni:2016ogt,David:2022jfd}.\footnote{As explained in appendix \ref{app:empm}, one could in principle renormalise the Fock vacuum of Maxwell theory on the corner to allow for such singular smearings of the radiation mode $E_T$. However, we feel this goes outside the basic framework of algebraic QFT in which smearings for a local region $\mathcal{O}$ are always test functions (i.e.\ smooth and compactly-supported inside $\mathcal{O}$) for operator-valued distributions. In our construction in this paper, we define a regional algebra of operators $\mathcal{A}_C(\mathcal{O})$ in eq.~\eqref{eq:Aout} whose space of smearing functions is from this class. What happens is that one of the operators, namely the new centre mode $E_L^\text{out}$, requires a corner renormalisation so that its smeared versions yield well-defined two-point functions. This is due to the double nature of the smearing, made explicit in eq.~\eqref{E_L^out}, which is smooth and compact in $\mathcal{O}$, but due to the Gauss law also involves a non-compact and non-smooth smearing of $\rho$ in the complement. In other words, after this renormalisation it becomes a valid operator-valued distribution like the other bulk operators in the algebra, even though it ``lives'' on the corner $\p\mathcal{O}$. In contrast,  $E_T$ is already an operator-valued distribution and demanding it to further give finite two-point function when smeared singularly (i.e.\ only on $\p\mathcal{O}$) is, we believe, an unjustified extra requirement.}

In our case, we show that, starting from local gauge-invariant operators (which in the matter sector are generated by quadratic operators), the longitudinal corner mode is labelled by compactly supported smooth smearings in the bulk of the region.  This mode does not survive the limit of vanishing coupling. It thus arises from a very different mechanism than the usual normal electric edge mode described above, which would also be present in free Maxwell theory. 

The main purpose of the paper is to provide a quantisation of this longitudinal corner mode,  such that it has well-defined correlation functions. In this endeavour, we stay strictly in the continuum throughout.  

\paragraph{Boundary renormalisation of the GNS state.}
In order to identify and analyse our corner operators, we proceeded sequentially. We first considered a Cauchy slice algebra ${\cal A}_{\textrm{phys}}(\Sigma)$ generated by local gauge invariant operators $\langle\, \rho(f), J(V), \vert \psi \vert^{2}(g), E_{T}(V), B(W)\, \rangle$.
We then showed that the covariant quasi-free Fock vacuum of scalar QED could be used to define a GNS state on ${\cal A}_{\textrm{phys}}(\Sigma)$. More in detail, pulling back the Fock vacuum onto the Cauchy slice and renormalising the $n$-point functions of the algebra helped us identify a (scheme-dependent) GNS state which can be used to quantise  ${\cal A}_{\textrm{phys}}(\Sigma)$. Positivity of the state puts a set of constraints on the class of counter-terms we can choose. We showed that existence of well-defined two-point functions generated by spatially locally smeared density operators along with positivity implies that the renormalised vacuum is that of an effective field theory (EFT) in which the QED Hamiltonian is deformed by a set of irrelevant operators containing fourth order spatial derivatives. This specific quantisation of a spatial gauge-invariant algebra of local operators was then our starting point to study the perturbative quantisation of the new longitudinal electric corner modes in a continuum QFT.

Fixing a given subregion $\mathcal{O} \subset\Sigma$, we then showed that an enlargement of ${\cal A}_{\textrm{phys}}(\Sigma)$ to ${\cal A}_{C}(\Sigma)=\mathcal{A}_{\rm phys}(\Sigma)\vee\mathcal{A}_{\rm out}(\mathcal{O})$, involving adding the extra generator $E_L^\text{out}(V),\, V\in C^\infty_c(T\cal O)$, is only possible in the quantum theory if the GNS vacuum obtained above is further renormalised \emph{on the corner}. The fact that the renormalisation required is entirely corner supported (even though our mode is smeared in the bulk) is the quantum version of the ``edgeness'' of $E_L^\text{out}$. This additional renormalisation of the two-point function of the new corner modes then defines a state which breaks spatial translation invariance, thus picking a specific subregion whose boundary supports them.  We point out that to check that our renormalised state is a valid GNS state, one would need to check for positivity, which, as discussed throughout the paper, has a non-trivial interplay with renormalisation. A preliminary discussion of how to prove this is given in appendix \ref{sec_positivity}, but a complete treatment lies beyond the scope of this work.  

As mentioned above, this vacuum state is defined for a single subregion at a time (or a union of subregions with disconnected, non-overlapping boundaries). It is not possible to associate our corner mode to every arbitrary subregion with the hope of incorporating it into a net of regional algebras. This is broadly consistent with recent efforts in covariant algebraic QFT to formalise edge modes (including corner electric fields), which too fixed a region \cite{Fewster:2025ijg}.  In our work, this point manifests itself in a precise technical sense, in the very definition of the GNS vacuum state. 
However, it is not only the region $\mathcal{O}$ that matters here, but the local degrees of freedom in the complement play an essential role in being the ones sourcing the longitudinal edge mode. Even in the vacuum, charge fluctuations in the complement of the region source a non-vanishing two-point function for our edge mode inside the region. 

\paragraph{\pah{Longitudinal corner} mode two-point function.}

Upon renormalisation, the kernel of the two-point function of $E_L^\text{out}(V)\,,\text{ for } V\in C^\infty_c(T\cal{O})$, has a leading UV behaviour which is supported on the corner. Namely, we obtain that (at leading order in the coupling)
\begin{align} \label{eqn:2ptfct UV}
    \Tilde{C}_{E_L^\text{out}E_L^\text{out}}^{\rm ren}\sim\tilde{C} _{\p \p } ^{\rm ren} (k) \sim  \frac{k^2}{48m^3 \pi} \ln \frac{k}{2m}
\end{align}
in the limit $k\gg m$, where $k$ stands for the two-dimensional wave number on the corner and $m$ is the mass of the scalar particle. We note that, due to the presence of a log-scaling, this behaviour is independent of the renormalisation scheme. Thus, our result is universal.\footnote{We note in passing an apparent connection with the work \cite{Casini:2019nmu} in which the mutual information across a thin shell of width $\epsilon$ changes behaviour at the scale of particle-antiparticle production, i.e.\ at $\epsilon\sim 1/ 2m$. We leave a more direct relationship to that setup as an interesting direction for future work.}

Although the purpose of the renormalisation was to make our corner mode a well-defined operator-valued distribution, meaning that it yields finite two-point functions when smeared with smooth, compactly-supported test vector fields, the explicit form of the kernel provides us with a norm, thereby allowing us to complete this space of test fields. More precisely, our $V\in C^\infty_c(T\cal{O})$ must be such that
\begin{align}
    ||F_V||_{\tilde{C} _{\p \p } ^{\rm ren}}\coloneqq\int\dt^2k \;|\Tilde{F}_V(k)|^2\; k^2\ln(k) < \infty  \,,
\end{align}
where we had that $F_V(x)=1_{\mathcal{O}^c}\int_{\mathcal{O}}\dt^3y\frac{1}{|x-y|}\nabla\cdot{V}(y) $. So our space of allowed test functions is enlarged to harmonic functions whose corner values belong to a \emph{generalised logarithmic Sobolev space}, noted by $H^{1,1/2}({S}^2)$, as argued in section~\ref{ssec_completecornersmearing}. This is a dense subspace of $L^2({S}^2)$.\footnote{Contrary to common lore, even after renormalisation we cannot smear our corner modes with arbitrary square-integrable functions on the sphere. This restriction is forced on us by the dynamics of the theory.} The space $H^{1,1/2}({S}^2)$ includes plenty of non-smooth, non-continuous and even singular functions on the sphere. But they are still much ``smoother'' than the square integrable functions, due to the stronger decay rate requirement at large $k$ imposed by convergence under this norm. 

Thus, our final quantitative characterisation of the quantum longitudinal corner mode operators is that they correspond to 
\begin{align}
    E_L^\text{out}(V)\,,\text{ such that } F_V\big|_{\partial\mathcal{O}}\in H^{1,1/2}(\pah{S^2})\,,
\end{align}
with the UV behaviour of their fluctuations governed by \eqref{eqn:2ptfct UV}.

\paragraph{Non-local centre.}

Despite being a bulk field, our new quantum corner mode is also central. This regional centre, however, has several unusual features stemming from its intrinsic non-locality. It is not a joint centre for the algebras of the region and its complement. There are local operators in the complement that have non-vanishing commutator with it and can thus ``shift'' its value inside the region. This is unlike the \pah{the situation with the usual electric edge mode encoding the normal electric field through the surface in classical and quantum continuum discussions \cite{Donnelly:2014fua,Donnelly:2015hxa,Soni:2016ogt,Casini:2015dsg,Casini:2019nmu,David:2022jfd,Fewster:2025ijg}  and its quantum lattice incarnations as ``electric centre'' \cite{Casini:2013rba,Soni:2015yga,VanAcoleyen:2015ccp,Delcamp:2016eya,Araujo-Regado:2025ejs}}. There, because the local electric flux is supported only on the corner, it commutes with local operators on either side. Hence shifting its value would require a non-local operator across the corner \cite{Araujo-Regado:2024dpr,Araujo-Regado:2025ejs}. In our case, local operations supported entirely in the complement are sufficient to move us across the regional superselection sectors.

\paragraph{Comments on interpretation.}

Our regional centre algebra $\mathcal{A}_\text{C}(\cal O)$ is \emph{not}  a subalgebra of the global Cauchy slice algebra  $\mathcal{A}_{\rm phys}(\Sigma)$. This is because, although the total longitudinal electric field is an element of the global algebra, the split into $E_L=E_L^\text{in} + E_L^\text{out}$ introduces new operators in the theory. For each to be well-defined, the corner renormalisation described above is necessary. Given this, we see two ways to understand the physics at hand, which we present now. More work would be needed to reach a final conclusion.

\begin{enumerate}
    \item The first interpretation is that the corner renormalisation is associated with the presence of a \emph{physical boundary} in the spacetime. We are thus studying a continuum effective field theory of a scalar field coupled to an Abelian gauge field on a modified background with something like a thin-shell defect placed in the middle of Minkowski spacetime.
    
    The vacuum of this theory would then be unitarily inequivalent from the Fock vacuum and the corner renormalisation would incorporate precisely that. Correspondingly, the Hamiltonian of the theory would also differ by some corrections supported on the corner.\footnote{There could also be further bulk corrections to the Hamiltonian. But, since at the level of the analysis done in this paper the new vacuum agrees with the Fock vacuum on the usual QED bulk observables, we expect that at leading order in $e$ the correction is only corner-supported.} The presence of this physical boundary then allows for the definition of new operators (including our $E_L^\text{out}$, but possibly more, see below), which were not allowed in the ``empty'' theory. Thus, the global algebra can be enlarged naturally to include $\mathcal{A}_\text{C}({\cal O})$ as a subalgebra. In this way of thinking, the boundary is \emph{essential} for the existence of \pah{our new} edge modes for the subregion. In contrast to common lore, this would mean that in empty space the QED edge modes are in some sense a ``fake'' construct, not corresponding to actual physical observables, since they cannot be defined on the Fock vacuum.

    This possibility raises the question about the relationship between boundary conditions on timelike boundaries (as used extensively in the edge mode literature: see for instance \cite{Carrozza:2021gju, Araujo-Regado:2024dpr,Hoehn:2025pmx, Ball:2024hqe, Dudal:2024xdu, Canfora:2024awy, Shimizu:2026zug,Shimizu:2026wjz}) and the corner renormalisation performed in this paper. In particular the works \cite{Dudal:2024xdu, Canfora:2024awy} have considered the effect of different boundary conditions on the Casimir energy between two parallel plates \cite{Casimir1948}. If imposing boundary conditions is indeed related to our corner renormalisation, then we would like to understand whether or not the presence of our new non-local edge mode induces a correction to the Casimir force, as well as to the Casimir-Polder effect, which quantifies the QED correction to the van der Waals force between a neutral atom and a plate \cite{CasimirPolder1948, Lifshitz1956}. We see this as an interesting future line of work with possible observational signatures of our proposal.

    \item The second interpretation is that the corner renormalisation is instead associated with \emph{tracing out} the  local degrees of freedom of the complement. This possibility is somewhat tangential to the construction presented in this paper and more speculative.  We include it because it represents an interesting future direction.
    
    In  algebraic QFT, given a state on the global algebra, the ``tracing out'' operation simply means to restrict the domain of the algebraic state to a given subalgebra associated to a subregion.  Given a subalgebra $\mathcal{A}(\cal O)\subset \mathcal{A}_\text{global}$ (at the $\star$-algebra level) and a global state $\omega$, the restriction to the local algebra leads to a local algebraic state $\omega_{\cal O}\coloneqq \omega|_{\mathcal{A}(\cal O)}$. Then, performing a local GNS construction of $\mathcal{A}(\cal O)$ on this state leads to a local representation $\pi_{\omega}(\mathcal{A}(\cal O))$ for the subregional algebra. Usually, even if one starts from different global states, the local representations one obtains are all locally unitarily equivalent \cite{Haag:1992hx}. The von Neumann completion in the local representation leads to the standard type III$_1$ factor algebra $\cal{M}(\cal{O})=\mathcal{A}(\cal O)''$, where the double commutant is taken in the local representation. Importantly, by unitary equivalence of the representations, the completion always lands on the same von Neumann algebra. 
    
    However, there is an interesting possibility worth exploring further in our model. Our corner mode $E_L^\text{out}$ commutes with every local gauge-invariant operator in the subregion. Thus, there is a sense in which it could be thought of as a \emph{stochastic background} for our subregion, sourced by local degrees of freedom in the complement. For the purpose of the following argument we restrict attention to the local gauge-invariant algebras ${\cal A}_{\text{loc}}({\cal O})\coloneqq \big\langle\rho(f), J(V_{1}), E_T(V_{2}), B(V_{3}), |\psi|^2(g)\big\rangle\,.$ We do not have a corner mode a priori; this will emerge instead from a von Neumann completion, as follows. 
    
    It being a background means that in a global state with non-zero charge density in the complement of the region, the one-point function of the electric field $E_T$ inside the region gets shifted by the corresponding Coulombic profile, sourced from the outside. Then, if we do the above procedure starting from two global states $\omega_1$ and $\omega_2$ differing by this expectation value inside the region, we may land on two local representations $\pi_{\omega_1}(\mathcal{A}_\text{loc}(\cal O))$ and $\pi_{\omega_2}(\mathcal{A}_\text{loc}(\cal O))$ which are \emph{unitarily inequivalent} instead. This is because the shift induced by the externally-sourced Coulombic profile on the local electric field has infinite norm and hence no unitary operator belonging to $\mathcal{A}_\text{loc}(\cal O)$ can be found that maps one local state to the other.\footnote{This intuition comes from the Shale-Stinespring theorem \cite{shale}, which when applied to CCR algebras (as would be the case here too if we restricted to the local Maxwell sector) states that a classical shift is unitarily implementable if and only if it has finite ``one-particle'' norm, which is equivalent to the norm induced by the two-point function in the Fock vacuum.} We would then reach a situation in which \emph{locally} we have a direct sum/integral of disjoint GNS representations. The von Neumann completions of $\mathcal{A}_\text{loc}(\cal O)$ in each representation would lead to distinct von Neumann algebras. The centre operator $E_L^\text{out}$ could then be obtained via the von Neumann completion in the total direct sum/integral representation. We think this is a promising direction for future work.

    To summarise, here the edge mode appears upon restricting attention to a local gauge-invariant algebra as a stochastic background sourced from the outside. It is not an operator in the global algebra.  It is stochastic because it has a non-zero variance induced from charge fluctuations in the complement. An interesting question is whether this picture would connect with the more explicit route taken in this manuscript, in which we add the longitudinal mode by hand, and especially how it would relate to the necessary corner renormalisation exhibited here.

    We note some instructive parallels and contrasts to the case of the asymptotic soft mode in quantum Maxwell theory \cite{He:2014cra, Strominger:2017zoo}. That is a situation in which the algebra of local gauge-invariant observables experiences a classical background shift by a coherent state of zero-energy photons. But there we have disjoint \emph{global} Fock representations, while all the local representations are unitarily equivalent. Here, we have unitarily equivalent global representations and disjoint \emph{local} ones. The difference comes from the fact that the non-implementability of a soft mode shift is a global \emph{IR effect}, whereas the non-implementability of a local shift of $E_L^\text{out}$ is a \emph{UV effect} coming from the presence of a sharp boundary. We remind the reader that we are working in the zero soft mode sector of the global theory and that our Coulombic field does not lead to any IR divergence due to the exponential fall-off of the charge density two-point function. 
    
\end{enumerate}

\paragraph{Comments on the von Neumann completion of ${\cal A}_\text{C}({\cal O})$.}

Our starting point is a polynomial $\star$-algebra (without a norm) of field operators smeared smoothly and compactly inside some region ${\cal O}\subset\Sigma$. This algebra involves brackets that are not CCR, because of the composite nature of the matter operators, forced on us by gauge invariance. Furthermore, the Fock vacuum of QED (or of the EFT with higher-derivative corrections) is not a quasi-free state on this algebra, for the same reason. Thus, we cannot straightforwardly apply the standard results on the von Neumann completions of Weyl algebras on quasi-free GNS representations, as pioneered in the works \cite{Araki1963, Araki1964_LocalObservables,Osterwalder1973, Hislop:1986at, Figliolini:1989vf}. Nevertheless, we can make a few observations based on the double commutant theorem, applied to the two interpretations outlined above.

Under the first interpretation, in which the physical boundary allows us to think of embedding our centre algebra in the global algebra $\mathcal{A}_C(\Sigma)$, we can use the fact that, as argued in section \ref{subsec:imposing the contraint}, the centre operators $E_L^{\rm out}(V)$ in $\mathcal{A}_C(\mathcal{O})$ do not commute with $J(V) \in \mathcal{A} _{\rm loc} (\mathcal{O}^c)$ to say that 
\begin{equation}
    \begin{aligned}
    \mathcal{A}_\text{C} (\mathcal{O})  ' =\; &\big\langle E_{T} (V),\, B(V) , \, \rho(f) \, \in \mathcal{A}_{\rm loc} (\mathcal{O}^c) \big\rangle\\
    &\vee\big\langle E_L^{\rm out}(V),\vert\psi\vert^2(f)\,\big|\,V\in C_c^\infty(T\mathcal{O})\,,f\in C_{r^{-1}}^\infty(\Sigma)\big\rangle\,,
    \end{aligned}
\end{equation}
which subsequently leads to
\begin{equation}\label{eq: double comm}
   \mathcal{A}_\text{C} (\mathcal{O}) '' = \mathcal{A}_\text{C} (\mathcal{O}) \vee \big\langle \rho (f),\, \forall f \in C_c ^\infty (\mathcal{O}^c) \big\rangle\vee\big\langle\vert\psi\vert^2(f)\,,f\in C_{r^{-1}}^\infty(\Sigma)\big\rangle\,,
\end{equation}
where the r.h.s.\ stands for the smallest von Neumann algebra generated by these algebras. We see that  the von Neumann completion of ${\cal A}_{C}({\cal O})$ contains all (bounded) elements generated by $\rho(f)$ with $\textrm{supp}(f) \in {\cal O}\, \sqcup\, {\cal O}^{c}$ \pah{and the global central mode $\vert\psi\vert^2(f)$, $f\in C^\infty_{r^{-1}}(\Sigma)$}. 

However, if $\o _0$ is a faithful GNS state for $\mathcal{A}_C(\Sigma)$, the double commutant should be equal to the strong/weak norm completion of the operators that we started with. In standard quasi-free representations of CCR $\star$-algebras, the von Neumann completion is obtained by completing the space of test functions in the norm provided by the Gaussian state, and then taking all bounded functions of operators smeared with these test functions. In a quasi-free representation of a CCR algebra, we thus have a strongly continuous extension of the Weyl operators on the closure of the space of smearing functions \cite{Hislop:1986at}. However, since ${\cal A}_{C}(\Sigma)$ is an algebra whose generators have  commutators that are not proportional to the identity, given a (perturbative) GNS state $\vert \Omega \rangle$ at leading order in the coupling, 
\begin{align}\label{afnaf}
\Vert (A(f_n)-A(f))\,\vert \Omega \rangle \Vert^{2}\, \rightarrow\, 0 
\end{align}
does not automatically imply that\footnote{In the case of a Weyl algebra, given a GNS state which satisfies  regularity and cyclicity conditions, this implication follows.}
\begin{align}
 e^{iA(f_n)}\, \rightarrow\, e^{iA(f)}\ \textrm{on}\ {\cal H}_{\Omega}\,.
\end{align}

To the best of our knowledge, such von Neumann completions have not been considered in the mathematical physics literature so far. 
If we conjecture that  our non-CCR algebras do not have a radically new fate under strong/weak completion, eq.~\eqref{eq: double comm} would imply that the completion of test functions for the centre algebra in the norm defined by our boundary kernel $C_{\partial \partial } ^{\text{ren}}$ contains \emph{any} $C_c ^\infty$ test function with support in the complement (in fact, due to $|\psi|^2$ even in $C_{r^{-1}}^\infty$). This is a rather non-trivial extension, because the space of test functions for $\mathcal{A}_{\rm out} (\mathcal{O}) $ corresponds to harmonic functions in $\mathcal{O}^c$. Thus, it would at first appear that completing the space of test functions should only involve completing the space of their boundary values (as we described in section~\ref{ssec_completecornersmearing}). That is, in this example it is not a priori clear that the von Neumann completion will coincide with the completion in the space of test functions as in the case of Weyl algebras alluded to above.

Given the double commutant theorem, it is natural to ask what one might have to \emph{add} to the global algebra $\mathcal{A}_C(\Sigma)$, and more precisely in the complement of the regional centre algebra $ \mathcal{A}_{C} (\mathcal{O})$, such that the double commutant of $\mathcal{A}_C(\mathcal{O})$ now no longer contains arbitrary smooth complement smearings of $\rho$ (and so for the von Neumann completion to possibly match the test function completion). In particular, to that end, we would need to add an operator $\varphi(g)$ to the complement algebra that satisfies 
\begin{equation}
    [\varphi(g), E_L^{\rm out} (V)] =0, \quad  [\varphi(g), \rho (f)] \neq 0, \quad g,f \in C_c ^\infty (\mathcal{O}^c)\,.
\end{equation}
This can be achieved if $g = \triangle g'$ for some $g' \in C_c ^\infty (\mathcal{O}^c)$, and for an operator-valued distribution $\varphi(x)$ that obeys
\begin{equation}
    [\varphi(x), \, \rho (y)] = i \d^3(y-x)\,.
\end{equation}
From the definition of $\rho$ we can work out that $\varphi (x)  \sim \ln \psi (x)$, for $\psi$ the linear charged scalar field operator. To make it gauge-invariant, we can replace $\psi$ with a dressed version $\psi ^\dr$ (defined in section \ref{subsec:subregion phase space}), which gives 
\begin{equation} \label{eq:GM}
   \varphi (x) = \ln \psi (x) + i e \Phi[A,W](x)
\end{equation}
for some choice of frame $\Phi[A]$. An example of this would be a Wilson line from the point $x$ to the asymptotic boundary. Just like $E_L ^{\rm out}$, such an operator is not compact and is non-locally defined. Including it in the algebra would require a careful examination of its two-point function. We will not do this here, but simply observe that \emph{if} such an operator could be made well-defined, its presence would make the local algebra we have constructed and its von Neumann completion, via the double commutant, possibly consistent with some completion in the space of test functions. The same way that the presence of the physical boundary allowed us to include $E_L^\text{out}$, it is at least not ruled out that it could allow for \pah{such an operator to be well-defined} as well, possibly after a further corner renormalisation step. 

Finally, we note that under the second interpretation, the notion of von Neumann completion only makes sense in the local GNS representations constructed from the reduced state on $\mathcal{A}_\text{loc}(\cal O)$ and \emph{not} in the global (renormalised) Fock representation. As a consequence, there is no possibility of adding new operators from the complement. Instead, the completion follows the expected route (modulo the non-CCR nature of the algebra) of completing the space of test functions in the norm provided by the reduced state and building bounded functions of the generators. The potential novelty here is the direct integral of unitarily inequivalent representations discussed earlier, which remains to be explored further.

\paragraph{Further directions.} An important next step is to attempt a covariant definition of our centre mode in the setup of local algebras associated to open sets in spacetime rather than space. As noted earlier, the renormalisation steps associated with evaluating the Fock vacuum pulled-back to the Cauchy slice on composite operators would not be required. However, we expect that the corner renormalisation needed to defined the edge mode remains necessary even for algebras associated to causal diamonds, because the corner remains sharp.

That being said, the main direction we would like to take this line of research in is in constructing an analogous definition of corner modes for subregions in perturbative quantum gravity. There is a vast literature on the so-called ``corner program'' of (quantum) gravity, e.g.\ see \cite{Donnelly:2016auv,Freidel:2023bnj,Freidel:2020xyx,Speranza:2017gxd,Geiller:2017xad,Ciambelli:2021nmv,Carrozza:2022xut,Ciambelli:2022vot,Freidel:2025ous,Freidel:2026gcp,Bonifacio:2026jik},  in which, very much in the spirit of gauge theories, corner charges associated to generators of corner-supported diffeomorphisms are assigned to the subregion. In our construction, we expect the corner mode to be constructed instead from the conformal mode sourced by stress-energy density in the complement via the Hamiltonian constraint. An exciting possibility is that the geometrical fluctuations of the subregion (and hence its corner) could render our corner mode well-defined already in the perturbative vacuum, without needing any corner renormalisation. 

\acknowledgments 

GAR, AL and BT thank the hospitality of ICTS, Bangalore, during the workshop ``Quantum Information, Quantum Field Theory and Gravity'' in the summer of 2024, where this collaboration was started. GAR and BT also thank the hospitality of the Chennai Mathematical Institute where this collaboration continued. AL thanks S Sundar for countably infinite discussions on operator algebras and his patient explanations over the years. AL thanks the hospitality of the Okinawa Institute of Science and Technology, where this collaboration started coming to fruition. AL thanks Justin David and Ronak Soni for insightful discussions on edge modes in gauge theories and the organisers of ``Advances in Black Hole Theory'' at the Indian Institute of Science, Education and Research (IISER), Pune, in June earlier this year, where part of this work was presented. This work was supported by the ID\# 62312 grant from the John Templeton Foundation, as part of the project \href{https://www.qiss.fr/}{``The Quantum Information Structure of Spacetime'' (QISS)} and through the \href{https://withoutspacetime.org}{``WithOut SpaceTime project'' (WOST)}, led by the Center for Spacetime and the Quantum (CSTQ), and funded by Grant ID\# 63683 from the John Templeton Foundation. The opinions expressed in this work are those of the authors and do not necessarily reflect the views of the John Templeton Foundation.

\appendix

\newpage
\section{Useful Fourier formulae}\label{app:fourier}
Throughout this paper, we have used estimates based on three-dimensional Fourier transforms which we summarise in this appendix for the benefit of the reader. To set the convention, recall that 
\begin{align}
\tilde{f}(k)\, &=\, \int \dt^{3}x\, e^{i k \cdot x} f(x)\nonumber\\
f(x)\, &=\, \int \frac{\dt^{3}k}{(2\pi)^{3}}\, e^{-i k \cdot x} \tilde{f}(k)\,.
\end{align}
Hence,
\begin{align}
&\text{FT}\l \delta^{3} \r (k) \, = \, 1 , \qquad \text{FT}\l \triangle \delta^{3}\r (k) \, =\, - \vert k \vert^{2}\,.
\end{align}
From here, we can define the Fourier transform of functions on which we have acted with the Laplacian or its inverse:
\begin{align}
    \text{FT}\l \triangle f \r (k) \, =\, - \vert k \vert^{2} \tilde{f}(k), \qquad   \text{FT}\l-  \triangle ^{-1} f \r (k) \, =\,\frac{\tilde{f}(k)}{|k|^2}\,.
\end{align}
The Fourier transform of the characteristic function of a ball of radius $R$ is given by:
\begin{align}
\tilde{\chi_R}(k)\, =\, \frac{3}{\vert k \vert^{3}}\left(\sin(R\vert k \vert) - R\vert k \vert \cos(R\vert k \vert)\right)\,,
\end{align} 
while for a general spherically symmetric function $f(r)$, we have
\begin{equation}
    \begin{split}
        \tilde{f} (k) = \frac{1}{2 \pi ^2} \int _0 ^\infty \ r^2 \dt r   \, f(r) j_0 (|k|r)\,,
    \end{split}
\end{equation}
where $j_l$ is the spherical Bessel function of the first kind. Since functions of the type $1/r^n$ for $n>2$ are not integrable, we cannot evaluate their Fourier transform in the usual way. However, using
\begin{equation}
    \triangle \l \frac{1}{r^{n-2}} \r = \frac{(n-2)(n-3)}{r^{n}}, \qquad \text{FT}\l \frac{1}{r^2} \r = \frac{2\pi^2}{p}\,,
\end{equation}
we can attribute to $1/r^n$ the following Fourier transform function in a distributional sense:
\begin{align}
\textrm{FT}\l \frac{1}{r^{n}}\r (k)\, =\, 2^{3-n}\, \pi^{\frac{3}{2}}\, \frac{\Gamma(\frac{3-n}{2})}{\Gamma(\frac{n}{2})}\, \vert k \vert^{n - 3}\qquad 
\textrm{FT}\big((-\, \triangle)^{-1}\, f\big)(k)\, =\, \frac{\tilde{f}(k)}{k^{2}}\,.
\end{align}

Next, the Fourier transform of $\mathrm{1} _{\mathcal{O}^c} f (x)$, where $\mathcal{O}$ is a spherically symmetric bounded region of radius $R$, and $f$ is harmonic in its complement is:
\begin{equation}
    \begin{split}
        \text{FT} \l \mathrm{1} _{\mathcal{O}^c} f  \r (k) & =  \frac{1}{(2 \pi)^3}\int _R ^\infty \int _{S^2} r^2 \dt r \dt \Omega _2 ^2 \, e^{i k \cdot \vec{r}} \, \sum _{l,m} f_{l,m} \, \frac{Y_{l,m} (\theta,\phi)}{r^{l+1}} \\
    & = \l 2 \pi \r ^{3/2}\sum _{l,m} i^l f_{l,m}  Y_{l,-m} (\hat{k}) \frac{j_{l-\frac{1}{2}} (kR) }{k^{3/2}}\,,
    \end{split}
\end{equation}
where we have used Rayleigh's plane wave expansion formula 
\begin{equation}
e^{i k \cdot {r}}  = 4 \pi \sum _{l,m} \, i ^l j_l (kr)  \,  Y_{l,-m} ({k})  Y_{l,m} ({r})\,.
\end{equation}

Finally, we need to evaluate the Fourier transform of $\ln ^n (mr)$ in 2d. Once again, these functions are not integrable, so we need some trick to define their Fourier transform in a distributional sense. We do that in the following way:
\begin{equation}
    \begin{split}
       I(s)  & =   \frac{1}{2\pi} \int _0 ^\infty \dt r\,  m^{s} r^{s-1} j_0(kr) \\
       \textrm{FT}_{2d} (\ln ^n mr) & \coloneqq \frac{1}{m^2} \lim _{s \to 2}\frac{\dt^n I(s) }{\dt s^n  } \,,
    \end{split}
\end{equation}
where some precaution is needed when dealing with the limit. With this we find that
\begin{equation}
    \begin{split}
         \textrm{FT}_{2d} (\ln ^2 mr) &=  \frac{ \ln \frac{k}{2m} + \gamma }{\pi  k^2}\\
          \textrm{FT} _{2d} (\ln ^3 mr)&= - 3 \frac{\l  \ln \frac{k}{2m} + \gamma \r ^2}{2\pi k^2}\,.
 \end{split}
\end{equation}

\section{Epstein--Glaser and differential renormalisation}\label{app:EG}
In this section, we review the basics of Epstein-Glaser and differential renormalisation, which has been used throughout this paper to define the GNS state on the Cauchy slice.

Although these approaches have primarily been developed in the context of the covariant perturbation theory, at the heart of them lies the theory of extensions of distributions to coincident points. Thus, the formalism is also applicable to the renormalisation of equal-time correlation functions, as has been employed throughout the paper. While the theory of renormalisation developed by Epstein and Glaser is mathematically rigorous, a ``physicist's version'' of this formalism which can be used to do concrete computations was in fact developed in a seminal paper by Freedman et al.~\cite{Freedman:1991tk} 
 and is known as differential renormalisation. We begin with a brief summary of the Epstein-Glasser (EG) approach, before turning to differential renormalisation. For more details on the EG approach to renormalisation and its implementation via differential renormalisation, we refer the reader to \cite{Epstein:1973gw,Prange:1997iy, Gracia-Bondia:2001hlz, Gracia-Bondia:2014zwa}.

\subsection{Position space renormalisation}
The basic idea of EG renormalisation (adopted to Euclidean correlators) is the following. Let ${\cal T}$ be a distribution on the space of smooth functions in ${\bf R}^{n}/\{0\}$
\begin{align}
{\cal T}(\varphi)\, =\, \int \dt^{n}x\, {\cal T}(x)\, \varphi(x)\, ,
\end{align}
where $\varphi$ is some smooth function.
Let $s$ be the scaling degree of ${\cal T}$ in the sense that
\begin{align}
s\, \coloneqq\, \inf\left\{s^{\prime}\, \middle|\,
\lim_{\lambda\, \rightarrow\, 0^{+}}\lambda^{s^{\prime}}{\cal T}(\lambda x)\, =\, 0\right\}
\end{align}
and let $\omega\, =\, s - n$ which is known as the singular order.\footnote{In momentum space, $\omega$ is precisely the superficial degree of divergence.} When $\omega \geq\, 0$, this implies  that ${\cal T}$ cannot be defined as a distribution on the space of functions over ${\bf R}^{n}$ and thus indicates a superficial divergence. The idea of Epstein and Glaser essentially is to view renormalisation as the possible extension $\overline{\mathcal{T}}$ of ${\cal T}$ over the entire ${\bf R}^{n}$.

More in detail, $\overline{{\cal T}}$ is a distribution defined as 
\begin{align}
\overline{{\cal T}}(\varphi)\, =\, \int_{{\bf R}^{n}}\dt^nx\, {\cal T}(x) \big(\,  \varphi({x})\, -\, \kappa(x)\, \big(\, \varphi(0)\, +\, \sum_{m=1}^{\omega}\, x^{\vec{\beta}}\, \partial_{\vec{\beta}}\, \varphi(0)\, \big){ \big)}\,,
\end{align}
where $\kappa(x)$ is any smooth function of compact support around the origin that satisfies the following conditions
\begin{align}\label{kappader0}
\kappa(0)\, =\, 1\,,\qquad\qquad\partial_{\vec{\alpha}}\kappa(0)\, =\, 0\,\quad \forall\, \vert \vec{\alpha} \vert \leq\, \omega\,.
\end{align}
In this manner, the Taylor series of the function truncated at the singular order is subtracted around the origin in such a way that the surviving higher-order contributions of $\varphi$ lie in the domain of the unextended $\mathcal{T}$.

The extension of ${\cal T}$ to $\overline{\cal T}$ is scheme-dependent and this dependence is contained in the choice of $\kappa(x)$. This can be seen as follows. Let ${\cal T}^{\prime}$ be a different extension of ${\cal T}$ such that
\begin{align}
{\cal T}^{\prime}(\varphi)\, =\, \int_{{\bf R}^{n}}\dt^nx\, {\cal T}(x)\big(\, \varphi(x)\, -\, \kappa^{\prime}(x)\, \big(\, \varphi(0)\, +\, \sum_{m=1}^{\omega}\, x^{\vec{\beta}}\, \partial_{\vec{\beta}}\, \varphi(0)\, \big)
\end{align}
Then
\begin{align}
( \overline{\cal T}\, -\, {\cal T}^{\prime} )(\varphi)\, &=\, \int_{{\bf R}^{n}}\dt^nx\, (\kappa^{\prime} - \kappa)(x)\, {\cal T}(x)\, \big(\, \varphi(0)\, +\, \sum_{m=1}^{\omega}\, x^{\vec{\beta}}\, \partial_{\vec{\beta}}\, \varphi(0)\, \big)\nonumber\\
&=\, \sum_{\vert \vec{\beta} \vert = 0}^{\omega}\, a_{\vec{\beta}}\, \partial_{\vec{\beta}}\, \varphi(0)\,,
\end{align}
where
\begin{align}\label{avecbeta}
a_{\vec{\beta}}\, =\, \int_{\vert x \vert > \delta}\dt^nx\, {\cal T}(x)\, (\kappa^{\prime} - \kappa)(x)\, x^{\vec{\beta}}\,.
\end{align}
The bound on the integral above is due to the fact that, thanks to eq.~(\ref{kappader0}), any two smoothing functions will always agree in the $\delta$-neighbourhood of the origin for some $\delta > 0$. Using the identity
\begin{align}
\partial_{\vec{\beta}}\varphi(0)\, =\, (-1)^{\vert \beta \vert}\, \delta^{(\vec{\beta})}(\varphi)\,,
\end{align}
where
\begin{align}
\delta^{(\vec{\beta})}(\varphi)\, =\, \int \dt^{n}x \,\varphi(x)\, \partial_{\vec{\beta}}\delta^{n}(x)\,,
\end{align}
we see that the choice of finite local counter-term is precisely contained in the choice of smoothening function $\kappa(x)$. 

As a simple example, consider a distribution ${\cal S}$ in ${\bf R}$ given by
\begin{align}
{\cal S}(x)\, =\, \frac{1}{x^{2}}\,.
\end{align}
Its scaling degree is $2$ with $\omega = 1$, while its domain is the space of all smooth functions such that 
\begin{align}
\lim_{x\to0^\pm}\varphi(x)\, {=}\, O(x^{2})\,.
\end{align}
For generic smooth functions, 
\begin{align}
\overline{\cal S}(\varphi)\, =\, \int\dt x\, {\cal S}(x)\left(\, \varphi(x) - \kappa(x)\left(\varphi(0)+x\varphi'(0)\right)\, \right)\,,
\end{align}
where $\kappa(x)$ is a  smoothing function that satisfies
\begin{align}
\kappa(0)\, =\, 1\,,\qquad\qquad
||\, \kappa ||^{2} = 1\,.
\end{align}
In this manner, the bracket is of order $O(x^2)$ near the origin and the divergences are removed.

\subsection{Differential Renormalisation}
A more practical method to implement the Epstein-Glaser method was proposed in a seminal paper by Freedman et al.~\cite{Freedman:1991tk} and goes under the name of differential renormalisation. We briefly review their idea as it has been our main tool to renormalise two-point functions on a fixed time slice.

We illustrate the idea with a simple example. Consider a distribution in ${\bf R}^{3}$, which is defined as follows on the space of all smooth functions with support away from the origin.\footnote{We assume that as $r\, \rightarrow\, \infty$, $\varphi(x)\, \sim\, r^{1-\epsilon}$ for $\epsilon > 0$. This ensures that the integral is convergent in the infra-red.} 
\begin{align}
{\cal T}_{0}(\varphi)\, =\, \int_{{\bf R}^{3}/\{0\}}\dt^3x\, \frac{1}{r^{4}}\, \varphi({{x}}) \,,
\end{align}
where $r=\vert x\vert$. In this case, $s=4$ and $\omega = 1$.

To extend the distribution so that its domain includes the space of all smooth functions, we can invoke the following identity $\forall\, a$
\begin{align}
\triangle \left(\frac{1}{r^{2}}\right)\, =\, 2 \frac{1}{r^{4}}\, + a\, \delta^{3}(x)\qquad \forall\, r > 0\,,
\end{align}
Hence, we define the renormalised distribution as 
\begin{align}
\overline{{\cal T}_{0}}(\varphi)\, \coloneqq\, {\frac{1}{2}}
\int_{{\bf R}^{3}}\, \dt^{3}x\, \frac{\triangle \varphi(x)}{r^{2}}\, +\, \frac{a}{2}\, \varphi(0)\,.
\end{align}
That this approach is simply a way to implement the Epstein-Glaser idea can be deduced as follows. 
We can write the integral (for $a =0$) as 
\begin{align}
\overline{\cal T}_{0}(\varphi)\, &=\, 
\lim_{\epsilon\, \rightarrow\, 0}\, \frac{1}{2} \int_{r > \epsilon}\, \dt^{3}x \frac{1}{r^{2}}\, \triangle\varphi({x})\, \nonumber\\
&=\, \lim_{\epsilon\, \rightarrow\, 0}\, 
\left(\int_{r > \epsilon}\dt^3x\, \frac{1}{r^{4}}\, \varphi({x})\, -\, \frac{1}{2} \, \int_{S^{2}_{\epsilon}}\, \dt^2\hat{r}\, \hat{r} \cdot\, \nabla\varphi(\epsilon\hat{r})\, +\, \frac{1}{2}\, \epsilon^{2}\int_{S^{2}_{\epsilon}}\, \dt^2\hat r\, \varphi(\epsilon\hat{r})\, \hat{r}\cdot\nabla\, \left(\frac{1}{r^{2}}\right)\right)\,,
\end{align}
where the second line follows from Green's second identity. To leading order in $\epsilon$, we have
\begin{align}
\int_{S^{2}}\, \dt^2\hat r\, \hat{r}\, \cdot \nabla \varphi(\epsilon\hat{r})\, =\, \partial_{i}(\varphi)(0)\, \int_{S^{2}} \dt^2\hat r\, \hat{r}^{i}\, =\, 0\,.
\end{align}
Similarly, to leading order in $\epsilon$, the second boundary term equals $-\frac{1}{\epsilon}\varphi(0)$.
Hence, we see that
\begin{align}
 \overline{\cal T}_{0}(\varphi)\, =\, 
\lim_{\epsilon\, \rightarrow\, 0}\, 
\left(\, \int_{r > \epsilon}\, \dt^3x\, \frac{1}{r^{4}}\, \varphi(x)\, - \frac{1}{2\, \epsilon} \varphi(0) \right)\,,
\end{align}
which can be identified with the EG renormalised distribution for a suitable choice of $\kappa(x)$.\footnote{We note that, even though $\omega = 1$ in this example, the term proportional to 
$\kappa(x)\, {\cal T}\, x^{i}\partial_{i}\varphi(0)$ in the EG renormalisation is trivial since the distribution is spherically symmetric 
\begin{align}
\frac{1}{r^{4}}\, \int_{S^{2}_{r}}\, r^{2}\, d^{2}\hat{r}\, \hat{r}^{i}\partial_{i}\varphi(0)\, = 0\,.
\end{align}}

\section{Exact boundary kernel on the sphere}\label{app:S2kernel}

In this appendix, we derive explicitly the expression for the renormalised boundary kernel of section~\ref{subsec:boundary renorm} in the special case where the subregion boundary is a sphere, i.e. without the planar limit approximation invoked there. 

As shown in the main text, the boundary kernel in \eqref{eq:tildeC} is a linear combination of the following three terms
\begin{equation}\label{eq:3terms}
    \tilde{C} _1= -\frac{1}{12 m^3 r^{2}}, \qquad  \tilde{C} _2 \coloneqq- \frac{-1+2 \gamma - \ln 4}{4m}  \ln m r,  \quad \tilde{C} _3 \coloneqq - \frac{1}{4m} \ln ^2 \l m r\r \,.
\end{equation}
Because all operators with which we will work depend only on the relative distance distance $r_{\parallel} =2 R^2 (1- \cos \gamma )$, where $\cos \gamma  = \cos \theta \cos \theta' - \cos(\phi-\phi') \sin \theta \sin \theta '$, we will express all of them as a sum of the Legendre polynomials $P_l(\cos \gamma )$. Specifically, we make use of the following identities:
\begin{equation}
    \begin{split}
          \ln (1-\cos \gamma) & =  -  \sum_{l=1} ^{\infty}   \frac{2l+1}{l(l+1)} P_l (\cos \gamma ) + \l \ln 2 -1\r\,,\\
           \frac{1}{1-\cos \gamma} &  =\sum _{l=1} ^{\infty} (2l+1) \underbrace{\l H (l)  - \frac{1-(-1)^{l}}{2} \r }_{=:\tilde{H}(l)}P_l  (\cos \gamma ) , \\
         \ln ^2 \l 1-\cos \gamma \r  & = \sum _{l=1} ^{\infty} \, -\frac{2l+1}{l(l+1)}\l 4H(l) + \frac{2}{l(l+1)} +2 \ln 2(1-(-1)^l)  \r \, P_l  (\cos \gamma ) \\
        & \qquad\qquad\qquad\qquad\qquad+4 +2\ln^2 2-2 \ln 4 \,,\\
         \frac{\ln  \l 1-\cos \gamma \r }{1-\cos \gamma }  & = \sum _{l=1} ^{\infty} \frac{2l+1}{2} \l  2  \tilde{H}^2(l) + \frac{\ln^2 2}{2}  \r P_l  (\cos \gamma ) + \frac{\ln ^2 2 }{4}\,,
    \end{split}
\end{equation}
where $ H (l) = \sum _{k=1} ^l \frac{1}{k} $. A word of explanation is needed here: if one na\''ively attempts to determine the coefficients in the expansion above by integrating the left hand side of the bottom three lines above against $P_l \cos \gamma $, one would obtain divergent results. In the expansions above, we have removed these divergences. Thus, they should be viewed as renormalised expressions of  $\frac{1}{(1-\cos \gamma)}$, $\ln ^2 (1-\cos \gamma )$ and $ \frac{\ln  \l 1-\cos \gamma \r }{(1-\cos \gamma) } $. Furthermore, for harmonic functions in $\mathcal{O}^c$ we have:
\begin{equation}
    f (x)\Big|_{S^2} = \sum _{l=0} ^{\infty} \sum _{m=-l} ^{l} \frac{f_{l,m}}{R^{l+1}} Y_{l,m} (\theta ,\phi), \qquad \nabla _\perp ^x  f(x)\Big|_{S^2}   = \sum _{l=1} ^{\infty} \sum _{m=-l} ^{l} -(l+1)\frac{f_{l,m}}{R^{l+2}} Y_{l,m} (\theta ,\phi) \,.
\end{equation}
With the formulas above, we can write the boundary part of the two point function of $E_L ^{\rm out} (V)$ from \eqref{eq:w(el)_intial} as the sum of the following rather long integral expressions coming from the three terms in eq.~\eqref{eq:3terms}:
\begin{equation}
    \begin{split}
       I_1 &\coloneqq\int _{S^2} \int_{S^2} \, \dt^2 y \dt^2 x \, \left( \nabla _\perp ^x f(x) \tilde{C}_1(x-y)  \nabla _\perp ^y g(y) - \nabla _\perp ^x f(x) \nabla _\perp ^y \tilde{C}_1(x-y)   g(y) \right.\\
       & \qquad \qquad \qquad \qquad \qquad  \qquad \left. -  f(x) \nabla _\perp ^x \tilde{C}_1(x-y)^y \nabla _\perp  g(y)+  f(x)  \nabla _\perp ^x \nabla _\perp ^y \tilde{C}_1(x-y)  g(y)  \right) \\
        &   = -\frac{1}{12 m^3}\sum _{l ,m }  f_{l,m} g_{l,-m}\int _{S^2} \int _{S^2} \, \dt \Omega _2 ^2 \dt \Omega _2 ^{'2} \, Y_{l,m}(\theta, \phi) Y_{l, -m} \l \theta', \phi ' \r  \l \frac{l(l-1)}{2} \r\frac{1}{2R^4(1-\cos \gamma )} \\
       &   = - \frac{ \pi }{24 m^3 R^4}  \sum _{l ,m } f_{l,m} g_{l,-m} \tilde{H}(l) l(l-1)\,,
    \end{split}
\end{equation}
and
\begin{equation}
    \begin{split}
       I_2 &\coloneqq\int _{S^2} \int_{S^2} \, \dt^2 y \dt^2 x \, \left( \nabla _\perp ^x f(x) \tilde{C}_2(x-y)  \nabla _\perp ^y g(y) - \nabla _\perp ^x f(x) \nabla _\perp ^y \tilde{C}_2(x-y)   g(y) \right.\\
       & \qquad \qquad \qquad \qquad \qquad  \qquad \left. -  f(x) \nabla _\perp ^x \tilde{C}_2(x-y)^y \nabla _\perp  g(y)+  f(x)  \nabla _\perp ^x \nabla _\perp ^y \tilde{C}_2(x-y)  g(y)  \right)  \\
         &
         =- \frac{-1+2 \gamma - \ln 4}{4m} \sum _{l,m} f_{l,m} g_{l,-m} \int _{S^2} \int _{S^2} \, \dt \Omega _2 ^2 \dt \Omega _2 ^{'2} \, Y_{l,m}(\theta, \phi) Y_{l, -m} \l \theta', \phi ' \r   \\
         & \qquad \qquad \qquad \times   \left(   \frac{(l+1)^2}{R^2}\ln (1-\cos \gamma )+ \frac{(l+1)^2 \ln 2mR^2+2(l+1)}{R^2} - \frac{1}{R^2 (1-\cos \gamma  )} \right) \\
    & = -  \frac{\pi \l 1-2 \gamma + \ln 4 \r }{m} \sum _{l,m} f_{l,m} g_{l,-m} \l   \frac{(l+1)}{l R^2} + \frac{ \tilde{H}(l)}{R^2} - \d(l) \frac{2\ln 4 mR^2 }{R^2} \r\,,
    \end{split}
\end{equation}
as well as
\begin{equation}
    \begin{split}
 \      I_3 &\coloneqq\int _{S^2} \int_{S^2} \, \dt^2 y \dt^2 x \, \left( \nabla _\perp ^x f(x) \tilde{C}_3(x-y)  \nabla _\perp ^y g(y) - \nabla _\perp ^x f(x) \nabla _\perp ^y \tilde{C}_3(x-y)   g(y) \right.\\
       & \qquad \qquad \qquad \qquad \qquad  \qquad \left. -  f(x) \nabla _\perp ^x \tilde{C}_3(x-y)^y \nabla _\perp  g(y)+  f(x)  \nabla _\perp ^x \nabla _\perp ^y \tilde{C}_3(x-y)  g(y)  \right)   \\
       & =- \frac{1}{4m} \sum _{l,m} f_{l,m} g_{l,-m} \int _{S^2} \int _{S^2} \, \dt \Omega _2 ^2 \dt \Omega _2 ^{'2} \, Y_{l,m}(\theta, \phi) Y_{l, -m} \l \theta', \phi ' \r \, \left( \frac{(l+1)^2}{R^2} \ln ^2 \l 1- \cos \gamma  \r  \right. \\
       &   \qquad \qquad+  \frac{(l+1)^2 \ln^2 2mR^2 + 4(l+1)\ln m 2mR^2 +2}{R^2}  \\
       &     \qquad \qquad \left.+\frac{2(l+1)^2 \ln (2mR^2) + 4(l+1)}{R^2} \ln(1-\cos \gamma) - \frac{2\ln (1-\cos\gamma) +2\ln 2mR^2 }{1-\cos \gamma} \right)  \\
      &  = \frac{\pi }{m} \sum _{l,m} f_{l,m} g_{l,-m} \left( \frac{l+1}{l R^2} \l 4H(l) + \frac{2}{l(l+1)} +2 \ln 2(1-(-1)^l) \r  + \frac{2(l+1) \ln (2mR^2) + 4 }{lR^2} \right.\\
      & \qquad  \qquad \qquad \qquad  \qquad \left. + 2\l  \tilde{H}^2(l)  + \frac{\ln^2 2}{2} \r +  2\ln 2mR^2 \tilde{H}(l)  +\delta(l) \alpha_0 \right)\,.
    \end{split}
\end{equation}
where $\alpha _0$ is some complicated function of $m$ and $R$, whose exact expression is not relevant for the analysis bellow.

We emphasise again that the expression resulting from the integral above is only valid for the renormalised distributions.

Now that we have all the terms explicitly we need to verify whether the renormalised boundary kernel is positive. A simple plot in Mathematica can verify that for each spherical mode $l>1$ it is true that:
\begin{equation}
    \l I_1 (l)  + I_2 (l) + I_3(l)  - \triangle_{S^2} \d^2 (l)\r > 0\,.
\end{equation}
Note that, upon Fourier transformation, the counter-term given by the sphere Laplacian which restores positivity is just $k^2$, thus the same that we found in the planar limit in eq.~\eqref{eq: positive corner 2pt} in the main body. Furthermore, we have complete freedom to specify the $l=0$ mode by adding a counter-term proportional to the delta function on the sphere.

This summarises the construction of the positive renormalised boundary kernel for a spherical corner without approximation.

\section{Electric centre edge modes in Maxwell theory}\label{app:empm}
In this section, we show that the two-point function of the transversal electric field in free Maxwell theory can be used to define an extension of the quasi-free state $\omega_{{\cal O}}^{EM}$ on the domain that includes electric centre edge modes on a codimension-one boundary $\partial{\cal O}$ that measure the normal flux through the surface. We first use differential renormalisation to build a finite two-point function for these normal electric centre mode operators and then show that by adding suitable counter-terms (which are localised on $\partial{{\cal O}}$) to the renormalised kernel we obtain $\omega_{{\cal O}}^{\textrm{EM}}$.

We recall that for an arbitrary pair of smooth vector fields $V, W\, \in\, C_{c}^{\infty}(T{\cal O})$ the equal time correlator reads
\begin{align}
\omega_{\rm free}^{EM}(E_{T}(V)\, E_{T}(W))\, =\, \int_{{\cal O}^{2}}\dt^3x\, \dt^3y\, V^{i}(x)\, C_{ij}(\vert x - y\vert)\, W^{j}(y)\,,
\end{align}
where $C_{ij}(\vert x - y \vert)$ is given in \eqref{eixejyff}. To distinguish it from the Fock vacuum in our scalar QED setup, we denote the free vacuum for pure Maxwell theory by $\omega_{\rm free}^{EM}$.

For concreteness, we consider ${\cal O}$ to be a spherical region in $\Sigma$ with radius $R$. We can define ``corner electric fields''/edge modes supported on $\partial{\cal O}$ by expanding the space of smearing vector fields from $C_{c}^{\infty}(T{\cal O})$ to the space of smooth vector fields which do not vanish on the boundary.
That is, consider a one-parameter family of smooth vector fields  $V^{<}_{\delta}$ in $\mathcal{O}$ defined as, 
\begin{align}
V^{<}_{\delta}(r, \hat{n})\coloneqq \begin{cases}
    \frac{1}{\delta}\, e^{-\frac{(r-R)^{2}}{\delta^{2}}}\, v^{<}(\hat{n})\, \hat{r}&  \textrm{for}\,\, r \leq R\,,\\
    \, 0\, &\textrm{for}\,\, r\, >\, R\,,
\end{cases} 
\end{align}
for some $\delta>0$, where $v^{<}(\hat{n})$ is a smooth function on $S^{2}$ and $\hat{r}$ is the unit normal vector to the sphere. In particular, we have $\lim_{\delta\to0}1/\delta \,e^{-(r-R)^2/\delta^2}=\sqrt{\pi}\,\delta(r-R)$.

Thus, given this region $\mathcal{O}$, we may define new electric edge modes living entirely on the corner $\p\mathcal{O}$ via the limit
\begin{equation}
  E_\p(v^<)\coloneqq  \lim_{\delta\to0}E_T(W_\delta^<)\,,\qquad\text{where}\qquad E_T(W^<_\delta)=\int_{\mathcal{O}}\dt^3x\, E_T^i(x)\,W^<_{\delta,i}(x)\,.
\end{equation}
These observables measure the electric flux through the corner. Let us denote the Abelian $\star$-algebra generated by these electric edge modes by $\mathcal{A}_{\rm edge}^{EM}(\p\mathcal{O})$. Then we may consider the centrally augmented regional algebra 
\begin{equation} 
\mathcal{A}_{\rm loc}^{EM}(\mathcal{O})\vee\mathcal{A}_{\rm edge}^{EM}(\partial\mathcal{O})\,,
\end{equation}
where $\mathcal{A}_{\rm loc}^{EM}(\mathcal{O})$ is the same as the pure Maxwell theory part of the algebra in eq.~\eqref{eq:AlocO}. By construction, $\mathcal{A}_{\rm edge}^{EM}(\p\mathcal{O})$ constitutes the centre of the augmented algebra. 

We can now define a new state $\omega_{\cal O}^{EM}$, which is a positive GNS-functional on this extended regional algebra such that
\begin{align}
\omega_{{\cal O}}^{EM}(\, E_{T}(V)\, E_{T}(W)\, )\, =\, \omega_{\textrm{free}}^{EM}(\, E_{T}(V)\, E_{T}(W))\qquad \forall\, V, W\, \in\, C_{c}^{\infty}(T{\cal O})    
\end{align}
and its restriction to the electric edge mode algebra is defined by renormalising 
\begin{equation}
\omega_{{\cal O}}^{EM}(\, E_{T}(W^{<}_{\delta})\, E_{T}(W^{<}_{\delta}))\,.
\end{equation}

\subsection{Renormalised edge mode two-point function in pure Maxwell theory}

Computing the two-point function of the corner supported transversal fields in the free state yields
\begin{align}\label{wdeltavdelta}
\omega_{\textrm{free}}^{EM}(E_{T}(W_{\delta}^{<})\, E_{T}(V_{\delta}^{<}))\, =\, 
\int_{{\cal O}^{2}}\, r^{2} \tilde{r}^{2} \dt r \dt\tilde{r} \dt^{2}\hat{n}\, \dt^{2}\hat{\tilde{n}}\, W_{\delta}^{<\, i}(r,\hat{n})\, C_{ij}(\vert r\hat{n} - \tilde{r}\tilde{n} \vert)\, V_{\delta}^{< j}(\tilde{r}, \tilde{n})\,,
\end{align}
where eq.~\eqref{eixejyff} implies that, away from the coincidence limit, 
\begin{align}
C_{ij}(\vert r\hat{n} - \tilde{r}\tilde{n} \vert)\, =\, - \frac{1}{(2\pi)^{2}}\, \frac{1}{\vert r \hat{n} - \tilde{r}\tilde{n} \vert^{2}}\,.
\end{align}
The $\delta\, \rightarrow\, 0$ limit of the r.h.s.\ of eq.~(\ref{wdeltavdelta}) is divergent, and as a result $\omega_{\textrm{free}}^{EM}$ is not a GNS state on the centrally extended local algebra that includes electric edge modes. We now use differential renormalisation to obtain a boundary kernel $C_{rr}^{\textrm{ren}}$ which leads to finite two-point functions of the normal electric corner mode operator.

It may be verified that the $\delta\, \rightarrow\, 0$ limit of the above expression takes the following form 
\begin{align}\label{omega0etet}
\lim_{\delta\, \rightarrow\, 0}\, \omega^{EM}_{\rm free}(E_{T}(W_{\delta}^{<})\, E_{T}(V_{\delta}^{<}))\, =:\, R^{4}\, \int_{S^{2} \times S^{2}}\, \dt^{2}\hat n \dt^{2}\tilde{n}\, v^{<}(\hat{n})\, C^{{\textrm{ren}}}_{rr}(R \vert \hat{n} - \tilde{n}\vert)\, w^{<}(\tilde{n})\,,
\end{align}
Using the explicit form of $C_{ij}(\vert x - y\vert)$ in eq.~\eqref{eixejyff}, one may check that, away from the coincidence limit, the unrenormalised kernel reads
\begin{align}
C_{rr}(\vert r\, \hat{n} - \tilde{r}\, \tilde{n}\vert)\Big\vert_{r = \tilde{r} = R}=\, -\, \frac{1}{(2\pi)^{2}\, R^{4}}\, \frac{1}{(1 - \hat{n} \cdot \tilde{n})^{2}}\, \qquad\textrm{for}\, \hat{n} \neq\, \tilde{n}\,.
\end{align}
Invoking the identity
\begin{align}
\frac{1}{(1 - \hat{n} \cdot\, \tilde{n})^{2}}\, =\, \left[\, \frac{1}{8}\, \triangle_{S^{2}}^{2}\, \ln^{2}(1 - \hat{n} \cdot \tilde{n})\, +\, \frac{1}{4}\, \triangle_{S^{2}}\, \ln(1 - \hat{n} \cdot \tilde{n})\, \right]\, \qquad\textrm{for}\, \hat{n} \neq\, \tilde{n}\,,
\end{align}
we now obtain the renormalised two-point function as
\begin{align}\label{crrR1nn}
C^{{\textrm{ren}}}_{rr}(R \vert \hat{n} - \tilde{n}\vert)=\, -\, \frac{1}{(2\pi)^{2}{R^4}}\, \left[\, \frac{1}{8}\triangle_{S^{2}}^{2}\, \ln^{2}(1 - \hat{n} \cdot \tilde{n})\, +\, \frac{1}{4}\, \triangle_{S^{2}}\, \ln(1 - \hat{n} \cdot \tilde{n})\, \right]\,,
\end{align}
but with the domain \emph{including} the coincidence limit. This kernel is now manifestly finite, since
\begin{enumerate}
\item 
$\triangle_{S^{2}}\, \ln(1-\hat{n} \cdot \tilde{n})\, =\, -1 + 4\pi \delta^{2}({\hat{n} - \tilde{n}})$. 
\item $\int_{0}^{\pi} \dt\,\cos\theta\, f(\theta)\, \ln^{m}(1-\cos\theta)$ is finite for any smooth function $f$ and for all $m\ge 0$.
\end{enumerate}
We note that the resulting kernel is scheme-dependent. Indeed, away from the coincidence limit, we could have chosen the following representation of $C_{rr}$:
\begin{align}
C_{rr}^{\textrm{ren},\, c}\, &\coloneqq\,  \frac{1}{(2\pi)^{2}{R^4}}\, \left[\, \frac{1}{8}\triangle_{S^{2}}^{2}\, \ln^{2}(c(1 - \hat{n} \cdot \tilde{n}))\, +\, \frac{1}{4}\, \triangle_{S^{2}}\, \ln(c(1 - \hat{n} \cdot\tilde{n}))\, \right]\nonumber\\
&=\, C_{rr}^{\textrm{ren}, c = 1} + \pi \ln c\, \triangle_{S^{2}} \delta^{2}(\hat{n}-\tilde{n})\,,
\end{align}
where $c$ is a dimensionless parameter which can also be written as $\mu\, R^{2}$, with $\mu$ being the renormalisation scale. This then makes the scheme dependence of the $C_{rr}^{\textrm{ren}}$ explicit.

Choosing the scheme $c=1$, we then have a new GNS state such that
\begin{align}
\lim_{\delta\, \rightarrow\, 0}\, \omega_{\textrm{free}}^{EM}(E_{T}(W_{\delta}^{<})\, E_{T
}(V_{\delta}^{<}))\, =:\, \omega_{{\cal O}}^{EM}(E_{\p}(v^<) E_{\p}(w^<))\,.
\end{align}
In the next subsection, we will prove that the renormalised boundary kernel $C_{rr}^{\textrm{ren}}$ is not positive. We then restore positivity by choosing a suitable local counter-term  on $\partial{{\cal O}}$.

\subsection{Restoring positivity}

We will now prove that although the r.h.s.\ of eq.~\eqref{omega0etet} is UV-finite, it is not positive.
This can be  seen in the partial wave basis, where in terms of Legendre polynomials we have 
\begin{align}
C_{rr} ^{ren}(l)\, &=\, \frac{1}{\pi}\, \left[\, -\, (l (l+1))^{2}\, \int_{-1}^{1}\dt x\, P_{l}(x)\, \ln^{2}(1-x)\, + \frac{1}{4} l(l+1)\, \int_{-1}^{1}\dt x\, P_{l}(x)\,\ln(1-x)\, \right]\nonumber\\
 &=:\, \frac{1}{\pi}\, \left[\, -\frac{1}{8} (l(l+1))^{2}\, I_{2}(l)\, +\, \frac{1}{4}\, l\, (l+1)\, I_{1}(l)\, \right]\,,
\end{align}
with the second line defining the integrals $I_1(l),I_2(l)$. One finds
\begin{align}
I_{1}(l)\, \coloneqq\, \int_{-1}^{1} \dt x P_{l}(x) \ln(1-x)
=\begin{cases}
   -\, \frac{2}{l(l+1)}\, &\forall\, l > 0\,,\\
   -\, 2 \ln 2\, &\textrm{for}\, l = 0\,.
\end{cases}
\end{align}
This implies that the second term is negative-definite $\forall\, l$.

We will now prove that $I_{2}(l) > 0\, \forall\, l$. Using 
\begin{align}
\int_{-1}^{1} P_{l}(x)\, P_{k}(x) = \delta_{k,l} \frac{2}{2l+1}\,,
\end{align}
we can expand $\ln(1-x)$ into Legendre polynomials with negative-definite coefficients.
It can be immediately seen that
\begin{align}
\ln(1-x)\, =\, (\ln 2 - 1)\, +\, \sum_{l=1}^{\infty}\, \left(-\, \frac{2l+1}{2 l (l+1)}\, \right)\, P_{l}(x)\, =:\, \sum\, \alpha_{l}\, P_{l}(x)\,.
\end{align}
Thus, $\alpha_{l} < 0\, \forall\, l$, and 
\begin{align}
I_{2}(l)\, &=\, \sum_{a < b}\, \alpha_{a}\alpha_{b}\, \int_{-1}^{1}\dt x\, P_{a}(x) P_{b}(x)\, P_{l}(x)\nonumber\\
&=\, {2}\, \sum_{a < b}\, \alpha_{a}\, \alpha_{b}\, \left(C_{0,0,0}^{a,b,l}\right)^2\,,
\end{align}
where $C_{0,0,0}^{a,b,l}$ is the Wigner $3j$-symbol with magnetic quantum numbers being zero. This proves positivity of $I_{2}(l)$, and altogether the negative-definiteness of $C^{ren}_{rr}(l)$ for $l>0$.

Accordingly, to ensure that the extended GNS state $\omega_{0}$ is positive on the edge mode operators, we need to add counter-terms of the form
\begin{align}
\sum_{n=0}^{n_{\rm max}}\, a_{n} \triangle_{S^{2}}^{n}\, \delta^{2}(\hat{n} - \tilde{n}) \,.
\end{align}
At the outset, we choose the renormalisation constant $a_{0}\, =\, 0$ which corresponds to the global constraint 
\begin{align}
C_{rr}^{ren}(l=0)\, =\, 0\,.
\end{align}
This is consistent with the fact that in Maxwell theory $\omega_{0}$ has a null vector which corresponds to the total electric flux on $S^{2}$.

We now need to determine $n_{\rm max}$ and for that we have to find the asymptotic form of $\vert C_{rr}(l)\vert$ as $l\, \rightarrow\, \infty$.\footnote{This suffices since $C_{rr}(l)$ is a monotonically decreasing function of $l$.}
We have
\begin{align}\label{I2lasymp}
I_{2}(l)=\, \int_{-1}^{1} \dt x P_{l}(x) \ln^{2}(1-x)\underset{l \rightarrow\, \infty}{=}\,c \frac{\ln l}{l^{2}} \,,
\end{align}
where $c$ is a $O(1)$-positive number. Hence, 
\begin{align}
\vert C_{rr} ^{ren} (l) \vert\, \underset{l\, \rightarrow\, \infty}{=}\vert c\vert\, l^{2} \ln l \,.
\end{align}
As a result, we see that to ensure positivity for all $l$ we have to choose
\begin{align}
n_{\textrm{max}}\, =\, 2 \,.
\end{align}

We have thus shown that even for pure Maxwell theory, there exists a positive renormalised GNS state in which the subregion algebra including the edge modes is quantised with renormalised correlation functions. We end this appendix with a few remarks. 

\begin{enumerate}
\item The renormalised state supports the centre modes on the boundary of a fixed subregion. Hence, this state breaks spatial translation invariance, as in the case of our longitudinal mode in the main body.
\item The electric centre in pure Maxwell theory is obtained via a ``singular smearing'' of the well-defined transversal electric field operator. It is thus localised sharply on the boundaries and, as a result, its fixed-time correlation function requires UV renormalisation. 
\item It is important to note that the fact that $C^{ren}_{rr}(\vert \hat{n} - \tilde{n} \vert)$ is not a positive kernel is because we are defining a correlation function between the boundary modes defined simply from singular smearings of $E_{T}(W_{\delta}^<)$ where $W\, \in\, C_{c}^{\infty}(T{\cal O})$. However, 
\begin{align}
\lim_{\delta\, \rightarrow\, 0}(\, \omega_{0}(E_{T}(W_{\delta}^{<})\, E_{T}(V_{\delta}^{>})\, )\, =\, 
-\, \int_{S^{2} \times S^{2}}\dt^2\hat n\,\dt^2\tilde n \,v^{<}(\hat{n})\, C_{rr}(R\vert \hat{n} - \tilde{n}\vert)\, w^{>}(\tilde{n}) 
\end{align}
generates a positive kernel $- C^{ren}_{rr}$, and the local counter-terms are no longer required to restore positivity. This is consistent with the fact that the entanglement of modes across $\partial{\cal O}$ features positive correlations. 

\item It is the renormalisation of $\omega_{\rm free}^{EM}$ that leads to the identification of the centre with edge modes. 
\item In pure Maxwell theory, the two-point function of the electric edge modes can be used to obtain the coefficient of the universal log-term in entanglement entropy. This coefficient is precisely the $a$-anomaly in a CFT (such as pure Maxwell theory) and as such, it is scheme independent \cite{David:2022jfd}.

We expect that even in scalar QED, which corresponds to the free field fixed point perturbed by marginal and relevant perturbations, the covariance of $E_{L}^{\rm out}$ is associated with the conformal anomaly. We hope to return to the detailed comparison of the renormalised GNS state and replica method used to obtain entanglement entropy in future works. 
\item The inner product induced on $v(\hat{n})$ by the renormalised state is 
\begin{align}
\vert\vert v \vert\vert^{2}\, =\, \int_{S^{2}\times S^{2}}\dt^2\hat n\,dt^2\tilde n\, \vert v(\hat{n}) \vert^{2}\, \pah{C_{rr}}(\hat{n}-\tilde{n})\,.
\end{align}
As the kernel is invariant under rotations, for the class of functions which satisfy $\partial_{\phi}\, v(\hat{n})\ = 0$ the above norm can be rewritten as
\begin{align}
\vert\vert v \vert\vert^{2}\, =\,\sum_{l} \vert v(l) \vert^{2}\, C_{rr}^{ren}(l)  \sim\, \sum_{\textrm{large } l} l^{4} \vert \tilde{v}(l) \vert^{2} \,.
\end{align}
Thus, the space of functions obtained by completing the space of smooth functions under this norm is defined as
\begin{align}
E(\partial {\cal O})\, \coloneqq\, \Big\{\, \tilde{v}(l)\, \Big\vert\, \sum_{l}\, l^{4}  \vert \tilde{v}(l) \vert^{2} < \infty\, \Big\}\,.
\end{align}
\item The von Neumann algebra completion of the Abelian $C^*$-algebra generated by the radial electric field on the corner $E_{r}(\partial {\cal O})$, is then a Type I algebra generated by 
\begin{align}
\int_{\p{\cal O}}\dt^2\hat n\, E_{r}(R, \hat{n})\, f(\hat{n})\, \qquad\forall\, f\, \in\, E(\partial {\cal O})\,.
\end{align}
\end{enumerate}

\section{Review of generalised Sobolev spaces}\label{app:sobolev}

Sobolev spaces are vector spaces of functions with a notion of ``smoothness'' that is generalised from the usual $C^\infty_c(\mathbb{R}^d)$ one, by being instead defined in terms of norm-convergence of the function under a modified version of the $L^p(\mathbb{R}^d)$-norm, for $1<p<\infty$. We will restrict in what follows to the Hilbert space case of $p=2$. The characterisation proceeds most cleanly in Fourier space. More specifically, we have the following definition \cite{dominguez2018function, booktriebel}
\begin{align}
    H^{s,b}(\mathbb{R}^d)&\coloneqq\{f\in L^2(\mathbb{R}^d)|\;||f||_{H^{s,b}(\mathbb{R}^d)}<\infty\}\,,\\
    ||f||_{H^{s,b}(\mathbb{R}^d)}&\coloneqq\int_{\mathbb{R}^d} d^dk \,(1+|k|^2)^s(1+\ln(1+|k|))^{2b}|\hat{f}(k)|^2\,,
\end{align}
where $s,b\in\mathbb{R}$. We see that setting $s=b=0$ we reduce to the $L^2(\mathbb{R}^d)$ space, after making use of Plancherel's theorem to write the norm in real space. The $H^{s,b}(\mathbb{R}^d)\subsetneq L^2(\mathbb{R}^d)$ $(s,b>0)$ spaces are strict subspaces of $L^2(\mathbb{R}^d)$ by imposing stronger decaying constraints on the behaviour of the functions at high momentum.
The standard Sobolev spaces $H^s(\mathbb{R}^d)$ are those for which $b=0$ and $s\in \mathbb{N}$ \cite{hormander2003analysis1, taylor2011pde1}. They have a straightforward real space interpretation in terms of functions whose derivatives up to order $s$ are square integrable.

We have the dense inclusion relations
\begin{align}
    C^\infty_c(\mathbb{R}^d) \subsetneq H^{s+\epsilon,0}(\mathbb{R}^d)\subsetneq H^{s,b}(\mathbb{R}^d)\subsetneq H^{s,0}(\mathbb{R}^d)\subsetneq L^2(\mathbb{R}^d)\,,
\end{align}
for any $s,b,\epsilon\in \mathbb{R}_+$. All of these spaces are dense in $L^2(\mathbb{R}^d)$. 

The generalised Sobolev spaces are also, in particular, Hilbert subspaces of $L^2(\mathbb{R}^d)$, with inner product
\begin{align}
    \langle f,g\rangle_{H^{s,b}(\mathbb{R}^d)}\coloneqq\int_{\mathbb{R}^d} d^dk \,(1+|k|^2)^s(1+\ln(1+|k|))^{2b}\,\hat{f}(k)\overline{\hat{g}(h)}\,.
\end{align}

\section{Preliminary  positivity discussion of the renormalised kernel}\label{sec_positivity}

In this paper we  have not established that $\omega_0$ is positive on \emph{all} of $\mathcal{A}_C(\Sigma)$, i.e.\ also on algebra elements built from elements of both $\mathcal{A}_{\rm phys}(\Sigma)$ and $\mathcal{A}_{\rm out}(\mathcal{O})$. Therefore, in this appendix, we provide a preliminary discussion of the type of technical questions to be answered, along with a few preliminary results. We discuss a \emph{necessary} condition for the existence of a renormalisation scheme used to define the quasi-free state $\omega_{0}$ on the enlarged  algebra ${\cal A}_{C}(\Sigma)$ such that $\omega_{0}$ is positive semi-definite as long as its restriction to ${\cal A}_{\text{phys}}({\Sigma})$ is positive. Before stating the main claim of this section, we recall what we have established so far.

We have chosen $\omega_F^R$ on $\mathcal {A}_{\textrm{phys}}({\Sigma})$ as the pull back of the quasi-free Fock vacuum in spacetime. As we explicitly showed in section \ref{subsec:restoring_positivity}, there exists a renormalisation scheme in which $\omega_F^R(\rho(f)^{\dagger}\rho(f))$, $ \omega_F^R(J(V)^{\dagger}J(V))$, $\omega_F^R(\l |\psi|^2 (f) \r ^\dagger |\psi| ^2(f))$ are positive. We have further assumed that $\omega_F^R$ is positive on the entire ${\cal A}_{\textrm{phys}}(\Sigma)$. The proof of positivity on any element in ${\cal A}_{\textrm{phys}}(\Sigma)$ is beyond the scope of this paper, because the state is not Gaussian on all generators of the algebra ${\cal A}_{\textrm{phys}}({\Sigma})$.

As we have shown in section \ref{sec:GNS A_C}, the extension of $\omega_F^R$ to $\omega_{0}$ to encompass $\mathcal{A}_C(\Sigma)$ (or $\mathcal{A}_C(\mathcal{O})$) required a further renormalisation which led to a boundary contribution in $\omega_{0}(E_{L}^{\rm out}(V)\, E_{L}^{\rm out}(W))$. Thus, even if the pull-back of the GNS Fock state to $\mathcal{A}_{\rm phys}(\Sigma)$ is renormalised and positive, its extension to the extended algebra $\mathcal{A}_C(\Sigma)=\mathcal{A}_{\rm phys}(\Sigma)\vee\mathcal{A}_{\rm out}(\mathcal{O})$  for a fixed region $\mathcal{O}$ is not guaranteed to be positive.

We now give arguments in support of a necessary condition for $\omega_{0}$ to be positive assuming its positivity on, $\mathcal{A}_{\rm phys}(\Sigma)$ (as we assumed above). We restrict our analysis of the positivity of $\omega _0$ to elements of ${\cal A}_{C}(\Sigma)$ which are linear in $E_{L}^{\textrm{out}}$. That is, we consider elements that can be written as
\begin{align}
A_{\lambda}(V, Q)\, \coloneqq\, \lambda\, E_{L}^{\rm out}(V)\, +\, Q, \quad Q\in \mathcal{A}_{\text{phys}}(\Sigma), \quad \lambda \in \mathbb{C}\,.
\end{align}
Under the assumption $\omega_{0}(Q^{\dagger}Q)\, \geq\, 0$, $\forall\, Q\, \in {\cal A}_{\text{phys}}({\Sigma})$, the positivity condition 
\begin{equation}
    \omega_{0}(A_{\lambda}(V, Q)^{\dagger}\, A_{\lambda}(V, Q))\geq\, 0, \quad \nonumber\, \forall\, \lambda  \in \mathbb{C}
\end{equation}
is, via the Cauchy-Schwarz inequality, equivalent to
\begin{align}\label{eq:gnsposnec}
\vert \omega_{0}(Q^{\dagger}\, E_{L}^{\rm out}(V)) \vert^{2}\, \leq\, \omega_{0}(Q^{\dagger}Q)\, \omega_{0}(E_{L}^{\rm out}(V)^{\dagger}\, E_{L}^{\rm out}(V))\,.
\end{align}

We will now argue why this is true for any $Q \in \mathcal{A} _{\text{phys}} (\Sigma)$.
Let $Q\, \coloneqq\, {\cal Q}({\bf T})$, where ${\bf T}$ is the smearing tensor whose rank depends on the tensorial nature of the operator valued distribution ${\cal Q}(x)$. Then  
\begin{align}
\omega_{0}\left((E_{L}^{\rm out}(V))^\dag\,Q\right)\, &=\, \int_{{\cal O}}\, \dt^{3}x\, {\bf T}(x)\, \int_{{\cal O}^{c}} \dt^{3}y\, F_{V}(y)\, \omega_{0}(\, {\cal Q}(x)\, \rho(y)\, )\nonumber\\
&=\, \int\, \frac{\dt^{3}k}{(2\pi)^{3}}\, \tilde{\bf T}(k)\, C_{{\cal Q}\rho}(k)\, \tilde{F}_{V}(-k)\,.
\end{align}
Thus, eq.~\eqref{eq:gnsposnec} will be satisfied if and only if
\begin{align}\label{eq:positiv_eq}
\int_{k, k^{\prime}} \vert \tilde{\bf T}(k) \vert^{2} C_{{\cal Q}{\cal Q}}(k)\, \vert \tilde{F}_{V}(k^{\prime})\vert^{2}\, C ^{\rm ren}_{\rho\rho}(k^{\prime})\geq \int _{k,\, k^{\prime}} \tilde{T}(k) \tilde{T}(-k^{\prime})\, \tilde{F}_{V}(-k) \tilde{F}_{V}(k^{\prime})\, C^{\rm ren}_{{\cal Q}\rho}(k)\, C^{\rm ren}_{{\cal Q}\rho}(-k^{\prime})\,,
\end{align}
where all the two-point kernels have been renormalised. 

If $\mathcal{Q}$ is a polynomial in $E_T$ and $B$, then  
\begin{align}
C_{Q\rho}\, \sim\, C_{Q}\, C_{\rho}\, =\, 0,
\end{align}
since in $\omega_{0}$ correlation functions involving matter operators and Maxwell fields factorise. Furthermore, consider $\mathcal{Q}$ of the from $\rho ^{m} \, J_i ^n ( |\psi |^2 )^k$, for arbitrary $k$, and any $m$ and $n$ such that $m+n$ is even.
In that case because $\rho $ and $J_i$ are $C$-odd, and $|\psi|^2$ is $C$-even at equal times we must have $\omega _0 \l \rho ^{2m+1} \, J_i ^n ( |\psi |^2 )^k \r =0 $, meaning that for these types of operators we also have $C_{\mathcal{Q} \rho} =0 $.

For all other operators $\mathcal{Q}$ for which $C^{\rm ren}_{{\cal Q}\rho}(k)$ does not vanish, more work needs to be done to establish rigorously the validity of \eqref{eq:positiv_eq}.

Finally, we note that because of the connected components of the $n$-point function of  $ E_{L}^{\rm out}(V)$, positivity of $\omega _0$ on operators linear in $ E_{L}^{\rm out}(V)$ does not guarantee positivity for the following types of operators:
\begin{align}
D \, \coloneqq\, \sum_{n=0}^{N}\, c_{n}\, \big(\, E_{L}^{\rm out}({V_{1}})\, \dots\, E_{L}^{\rm out}(V_{n})\, \big)\, Q\,.
\end{align}
To show that renormalisation and an appropriate choice of counter-terms would establish positivity of $\omega_{0}(D^{\dagger}\, D)$ is beyond the scope of this paper.\footnote{We suspect that  the finiteness of the connected $n$-point functions $\omega_{0}(\prod_{i=1}^{n} \rho(f_{i}))$ should render a detailed positivity analysis of $\omega_0$ feasible; however, we leave this analysis for elsewhere.}

\bibliographystyle{jhep}
\bibliography{ref}

\end{document}